\documentclass[aps,prd,reprint,superscriptaddress,amsmath,amssymb,nofootinbib]{revtex4-1}
\usepackage[final]{graphicx}
\usepackage{verbatim}
\usepackage{color}

\begin{document}

\title{The HAYSTAC Axion Search Analysis Procedure}

\author{B.~M.~Brubaker}\email{benjamin.brubaker@yale.edu}
\author{L.~Zhong}
\author{S.~K.~Lamoreaux}
\affiliation{Department of Physics, Yale University, New Haven, Connecticut 06511, USA}
\author{K.~W.~Lehnert}
\affiliation{JILA and the Department of Physics, University of Colorado and National Institute of Standards and Technology, Boulder, Colorado 80309, USA}
\author{K.~A.~\surname{van Bibber}}
\affiliation{Department of Nuclear Engineering, University of California Berkeley, Berkeley, California 94720, USA}

\date{\today}

\begin{abstract}
We describe in detail the analysis procedure used to derive the first limits from the Haloscope at Yale Sensitive to Axion CDM (HAYSTAC), a microwave cavity search for cold dark matter (CDM) axions with masses above $20~\mu$eV. We have introduced several significant innovations to the axion search analysis pioneered by the Axion Dark Matter eXperiment (ADMX), including optimal filtering of the individual power spectra that constitute the axion search dataset and a consistent maximum likelihood procedure for combining and rebinning these spectra. These innovations enable us to obtain the axion-photon coupling $|g_\gamma|$ excluded at any desired confidence level directly from the statistics of the combined data.
\end{abstract}

\maketitle
\section{Introduction}\label{sec:intro}
The axion~\cite{PQ1977a,*PQ1977b,weinberg1978,*wilczek1978} is a hypothetical pseudoscalar field originally postulated to explain the absence of CP violation in the theory of quantum chromodynamics (QCD); light axions ($m_a \lesssim 1$~meV) have since been recognized as attractive candidates for a microscopic description of cold dark matter (CDM)~\cite{pww1983,*as1983,*df1983}. Axions constituting our galactic halo with masses in the range $1 \lesssim m_a \lesssim 50~\mu$eV may be detected via their resonant conversion into nearly monochromatic microwave photons in an ``axion haloscope:'' a high-$Q$ cryogenic cavity immersed in a strong magnetic field and coupled to a low-noise receiver~\cite{sikivie1985}. All haloscope detectors to date have used spectrally resolved coherent receivers, in which an axion signal would appear as an extremely weak but spectrally sharp persistent power excess over the noise floor at frequency $\nu_a = m_ac^2/h$. In practice, the axion mass is unknown, so the cavity must be tunable. It is typical to assume that the halo axions are virialized, in which case the spectral distribution of the conversion power is inherited from the halo's kinetic energy distribution, with fractional width of order $\left<v^2\right>/c^2 \sim 10^{-6}$. 

One example of a haloscope detector is the Haloscope At Yale Sensitive To Axion CDM (HAYSTAC), which recently demonstrated cosmological sensitivity to halo axions with $m_a > 20~\mu$eV for the first time~\cite{PRL2017}. The HAYSTAC detector is described in detail in Ref.~\cite{NIM2017}; the purpose of the present paper is to provide a detailed pedagogical account of the analysis procedure used to generate the exclusion limit reported in Ref.~\cite{PRL2017}. The basic framework of our analysis owes much to the procedure developed by the Axion Dark Matter eXperiment (ADMX)~\cite{ADMX2001}; we have introduced a number of refinements that collectively enable us to obtain the relationship between search sensitivity and confidence directly from the statistics of the combined data without recourse to Monte Carlo. These innovations can easily be adapted to the analysis of data from other haloscope detectors such as ADMX and CULTASK~\cite{CAPP2016} and perhaps also to ``dielectric haloscopes'' like MADMAX~\cite{MADMAX2017} and resonant hidden photon detectors like DM Radio~\cite{chaudhuri2015}. 

The remainder of the paper is organized as follows. Section~\ref{sec:experiment} briefly reviews the aspects of the HAYSTAC detector most relevant to understanding the analysis and describes the axion search data set. Section~\ref{sec:analysis_overview} presents a big-picture overview of the analysis procedure, whose distinct stages are discussed in greater detail in Sec.~\ref{sec:cuts} -- \ref{sec:rescan}. In Sec.~\ref{sec:conclusion} we present our limit and conclude with a summary of our main innovations. Various tangential topics that are nonetheless important to a full understanding of the analysis procedure are discussed in appendices. 

\section{Experiment}\label{sec:experiment}
\subsection{Detector}\label{sub:detector}
HAYSTAC is sited at the Wright Laboratory of Yale University, and housed within a cryogen-free dilution refrigerator integrated with a $9$~T superconducting solenoid. The cavity hangs in the center of the magnet bore from a gold-plated copper gantry anchored to the dilution refrigerator's mixing chamber plate at temperature $T_C =127$~mK.

Our current cavity is a 2~L copper-plated stainless cylinder whose axion-sensitive TM$_{010}$ mode may be tuned over the range $3.6< \nu_c < 5.8$~GHz via rotation of an off-axis copper rod occupying 25\% of the cavity volume. We can also independently adjust the insertion into the cavity of a thin dielectric shaft and a coaxial antenna, used to fine-tune the mode's frequency and control its coupling to the receiver, respectively.

The most notable feature of the HAYSTAC receiver is its use of a tunable Josephson parametric amplifier (JPA) as a preamplifier. The JPA is essentially a nonlinear $LC$ circuit that exhibits parametric gain when driven with a sufficiently strong microwave pump tone near its resonant frequency. For a small signal detuned completely to one side of the pump, a JPA acts like a phase-insensitive linear amplifier whose added noise is close to the fundamental limits imposed by quantum mechanics~\cite{caves1982}. Our current JPA may be tuned over the range 4.5--6.4 GHz via application of a small DC magnetic flux bias. 

The first element in the receiver signal path is a microwave switch that allows us to calibrate the cavity noise by comparison with a known blackbody source at $T_H = 775$~mK, the temperature of the dilution refrigerator's still plate. Signals at the JPA output are amplified further at 4~K and room temperature, and downconverted to an intermediate frequency (IF) band using an IQ mixer whose local oscillator (LO) is set 780~kHz above the cavity resonance. After further amplification and filtering the IF signals are digitized at 25~MS/s.

For the first HAYSTAC data run we scanned over the range 5.7--5.8~GHz in two continuous passes followed by several shorter scans to compensate for nonuniform tuning. This nonuniformity was a consequence of fine tuning with the dielectric shaft and moving the copper rod less frequently to mitigate imperfection in the rotary tuning system.

\subsection{Axion search data}\label{sub:data}
A haloscope axion search consists of a sequence of iterations separated by discrete tuning steps, with a cavity noise measurement of duration $\tau$ and various auxiliary measurements at each iteration.\footnote{$\tau$ is the axion-sensitive averaging time, not the total data collection time including inefficiency.} We construct and average power spectra in parallel with acquisition of the cavity noise timestream data from the HAYSTAC detector, so only a single heavily averaged power spectrum is written to disk at each iteration. The auxiliary data consists of vector network analyzer (VNA) measurements of the cavity mode and JPA gain profile at each step and periodic $Y$-factor measurements to calibrate the noise; temperatures and pressures at various points in the cryogenic system are also logged independently. The purpose of the auxiliary data is to characterize detector parameters that can vary during the run, both to define data quality cuts (Sec.~\ref{sub:badscans}) and optimally rescale spectra (Sec.~\ref{sub:rescale}).

\begin{figure}[t]
\includegraphics[width=0.5\textwidth]{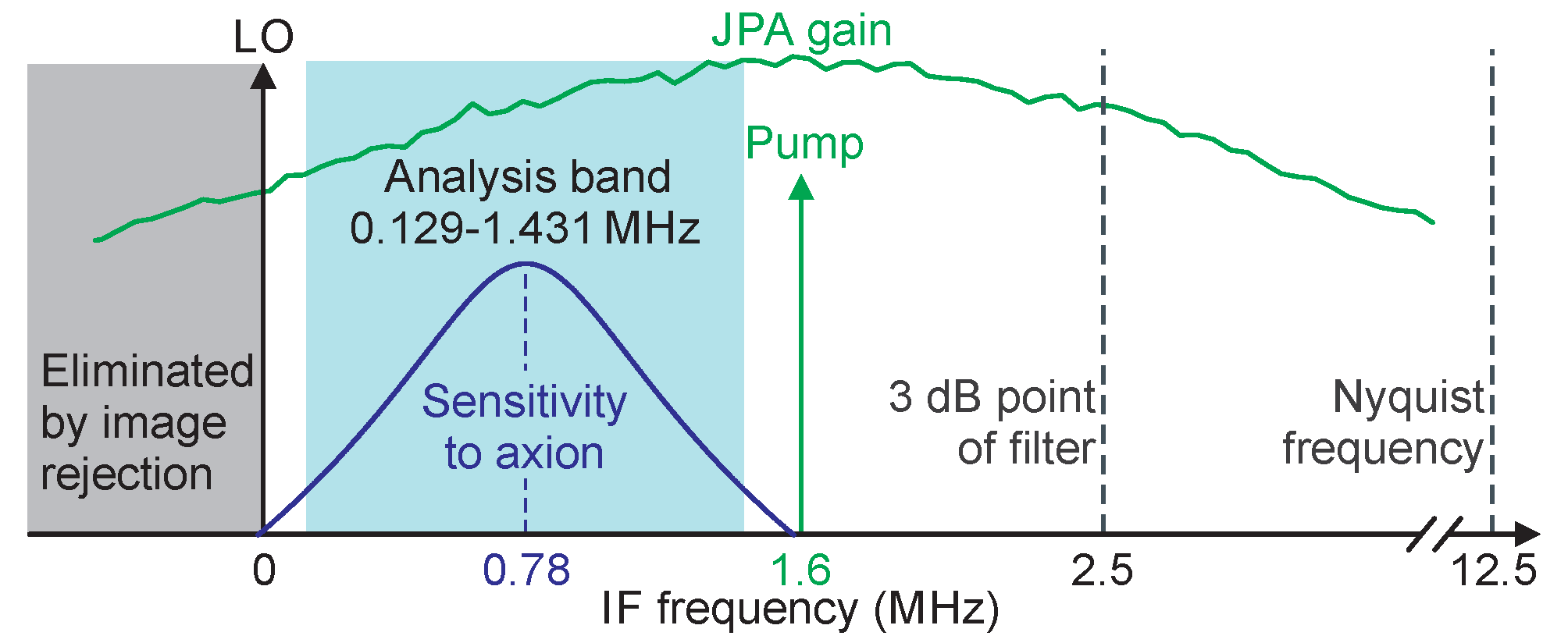}
\caption{\label{fig:layout} Schematic layout illustrating regions of interest in a typical HAYSTAC power spectrum, reprinted from Ref.~\cite{NIM2017} with some inessential elements omitted for clarity. Note that increasing IF frequency corresponds to decreasing RF frequency because the LO is at higher frequency than the Fourier components of interest in each spectrum. The JPA gain profile and Lorentzian cavity mode profile are plotted using real data and a fit to real data, respectively. Both plots have logarithmic y-axes; the relative vertical scale is not meaningful.}
\end{figure}

The principal data from the first run consisted of 6936 power spectra with bin width $\Delta\nu_b = 100$~Hz, each obtained from $\tau = 15$~minutes of averaging. We acquired the first 2244 spectra in winter 2016, and the rest in summer 2016 following a power outage that damaged the system and disrupted operations. Filters limit the usable IF bandwidth of each spectrum to roughly 2.5~MHz, well below the 12.5~MHz Nyquist frequency. Fig.~\ref{fig:layout} shows a schematic layout of the regions of interest in each spectrum. 

It may be useful at this point to summarize the relations between the various frequency scales that will play a role in our subsequent discussions. When appropriately biased, the JPA has about 21 dB peak gain in a bandwidth $\Delta\nu_{\text{JPA}} \approx 2.3$~MHz centered on the pump tone. $\Delta\nu_{\text{JPA}}$ is larger than the typical cavity linewidth $\Delta\nu_c \approx 500$~kHz, which ensures that the total noise referred to the JPA input remains low over all frequencies of interest in each spectrum. The cavity linewidth $\Delta\nu_c$, which sets the width of the axion-sensitive region in each spectrum, is in turn much larger than the typical axion linewidth $\Delta\nu_a \sim \nu_a\left<v^2\right>/c^2 \approx 5$~kHz for a virialized axion in the initial HAYSTAC scan range. Finally, $\Delta\nu_a \gg \Delta\nu_b$, which helps us reject spurious single-bin features (Sec.~\ref{sub:badbins}) and take the axion lineshape into account in our analysis (Sec.~\ref{sub:grand_spectrum}). In principle fine frequency resolution also enables us to search for non-virialized structure in the axion energy spectrum (see Sec.~\ref{sub:lineshape}); for the present analysis, we restrict our focus to virialized axions.

As illustrated in Fig.~\ref{fig:layout}, the dependence of the haloscope signal power on the detuning $\delta\nu_a = \nu_c - \nu_a$ is Lorentzian with FWHM $\Delta\nu_c$; at $\delta\nu_a = \Delta\nu_c$, the signal power is thus down from its peak value by a factor of 5. Ultimately the quantity we care about is the signal-to-noise ratio (SNR) throughout the tuning range, to which individual spectra will contribute in quadrature [See Eq.~\eqref{eq:snr_c}]. Including bins further than $\Delta\nu_c$ from the cavity mode in each spectrum in our analysis would improve the SNR only very marginally. Thus we can restrict our focus to an \textit{analysis band} of full width $\approx2\Delta\nu_c$ centered on the cavity mode in each spectrum without appreciably affecting our sensitivity.

During the data run we fit the TM$_{010}$ resonance in transmission after each tuning step, and set the LO frequency by adding 780~kHz to the measured mode frequency and rounding to the nearest 100~Hz.\footnote{Coercing the LO frequency to the nearest 100~Hz ensures that the bin boundaries in different spectra are always aligned. As a result the analysis band is not exactly centered on $\nu_c$ in each spectrum, but the maximum offset is always $<\Delta\nu_b$.} We then set the JPA pump frequency 1.6~MHz below the LO. Setting the JPA pump frequency at a fixed offset from the LO instead of the cavity resonance ensures that the $1/\Delta\nu_b = 10$~ms integration time of each subspectrum is an integer number of periods at the pump frequency, and thus minimizes spreading of the pump power throughout the spectrum.\footnote{Sinusoidal signals of arbitrary frequency will generally not be confined to single bins in the spectrum because we do not apply a window function to the timestream data in the process of computing the power spectrum of each 10~ms record. The ``rectangular window'' (equivalent to not windowing at all) is the correct choice for a haloscope search as it has the smallest equivalent noise bandwidth. Given the constraint of the rectangular window, a small bin width $\Delta\nu_b\ll\Delta\nu_a$ also ensures that distortion of the axion signal lineshape by the FFT is negligible.} The analysis band is defined as the set of bins between 129 kHz and 1.431 MHz in each spectrum; this is a conservative choice that accounts for variation of $\Delta\nu_c$ over the scan range. 

\section{Analysis overview}\label{sec:analysis_overview}
The goal of a haloscope analysis is to combine a set of overlapping axion-sensitive power spectra to produce a single spectrum that optimizes the SNR throughout the scan range. Put another way, if there exists an axion with $\nu_a$ within the scan range and photon coupling $|g_\gamma|$ sufficiently large, the conversion power should almost always result in a large excess relative to noise in the bin corresponding to $\nu_a$ in the final spectrum. The minimum coupling $|g^\text{min}_\gamma|$ for which this statement will hold is set primarily by the detector design, but we must still understand how much the analysis procedure degrades this intrinsic sensitivity. The analysis should ideally allow us to write down an explicit expression for $|g^\text{min}_\gamma|$ as a function of the desired confidence level (which quantifies the ``almost always'' in the informal description above).

When we consider how best to combine spectra, one issue that immediately arises is that the shape and normalization of each spectrum depend both on quantities that affect the SNR (e.g., the system noise temperature), and quantities that do not (e.g., the net gain of the receiver chain, including the frequency-dependent attenuation of all room-temperature components). Rather than try to tease apart the relevant and irrelevant contributions, we can remove the spectral baseline entirely using a fit or filter, then rescale the resulting spectra using parameters extracted from the auxiliary data. In this way we can properly account for variation in sensitivity among spectra and within each spectrum. 

After baseline removal the bins in each spectrum may be regarded as samples drawn from a single Gaussian distribution.\footnote{The spectra are approximately Gaussian because each spectrum saved to disk is the average of a large number of subspectra, so the bin variance is much smaller than the mean squared bin amplitude. This point is discussed further in Sec.~\ref{sub:stats}.} This is a convenient reference point for understanding the effects of subsequent processing on the statistics of the spectra. Of course, we need to make sure that the baseline removal procedure does not fit out bumps in the spectra on frequency scales comparable to $\Delta\nu_a$, or we will significantly degrade the axion search sensitivity. This point suggests that baseline removal is more fruitfully regarded as a problem in filter design than a fitting problem, as it has been described in previous ADMX analyses. The filter perspective will turn out to be quite useful in understanding the statistics of the spectra. 

The task of removing the spectral baseline without appreciably attenuating any axion signal is made tractable by their different characteristic spectral scales, or in other words by $\Delta\nu_c \gg \Delta\nu_a$. It is worth noting that this inequality is ultimately a consequence of the difficulty of achieving high cavity $Q$ factors with normal metals at GHz frequencies; a detector with higher cavity $Q$ and thus $\Delta\nu_c \approx \Delta\nu_a$ would in principle be more sensitive. Because such a detector has yet to be built, we can exploit the fact that $\Delta\nu_c \gg \Delta\nu_a$ where it simplifies the analysis. 

Because our spectra have $\Delta\nu_b \ll \Delta\nu_a$, the analysis procedure will generally involve taking appropriately weighted sums both ``vertically'' (i.e., combining IF bins from different spectra corresponding to the same RF bin) and ``horizontally'' (i.e., combining adjacent bins in the same spectrum). One of the main innovations of our analysis procedure is that we use the same maximum likelihood principle to obtain the optimal weights in both cases. Various statistical subtleties arise in the latter case because nearby bins in the same spectrum can be correlated. We will demonstrate below that we understand the origin of these correlations sufficiently well to obtain the relationship between  $|g^\text{min}_\gamma|$ and the confidence level from the statistics of the combined data, rather than from Monte Carlo as in previous ADMX analyses.

In the preceding paragraphs we have emphasized what we regard as the main themes of this paper, which may be helpful to keep in mind as we work through the details. For ease of reference, we have outlined the steps of our procedure below, and indicated the section of the paper in which each step is discussed more thoroughly.
\begin{enumerate}
\item Use the auxiliary data to identify spectra that appear to be compromised and cut them from further analysis (Sec.~\ref{sub:badscans}).
\item Average the remaining \textit{raw spectra} together aligned according to IF frequency to identify compromised IF bins and cut them from further analysis (Sec.~\ref{sub:badbins}). This procedure also yields an estimate of the average shape of the spectral baseline in the analysis band.
\item Normalize the analysis band in each raw spectrum to the average baseline, then use a Savitzky-Golay (SG) filter to remove the remaining spectral structure in each \textit{normalized spectrum} (Sec.~\ref{sub:sg_filter}). Then subtract 1 from each spectrum to obtain a set of dimensionless \textit{processed spectra} described by a single Gaussian distribution (Sec.~\ref{sub:stats}).
\item Multiply each processed spectrum by the average noise power per bin and divide by the Lorentzian axion conversion power profile to obtain a set of \textit{rescaled spectra} (Sec.~\ref{sub:rescale}). Construct a single \textit{combined spectrum} across the whole scan range by taking an optimally weighted sum of all the rescaled spectra (Sec.~\ref{sub:combine}).
\item Rebin the combined spectrum via a straightforward extension of the optimal weighted sum from the previous step to non-overlapping sets of adjacent combined spectrum bins (Sec.~\ref{sub:rebinned_spectrum}). Then, taking into account the expected axion lineshape (Sec.~\ref{sub:lineshape}), construct the \textit{grand spectrum} by adding an optimally weighted sum of adjacent bins to each bin in the \textit{rebinned spectrum} (Sec.~\ref{sub:grand_spectrum}).
\item After correcting for the effects of the SG filter on both the statistics of the grand spectrum (Sec.~\ref{sub:correlations}) and the SNR (Sec.~\ref{sub:axion_atten}), set a threshold $\Theta$ for which some desired fraction of axion signals with a given SNR would result in excess power $>\Theta$. Then flag all bins with excess power larger than $\Theta$ as rescan candidates (Sec.~\ref{sub:target_confidence}).
\item Acquire sufficient data around each rescan candidate to reproduce the sensitivity at that frequency in the original grand spectrum (Sec.~\ref{sub:rescan_daq}). Follow the procedure above, with a few minor differences, to construct a grand spectrum for the rescan data, and determine if any candidate exceeds the corresponding threshold (Sec.~\ref{sub:rescan_analysis}). If no candidate exceeds the second threshold, the corrected SNR obtained in step 6 sets the exclusion limit. Any persistent candidates can be interrogated manually.
\end{enumerate}
A great deal of notation is introduced in the sections to follow; we have attempted to simplify it wherever possible by adopting consistent notational conventions. The notation used throughout the paper is summarized in Appendix~\ref{app:notation} for ease of reference.

\section{Data quality cuts}\label{sec:cuts}
\subsection{Cuts on spectra}\label{sub:badscans}
Our first task is to flag and cut any spectra whose sensitivity to axion conversion we cannot reliably calculate, due to e.g., large changes in the TM$_{010}$ mode frequency $\nu_c$ or the JPA gain during a noise measurement. We had reason to anticipate both of these effects in the first HAYSTAC data run: imperfections in the rotary tuning system noted in Sec.~\ref{sub:detector} resulted in a slow drift of $\nu_c$ following actuation of the tuning rod, and the JPA gain is very sensitive to changes in the local magnetic flux. 

We sought to mitigate both gain and mode frequency drifts in the design of the data acquisition procedure (for example, by controlling the JPA's flux bias with feedback as described in Ref.~\cite{NIM2017}). However, the mode frequency still occasionally drifts sufficiently far during a single noise measurement to systematically distort the subsequent weighting of the spectrum by the Lorentzian profile of the cavity mode (see Sec.~\ref{sub:rescale}). Likewise, the flux occasionally drifts sufficiently far that the feedback system is unable to correct for it; the average JPA gain in such iterations is reduced and thus the input-referred noise is systematically higher than what we infer from periodic \textit{in situ} noise calibrations.

Cutting measurements compromised by mode frequency drift is straightforward, because we make VNA measurements of the cavity mode in transmission both before and after the cavity noise measurement at each iteration during the data run. Our analysis routine fits both measurements to Lorentzians and cuts iterations exceeding the conservative threshold $|\nu_{c1}-\nu_{c2}|>60~\text{kHz}\approx\Delta\nu_c/10$ from subsequent analysis. 

We flag iterations compromised by gain drifts using the spectra themselves. The average level of each spectrum in a 100~kHz window close to the JPA pump is a good proxy for the average JPA gain during the noise measurement, though of course it will also reflect other changes in the net receiver gain. Another measure of the average gain accessible in the spectrum is the weak CW tone used to provide a signal for our flux feedback system. We set thresholds for both measures of the average JPA gain empirically to separate obvious outliers from the normal variation among spectra. In both cases, the thresholds were approximately 1~dB below the typical power averaged across all spectra.\footnote{These thresholds are consistent with independent measurements indicating that flux feedback holds the JPA gain constant to within 10\% on timescales comparable to $\tau$. Gain fluctuations during a cavity noise measurement will cause the normalization of each 10~ms subspectrum averaged by the \textit{in situ} processing code to differ, but this variation is correlated across all the bins in each subspectrum; it affects the precision with which we can measure the mean noise power, but not the variance of the noise power within each spectrum, which is the quantity that determines our sensitivity to excess power on small spectral scales $\Delta\nu_a \ll \Delta\nu_c$. Thus absolute gain stability is not a critical parameter for haloscope experiments. At our operating gain, the effect of such small fluctuations on the system noise temperature is small compared to the uncertainty.}

We also scanned the rest of the auxiliary data for any other anomalies that might motivate a cut, and observed a narrow ($\approx60$~kHz) notch around 5.7046~GHz superimposed on measurements of the cavity response in transmission and reflection near this frequency. The absence of any analogous feature in the corresponding JPA gain profiles indicates that the notch originates in the cavity, most likely due to the TM$_{010}$ mode crossing with an ``intruder'' TE or TEM mode practically uncoupled to our antenna. The observation that the precise notch frequency depends on the insertion depth of the dielectric tuning shaft supports this interpretation. Because we used the dielectric shaft for fine tuning in our first data run, the notch frequency appeared to wander back and forth over a range of a few hundred kHz. 

We noticed that the notch was also visible in the spectra around the same frequency, which suggests that the effective temperature of the intruder mode was lower than that of the TM$_{010}$ mode.\footnote{As discussed in Ref.~\cite{NIM2017}, the TM$_{010}$ mode temperature was actually higher than the fridge temperature during our first data run due to a poor thermal link to the copper tuning rod.} All of these measurements collectively indicate that our basic assumption of the axion interacting with a single cavity mode fails around the intruder mode, and neither the VNA measurements of the cavity nor noise calibrations are likely to be reliable here. To be conservative, we simply cut all spectra containing any sign of the intruder mode.

Other auxiliary data (such as the JPA-off receiver gain measurement at each iteration and the fridge temperature records) did not not prompt us to define additional cuts. Overall, of the 6936 spectra obtained during our first data run, we cut 170 from the subsequent analysis, of which 128 were cut in connection with the intruder mode. Of the remaining 42 spectra, 33 were cut because of JPA gain drifts, and 9 because of mode frequency drifts.

\subsection{Cuts on IF bins}\label{sub:badbins}
Narrowband interference can contaminate individual bins in spectra that are otherwise sensitive to axion conversion. Insofar as the intrinsic linewidth of these interference features is $\ll\Delta\nu_b$, a smaller bin width $\Delta\nu_b$ helps reduce the number of contaminated bins that we fail to flag, whose collective effect is to distort the statistics of the spectra.

It is useful to distinguish IF interference (resulting in excess power in the same bins in each spectrum) from RF interference (which would appear to propagate through spectra from adjacent tuning steps). RF interference is more insidious in that it can mimic an axion signal, and small excesses will be hard to flag until we have already combined the contributing spectra. Empirically, all of the most prominent sharp features in HAYSTAC power spectra are due to IF interference.

The various IF features we observe have no single common origin. Some prominent features we eliminated during detector commissioning were associated with ground loops, others with switching power supplies in stepper motor drivers and other room-temperature electronics. Other features only appear when the system is cold, suggesting that cryocooler motors may be responsible. Single-bin IF features can also arise from small RF signals at fixed detuning from the LO or pump tones.

We flag the ``bad bins'' contaminated by IF interference using the following procedure. First, we divide the set of spectra (ordered chronologically) into three approximately equally sized groups. We truncate each spectrum to the analysis band plus $W=500$~bins (50~kHz) on either side. We then average all truncated spectra within each group aligned according to IF frequency; this averaging reveals many sharp features due to IF interference too small to be visible above the noise floor of individual spectra. We apply an SG filter with polynomial degree $d=10$ and half-width $W$ to the averaged spectrum to obtain an estimate of the spectral baseline. The SG filter is described in more detail in Sec.~\ref{sub:sg_filter}; for our present purposes it is sufficient to regard it as a low-pass filter with a very flat passband (i.e., it perfectly preserves features on large spectral scales).

Dividing the averaged spectrum by the SG filter output and subtracting 1 produces a spectrum whose statistics (in the absence of IF interference) are Gaussian, with mean 0 and standard deviation $\sigma^{\text{IF}}=(M_{\text{IF}}\Delta\nu_b\tau)^{-1/2}$, where $M_{\text{IF}}\approx2200$ is the number of spectra in the group.\footnote{The procedure used here to flag IF interference is similar to the baseline removal procedure described in Sec.~\ref{sec:baseline} with a few key differences. Here the SG filter is applied to a spectrum that is more heavily averaged by a factor of $M_{\text{IF}}$, and we do not divide out the average shape of the spectrum before applying the SG filter. Both effects imply that the polynomial degree $d$ of the SG filter must be higher here than in the main analysis.} The most obvious effect of IF interference is to produce a surplus of bins with large positive power excess. We flag all bins that exceed a threshold value of $4.5\sigma^{\text{IF}}$; in the 14020 bins of the truncated spectrum, we expect on average only 0.05~bins exceeding this threshold due to statistical fluctuations. As noted in Sec.~\ref{sub:data}, the fact that we do not apply any windowing in the construction of HAYSTAC power spectra implies that the excess power associated with narrowband IF interference will not be entirely confined to isolated bins. To be conservative, for every set of contiguous bins exceeding the threshold, we add the 3 adjacent bins on either side to the list of bad bins. Empirically, while many features due to IF interference are indeed quite sharp, others consist of $\sim30$ consecutive bins exceeding the threshold. Averaging smaller numbers of adjacent spectra reveals that these broader features are the result of narrow IF peaks that wander back and forth across a range of a few kHz over the course of the data run. 

\begin{figure*}[t]
\includegraphics[width=1.0\textwidth]{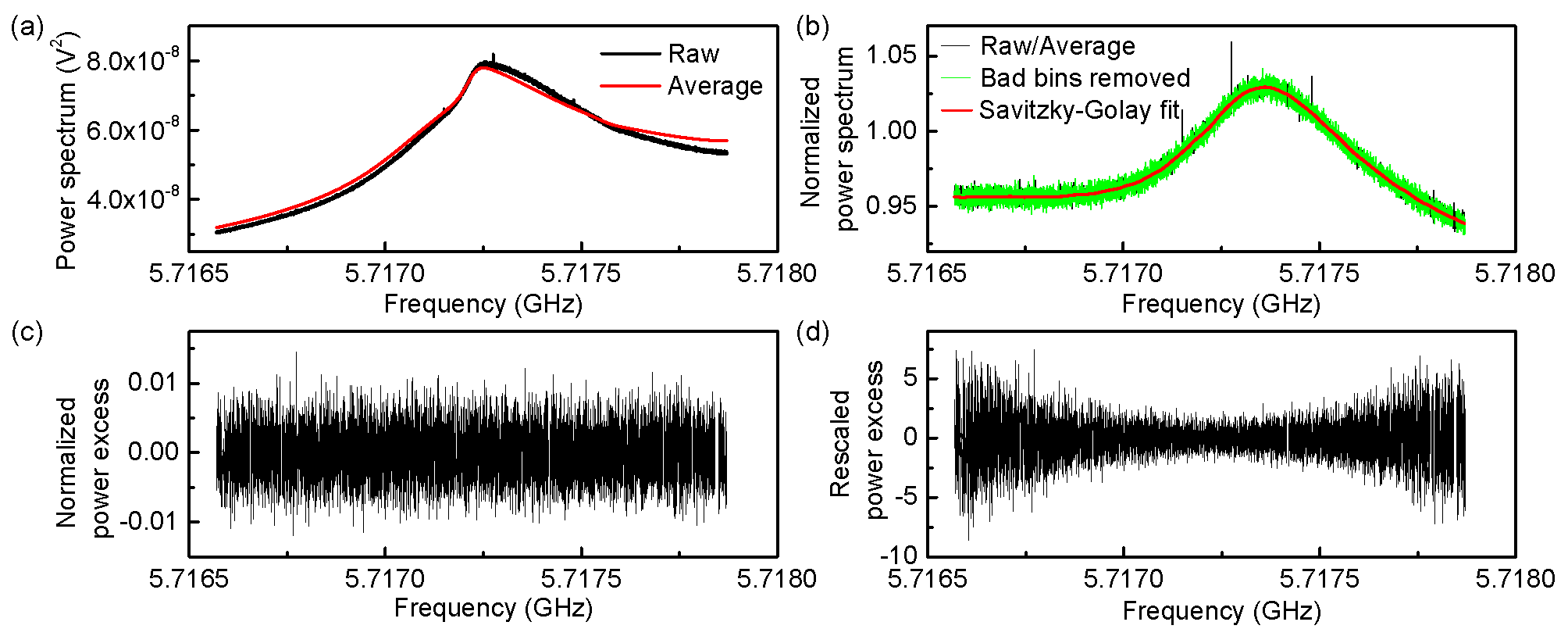}
\caption{\label{fig:sample} The analysis band in a representative power spectrum at various stages of processing. (a) Black: the raw spectrum, whose shape is determined by the cavity noise spectrum and the net gain of the receiver. Red: the average baseline, which is the output of a Savitzky-Golay filter applied to the average of many such spectra (Sec.~\ref{sub:badbins}). (b) Black: the normalized spectrum obtained by dividing the raw spectrum by the average baseline. Green: the same spectrum after removing bins contaminated by IF interference. Red: the Savitzky-Golay fit to the residual baseline of this spectrum (Sec.~\ref{sub:sg_filter}). (c) The processed spectrum obtained by dividing the normalized spectrum by the Savitzky-Golay fit and subtracting 1 (Sec.~\ref{sub:stats}). Gaps in the spectrum are the result of removing the contaminated bins. (d) The rescaled spectrum obtained by multiplying the processed spectrum by $k_BT\Delta\nu_b/P$, where both the noise temperature $T$ and signal power $P$ vary with frequency (Sec.~\ref{sub:rescale}). The combined spectrum is given by a maximum-likelihood weighted sum of the complete set of rescaled spectra.}
\end{figure*}

A second, more subtle effect of IF interference is to distort the local estimate of the spectral baseline around any sufficiently large power excess. To mitigate this effect we repeat the process described above iteratively. We remove all flagged bins from the averaged spectrum and apply the SG filter again to obtain a refined baseline estimate; using this improved baseline we generally find some additional bins with values exceeding the $4.5\sigma$ threshold; again 3 bins on either side are also flagged. In practice this procedure takes only 2 or 3 iterations to converge. The output of this iterative process is a list of bad bins within the truncated spectrum for each group of spectra; we also obtain an estimate of the average spectral baseline that we will use in the next stage of the analysis procedure.

The bad bin lists we obtain from our three distinct groups of spectra are quite similar: roughly 75\% of the bins that appear on each list also appear on the other two, and most discrepancies amount to shifting the boundaries of contiguous sets of bad bins. Because the three lists appear to describe IF interference that does not change throughout the run, we combined them into a single final list of bad bins to be cut from every spectrum. Any minimal group of 7 consecutive bins is included in the final list if it appears in two of the three lists and excluded if it appears on only one list. For all other features the final list is the union of the three lists. 11\% of the bins in the analysis band (1456 bins) appear on this final list.

Finally, we also want to flag narrowband interference that would average out in the procedure described above, so we set an additional threshold in each processed spectrum in units of the standard deviation $\sigma^\text{p}$ (Sec.~\ref{sub:stats}). We cannot afford to be as aggressive in cutting such features because Gaussian statistics dictates that roughly 300 bins will exceed 4.5$\sigma^\text{p}$ across all 6766 processed spectra. Thus we set the processed spectrum threshold at $6\sigma^\text{p}$, resulting in an additional 0--30 bins cut from each processed spectrum. The distribution of these bins throughout the spectra implicates temporally intermittent IF interference rather than RF interference.

\section{Removing the spectral baseline}\label{sec:baseline}
A typical raw power spectrum from the HAYSTAC detector, truncated to the analysis band, is shown in black in Fig.~\ref{fig:sample}(a). As emphasized in Sec.~\ref{sec:analysis_overview}, the spectral baseline is in principle the product of the total input-referred noise (which affects the sensitivity of the axion search) and the net gain of the receiver (which does not). On large spectral scales the shape of the baseline is mainly due to three effects. Rolloff at the low-RF (high-IF; see Fig.~\ref{fig:layout}) end of the spectrum is due to room-temperature IF components, rolloff on the high-RF side comes from the JPA gain profile, and the intermediate region around the cavity resonance is enhanced by the heightened temperature of the tuning rod~\cite{NIM2017}. We see that there can be as much as $\sim4$~dB variation in the ``gain'' within a single spectrum.

An average baseline obtained via the process described in Sec.~\ref{sub:badbins} is shown in red in Fig.~\ref{fig:sample}(a). Systematic deviations of the raw spectrum from the average baseline indicate that the spectral baseline can change from one iteration to the next. Such variation is not surprising, as the JPA is a narrowband amplifier for which gain fluctuations imply bandwidth fluctuations. The excess noise on resonance also depends on frequency-dependent parameters of the cavity mode, and there may be many other effects that can cause the spectral baseline to vary.

Nonetheless, normalizing each raw spectrum to the average baseline does reduce the typical variation across each spectrum from $\sim4$~dB to $\sim0.5$~dB; the normalized spectrum (which is now dimensionless) is shown in black in Fig.~\ref{fig:sample}(b). At this point we also remove all the bins compromised by IF interference from each spectrum. The normalized spectrum with bad bins removed is shown in green in Fig.~\ref{fig:sample}(b). Although only the analysis band is shown in Fig.~\ref{fig:sample}, we actually apply the above steps to the analysis band plus 500 bins on either side. These extra bins essentially serve as buffer regions for the SG filter that we now employ to remove the residual baseline of each spectrum.

\subsection{The Savitzky-Golay filter}\label{sub:sg_filter}
The simplest way to understand the SG filter is as a polynomial generalization of a moving average characterized by two parameters $d$ and $W$. For each point $x_0$ in the input sequence (assumed to be much longer than $W$), we fit a polynomial of degree $d$ in a $2W+1$-point window centered on $x_0$. The value of the SG filter output at $x_0$ is defined to be the least-squares-optimal polynomial evaluated at the center of the window, and this process is repeated for each $x_0$; thus the filter output is a smoothed version of the input sequence, with edge effects within $W$ points of either end.

Savitzky and Golay~\cite{sg1964} showed that this moving polynomial fit is equivalent to a discrete convolution of the input sequence with an impulse response that depends only on $d$ and $W$. This correspondence implies that we can fruitfully think about least-squares-smoothing from the perspective of filtering rather than fitting. The even symmetry of the SG filter impulse response implies that only even values of $d$ generate unique filters. We can gain further insight into the properties of SG filters by considering their performance in the frequency domain~\cite{schafer2011}. In the haloscope analysis considered here, we convolve the SG filter impulse response with an input sequence which is itself a power spectrum. Describing the Fourier transform of the SG impulse response as the filter's ``frequency response'' may thus be misleading; we will instead refer to this Fourier transform as a transfer function in the ``inverse bin domain.''

\begin{figure}[t]
\includegraphics[width=0.5\textwidth]{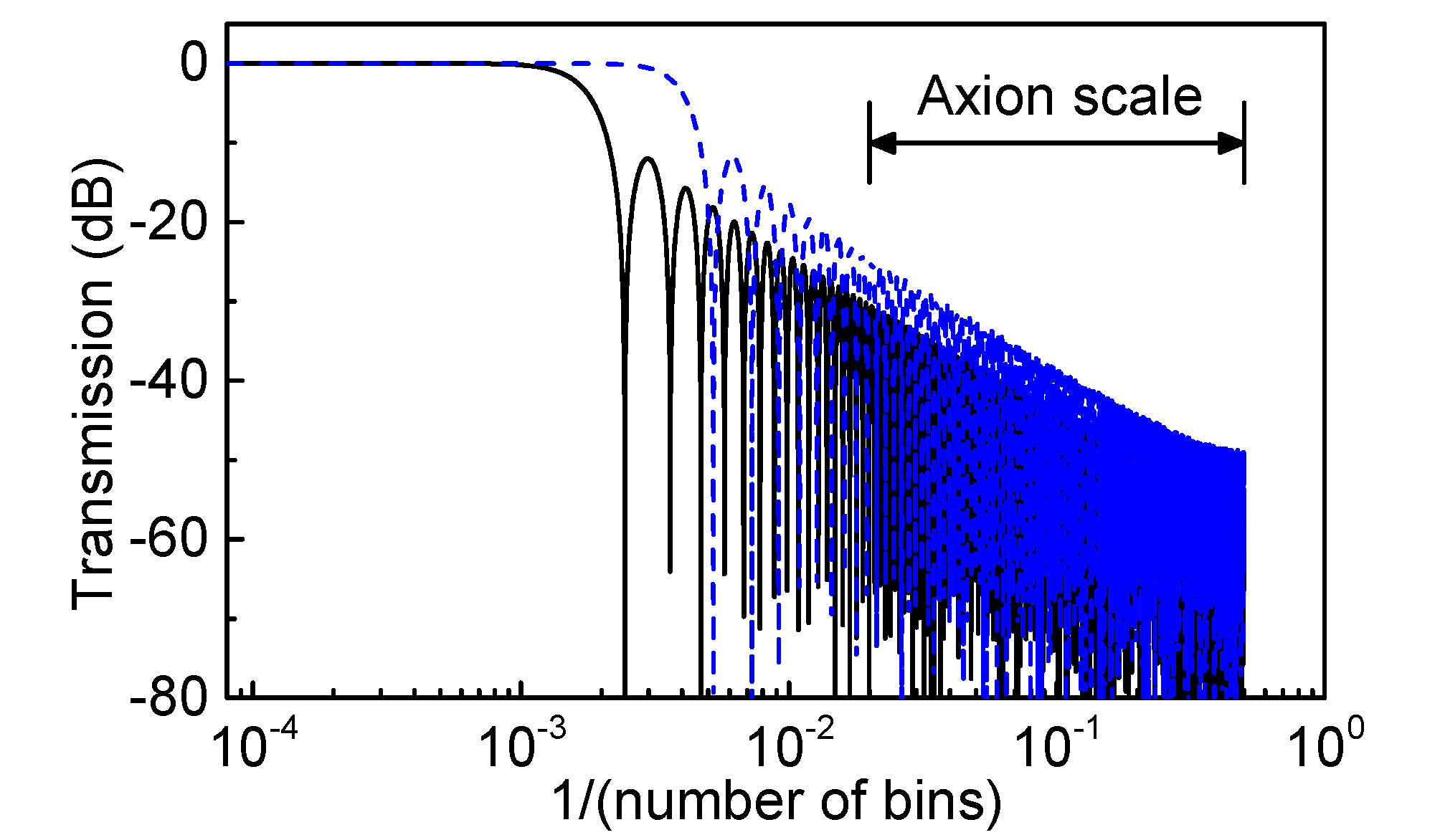}
\caption{\label{fig:filter} Transfer functions of the Savitzky-Golay filters used in our analysis. The solid black curve depicts the filter with $W=500$ and $d=4$ used in the initial scan analysis; the dashed blue curve depicts the filter with $W=300$ and $d=6$ used in the rescan analysis. We exploit the very flat passband of the filter on large spectral scales for baseline removal. The behavior of the filter on small spectral scales of 2-50 bins determines its effects on the axion signal and the statistics of the grand spectrum.}
\end{figure}

Two SG filter transfer functions used in the HAYSTAC analysis are plotted in Fig.~\ref{fig:filter}. In general, SG filters are low-pass filters with extremely flat passbands and mediocre stopband attenuation. The 3~dB point that marks the transition between these two regions scales approximately linearly with $d$ and approximately inversely with $W$. In particular, the 3~dB point for an SG filter with $d=4$ and $W=500$ (black solid line in Fig.~\ref{fig:filter}) is $\approx 1/(517~\text{bins})$. Thus when this filter is applied to one of the normalized spectra discussed above, features in the residual baseline on spectral scales sufficiently large compared to 51.7~kHz will be essentially perfectly preserved in the filter output, and features on smaller spectral scales are suppressed to varying degrees. The output of the SG filter applied to the normalized spectrum in Fig.~\ref{fig:sample}(b) is shown in red on the same plot. After dividing each normalized spectrum by the corresponding SG filter output to remove the residual baseline, we can discard the 500 bins at either edge of each spectrum, whose only purpose has been to keep edge effects out of the analysis band; all subsequent processing is applied to the analysis band of each spectrum only.

The design of any digital filter involves some tradeoff between passband and stopband performance, and we have seen that SG filters generally sacrifice some stopband attenuation to optimize passband flatness. It remains to be shown that this is the correct choice for a haloscope analysis. To see this, note that appreciable passband ripple implies the presence of systematic structure on large scales in the processed spectra. Such structure in turn implies that we cannot assume all processed spectrum bins are samples drawn from the same Gaussian distribution (see Sec.~\ref{sub:stats}); thus we cannot construct a properly normalized measure of excess power in an arbitrary IF bin, which is a central assumption of the rest of the analysis. 

Imperfect stopband attenuation, on the other hand, implies that features and fluctuations on small spectral scales are slightly suppressed when we divide each normalized spectrum by the SG filter output; equivalently, the SG filter slightly attenuates axion signals and imprints small negative correlations between processed spectrum bins. We will show that we can quantify both the filter-induced signal attenuation (Sec.~\ref{sub:axion_atten}) and the effects of correlations on the statistics of the grand spectrum in which we ultimately conduct our axion search (Sec.~\ref{sub:correlations}). Computing the axion search sensitivity directly from the statistics of the spectra requires a thorough understanding of both effects.\footnote{The application of SG filters to spectral baseline removal in a haloscope search was first explored by Ref.~\cite{malagon2014}, which did not however adopt our frequency-domain approach or consider the effects of filter-induced correlations. See Ref.~\cite{YMCE2015} for further discussion of this experiment.}

\begin{figure*}[t]
\includegraphics[width=1.0\textwidth]{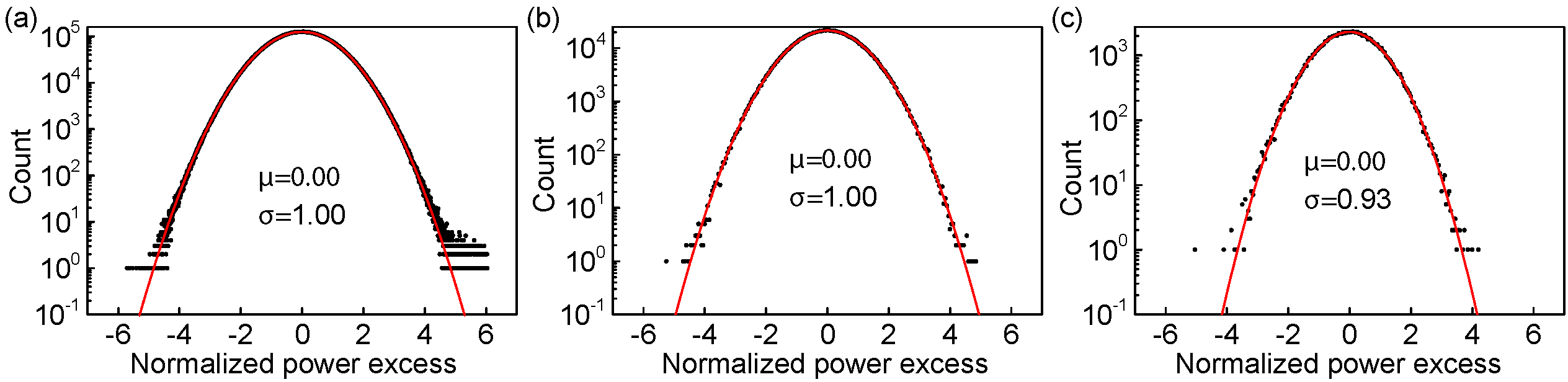}
\caption{\label{fig:data_hist} Histograms of HAYSTAC power spectra at various stages of the processing, with each bin in each spectrum normalized to its expected standard deviation. In each plot, the histogram (black circles) is fit with a Gaussian (red curve), and the mean $\mu$ and standard deviation $\sigma$ obtained from the fit are displayed. (a) Histogram of all bins $\delta^\text{p}_{ij}/\sigma^\text{p}$ from all processed spectra (Sec.~\ref{sub:stats}). There is a surplus of bins at large positive excess power (with a cutoff at $6\sigma^\text{p}$) due to narrowband IF interference (Sec.~\ref{sub:badbins}). Otherwise, the distribution of bins is Gaussian with the expected standard deviation. (b) Histogram of all combined spectrum bins $\delta^\text{c}_k/\sigma^\text{c}_k$ (Sec.~\ref{sub:combine}), demonstrating Gaussian statistics with the expected standard deviation. (c) Histogram of all grand spectrum bins $\delta^\text{g}_\ell/\sigma^\text{g}_\ell$ (Sec.~\ref{sub:grand_spectrum}). The statistics of the spectrum are still Gaussian, but the standard deviation is reduced by a factor $\xi=0.93$ due to small-scale correlations ultimately traceable to the imperfect SG filter stopband attenuation (Sec.~\ref{sub:correlations}).}
\end{figure*}

The above discussion implies that passband flatness is a more important consideration than stopband attenuation for estimating spectral baselines in a haloscope analysis, and thus the SG filter is a good choice.\footnote{An optimal Chebyshev filter with coefficients obtained from the Parks-McClellan algorithm may be able to achieve better attenuation than the SG filter in the relevant part of the stopband while retaining the requisite passband flatness. We did not explore this approach for the present analysis.} Acceptable values of the filter parameters $d$ and $W$ are constrained by the integration time at each tuning step. Longer integrations make us sensitive to smaller-amplitude systematic structure in the baseline on smaller spectral scales, and we must push the 3~dB point of the SG filter up towards smaller scales to ensure that this structure remains confined to the passband (see Appendix~\ref{app:sg_params} for a more detailed discussion). We will see in Sec.~\ref{sub:rescan_analysis} that different values of $d$ and $W$ are appropriate for the analysis of rescan data.

\subsection{Statistics of the processed spectra}\label{sub:stats}
At each data run iteration, the total noise referred to the receiver input is statistically equivalent to thermal noise at some effective (possibly frequency-dependent) temperature; thus the noise voltage distribution is Gaussian, and the fluctuations in each Nyquist-resolution subspectrum will have a $\chi^2$ distribution of degree 2. During data acquisition we average $\Delta\nu_b\tau = 9\times10^4$ such subspectra together, so the noise power fluctuations about the slowly varying baseline of each raw spectrum will be Gaussian by the central limit theorem. 

The baseline removal procedure described above should thus yield a set of flat dimensionless spectra, each with small Gaussian fluctuations about a mean value of 1. Ultimately, we are interested in \textit{excess power} (which may be positive or negative) relative to the average noise power in each bin, so we subtract 1 from each spectrum after dividing out the SG filter output. We refer to the set of spectra obtained this way as the processed spectra; a representative processed spectrum is shown in Fig~\ref{fig:sample}(c). 

In the absence of axion conversion, the bins in each processed spectrum should be samples drawn from a single Gaussian distribution with mean $\mu^\text{p}=0$ and standard deviation $\sigma^\text{p}=1/\sqrt{\Delta\nu_b\tau}=3.3\times10^{-3}$. In Fig.~\ref{fig:data_hist}(a) we have histogrammed all IF bins from all processed spectra together in units of $\sigma^\text{p}$. The excess power distribution is indeed Gaussian out to $\approx5\sigma$, and the excess above $5\sigma$ is likely due to intermittent IF interference slightly too small to exceed our $6\sigma^\text{p}$ threshold (Sec.~\ref{sub:badbins}). These large single-bin power excesses will be significantly diluted when we combine and rebin spectra.

Fig.~\ref{fig:data_hist}(a) indicates that each bin in each processed spectrum may be regarded as a random variable drawn from the same Gaussian distribution, and this is an important check on our baseline removal procedure. It does not follow that each spectrum is a sample of Gaussian white noise, because nearby bins in each spectrum will be correlated due to the imperfect stopband attenuation of the SG filter. 

We can observe effects of these correlations if we regard each spectrum (rather than each bin) as a sample of the same Gaussian process. Let $\delta^\text{p}_{ij}$ represent the value of the $j$th IF bin in the $i$th processed spectrum, for $i=1,\dots,M$ and $j=1,\dots,n^\text{p}$; $M=6766$ and $n^\text{p}=11564$ for the first HAYSTAC run after the cuts discussed in Sec.~\ref{sec:cuts}. The $i$th processed spectrum has sample mean 
\begin{equation}
\mu^\text{p}_i = \frac{1}{n^\text{p}}\sum_j\delta^\text{p}_{ij}
\label{eq:mu_p}
\end{equation}
and sample variance
\begin{equation}
\left(\sigma^\text{p}_i\right)^2 = \frac{1}{n^\text{p}-1}\sum_j\left(\delta^\text{p}_{ij}-\mu^\text{p}_i\right)^2.
\label{eq:sigma_p}
\end{equation}
In the absence of correlations, the set of sample means should be Gaussian distributed about $\mu^\text{p}$ with standard deviation $\sigma_\mu = \sigma^\text{p}/\sqrt{n^\text{p}}$, and the set of sample variances should be Gaussian distributed about $(\sigma^\text{p})^2$ with standard deviation $\sigma_{\sigma^2} = \sqrt{2/(n^\text{p}-1)}(\sigma^\text{p})^2$, again by the central limit theorem. The presence of negative correlations on small spectral scales will reduce $\sigma_\mu$ substantially and also increase $\sigma_{\sigma^2}$ slightly, without appreciably changing the mean value of either distribution. Empirically, we find that $\sigma_\mu$ is smaller than the above estimate by an order of magnitude, and $\sigma_{\sigma^2}$ is larger by about 8\%. 

The distortions of the sample mean and variance distributions noted above do not themselves affect the axion search sensitivity. But the correlations responsible for them are still important, since the remainder of our analysis procedure will involve taking both horizontal and vertical weighted sums of processed spectrum bins. A weighted sum of any number of \textit{independent} Gaussian random variables is another Gaussian random variable, with mean given by the weighted sum of component means, and standard deviation given by the quadrature weighted sum of component standard deviations. If instead the random variables are jointly normal but \textit{correlated}, the sum is still Gaussian and has the same mean, but computing the variance of the sum requires knowledge of the full covariance matrix. We will return to this point in Sec.~\ref{sub:correlations}.

\section{Combining spectra vertically}\label{sec:rescale_combine}
The $Mn^\text{p}$ processed spectrum bins $\delta^\text{p}_{ij}$ correspond to $n^\text{c} < Mn^\text{p}$ unique RF bins ($n^\text{c}\approx1.07\times10^6$ for the first HAYSTAC data run). For notational convenience we define the symbol $\Gamma_{ijk}=1$ if the $j$th IF bin in the $i$th spectrum is one of the $m_k$ bins corresponding to the $k$th RF frequency ($\Gamma_{ijk}=0$ otherwise). Our next task is to construct a single combined spectrum by taking an optimally weighted vertical sum of all $m_k$ IF bins corresponding to each RF bin $k$. The $m_k$ bins in each sum will be statistically independent, since each processed spectrum contains at most one IF bin to corresponding to any given RF bin $k$.

To gain insight into the form of the optimally weighted sum, let us consider the simple case where all axion conversion power is confined to a single RF bin $k'$. Then each processed spectrum bin with $\Gamma_{ijk'}=1$ may be regarded as a sample from a Gaussian distribution whose mean is nonzero. We will initially assume that all of these bins have the same mean $\mu_{k'}=1$ but possibly different standard deviations; of course, all bins with $\Gamma_{ijk'}=0$ also share a mean value, namely 0.  

This assumption allows us to formulate the requirement for an optimally weighted vertical sum more precisely: for each $k$ we will choose weights that yield the maximum likelihood (ML) estimate of the true mean value $\mu_k$ shared by all the contributing bins. ML estimation is briefly summarized in Appendix~\ref{app:mle}. In Sec.~\ref{sub:combine} we will see that ML weighting maximizes the SNR among all choices that yield unbiased estimates of the power excess.

In practice, the sensitivity of any given processed spectrum bin to axion conversion will generally depend on both $i$ and $j$, so each of the bins with $\Gamma_{ijk'}=1$ is actually a Gaussian random variable with a \textit{different} nonzero mean. Moreover, we saw in Sec.~\ref{sub:stats} that each bin in each processed spectrum has the \textit{same} standard deviation $\sigma^\text{p}$ -- we did not consider axion signals when discussing the statistics of the processed spectra, but we should expect the fluctuations of the noise power to be independent of the presence or absence of axion conversion power. 

Evidently the assumption we used above to motivate the ML-weighted vertical sum was precisely backwards. We can cast the problem into a form amenable to ML weighting by rescaling the processed spectra so that axion conversion would yield the same mean power excess in any rescaled spectrum bin. Determining the appropriate rescaling factor is the subject of the next section. After rescaling the spectra, we can meaningfully define ML weights and thus construct the combined spectrum.

\subsection{The rescaling procedure}\label{sub:rescale}
We rescale the processed spectra by multiplying each spectrum by the mean noise power per bin and dividing by the signal power. The $j$th bin in the $i$th rescaled spectrum is then
\begin{equation}
\delta^\text{s}_{ij} = \frac{k_BT_{ij}\Delta\nu_b\delta^\text{p}_{ij}}{P_{ij}},
\label{eq:delta_s}
\end{equation}
where $T_{ij}$ is the system noise temperature referred to the receiver input,\footnote{We follow the convention of haloscope papers in using ``system noise temperature'' to denote the total noise power per unit bandwidth, including whatever thermal noise enters the receiver along with the axion signal.} and $P_{ij}$ is the total conversion power we expect from an axion signal confined to the $j$th bin of the $i$th spectrum.\footnote{Note that to set a definite normalization for the rescaling factor we need to assume specific values for the theory parameters we hope to constrain; the assumption of single-bin signals likewise amounts to a simple but physically implausible choice of normalization for $P_{ij}$. The exclusion limit which is the final product of our analysis will not depend on either arbitrary choice of normalization.}

It may be helpful to discuss qualitatively why Eq.~\eqref{eq:delta_s} is the appropriate form for the rescaling factor. An axion signal with any given conversion power will be relatively suppressed by baseline removal if it happens to appear in a noisier spectrum or a noisier region of a given spectrum; multiplying by $T_{ij}$ undoes this suppression. Dividing by the signal power undoes the relative suppression of conversion power in spectra that are less sensitive overall due to e.g. smaller cavity $Q$, and undoes the Lorentzian suppression of the conversion power for axions at nonzero detuning from the cavity resonance.

The net result is that in the absence of noise, the hypothetical single-bin axion signal we have considered will yield $\delta^\text{s}_{ij}=1$ in each bin with $\Gamma_{ijk'}=1$. In the presence of noise, each of these bins is a Gaussian random variable with mean $\mu^\text{s}_{ij}=1$ and every other bin is a Gaussian random variable with $\mu^\text{s}_{ij}=0$. The rescaled spectra are no longer flat: each bin has a standard deviation
\begin{equation}
\sigma^\text{s}_{ij} = \frac{k_BT_{ij}\Delta\nu_b\sigma^\text{p}_{i}}{P_{ij}}.
\label{eq:sigma_s}
\end{equation}
Note that $\sigma^\text{s}_{ij} = (R^\text{\,s}_{ij})^{-1}$, where $R^\text{\,s}_{ij}$ is the SNR for our hypothetical single-bin axion signal (c.f. Eq.~(3) in Ref.~\cite{NIM2017}); this is completely equivalent to the statement that an axion signal in any bin of any rescaled spectrum produces a mean power excess of 1. A representative rescaled spectrum is shown in Fig.~\ref{fig:sample}(d). Its overall shape is primarily due to the Lorentzian cavity mode profile.

We have not yet addressed how we actually obtain values for $P_{ij}$ and $T_{ij}$. The axion conversion power~\cite{NIM2017} may be expressed as 
\begin{equation}
P_{ij} = U_0\left(\nu_{ci}\frac{\beta_i}{1+\beta_i}C_i\frac{Q_{Li}}{1+\left[2(\nu_{ij}-\nu_{ci})/\Delta\nu_{ci}\right]^2}\right),\label{eq:power}
\end{equation}
where 
\begin{equation}
U_0=g_\gamma^2\frac{\alpha^2}{\pi^2}\frac{\hbar^3c^3\rho_a}{\Lambda^4}\frac{2\pi}{\mu_0}\eta_LB_0^2V\label{eq:U_0}
\end{equation}
is a constant with dimensions of energy. 

The factors we have absorbed into the definition of $U_0$ are independent of both $i$ and $j$ and thus only affect the overall normalization of the rescaled spectra. Here $g_\gamma$ is a dimensionless number characterizing the strength of axion-photon coupling in a particular axion model, $\alpha$ is the fine-structure constant, $\rho_a$ is the local energy density of dark matter axions, $\Lambda = 77.6$~MeV is a fixed parameter that encodes the dependence of the axion mass on hadronic physics,\footnote{The value of $\Lambda$ used in our analysis comes from a calculation in chiral perturbation theory (see Ref.~\cite{sikivie1985}). Note also that $\Lambda^4 = \chi(T = 0)$, where $\chi$ is the QCD topological susceptibility that may be calculated on the lattice. A recent lattice calculation reported in Ref.~\cite{lattice2016} obtained $\Lambda = 75.6$~MeV. With the latter value the haloscope signal power would be enhanced by 11\%.} $\eta_L$ is a signal attenuation factor (see below), $B_0 = 9$~T is the applied magnetic field, and $V = 1.545$~L is the cavity volume excluding the tuning rod.\footnote{Of course $B_0$ can change in principle, but we operate our magnet in persistent mode so in practice it is extremely stable over the course of the run.} 

The parameters that experiment can constrain are $|g_\gamma|$ and $\rho_a$; it is conventional to fix $\rho_a=0.45\text{ GeV/cm}^3$ and cite the results of any given haloscope search as constraints on $|g_\gamma|$. To set a definite normalization for the rescaled spectrum, we need to temporarily fix both parameters, so we set $|g_\gamma|=|g^\text{KSVZ}_\gamma| = 0.97$, corresponding to the standard KSVZ model~\cite{kim1979,*SVZ1980}.

The remaining factors in Eq.~\eqref{eq:power} are all properties of the TM$_{010}$ mode of the cavity that can vary as it is tuned. The mode has resonant frequency $\nu_{ci}$, bandwidth $\Delta\nu_{ci}$, and quality factor $Q_{Li} = \nu_{ci}/\Delta\nu_{ci}$. Its coupling to the receiver is parametrized the dimensionless number $\beta_i$, defined implicitly by $Q_{Li}=Q_{0i}/(1+\beta_i)$, where $Q_{0i}$ is the unloaded quality factor. The form factor $C_i$ parametrizes the overlap between the spatial profile of the mode's electric field and the applied magnetic field. Finally, $\nu_{ij}$ is the RF frequency of the $j^\text{th}$ bin in the $i^\text{th}$ spectrum. 

We use the auxiliary data to obtain values for all these parameters except the form factor $C_i$, whose frequency dependence is obtained from simulations of the cavity mode. As discussed in Sec.~\ref{sub:badscans}, we made VNA measurements of the cavity mode in transmission both before and after each cavity noise measurement to cut iterations with excessive drift. For the remaining iterations, the ``before'' and ``after'' measurements are very similar, so we average them and fit the average to a Lorentzian to obtain $\nu_{ci}$ and $Q_{Li}$. We also used the VNA to measure the cavity mode in reflection: the magnitude of the reflection coefficient on resonance and the net resonant phase shift together determine $\beta_i$.

The system noise temperature $T_{ij}$ may be parametrized in units of quanta as
\begin{equation}
k_BT_{ij} = h\nu_{ci}\left[N_T + (N_\text{cav})_{ij} + (N_A)_{ij}\right],
\label{eq:noise}
\end{equation}
where $N_T$ is thermal noise at the known mixing chamber temperature $T_C$, $N_\text{cav}$ is the \textit{excess} thermal noise associated with the elevated tuning rod temperature, and the receiver added noise $N_A$ includes the added noise of the JPA preamplifier, small contributions from subsequent amplifiers, and effective noise associated with loss between the microwave switch and the JPA.\footnote{Technically, $N_T$ is a function of frequency evaluated at $\nu_{ci}$, but it changes negligibly over our tuning range, so we suppress its $i$-dependence. $j$-dependence due to the finite width of the analysis band is of course much smaller still.} We calibrate the noise using $Y$-factor measurements (discussed in detail in Ref.~\cite{NIM2017}); in our first data run, $Y$-factor measurements were repeated every 10 iterations (roughly 3.5 hours).

Assuming a cavity in thermal equilibrium with the mixing chamber plate of the dilution refrigerator (i.e., $N_\text{cav}=0$), $Y$-factor measurements are ideal for the haloscope search noise calibration because they naturally measure $N_A$ as defined above, in contrast with measurements of the SNR improvement from switching on the preamplifier, which are not sensitive to the loss contribution. Neither method is sensitive to losses between the cavity and the microwave switch (~$\approx0.6~\text{dB}$ throughout the initial HAYSTAC tuning range), which nonetheless degrade the axion search SNR. Thus the factor $\eta_L=10^{-0.6/10}=0.87$ must be included explicitly in Eq.~\eqref{eq:U_0}.

As already noted above, the cavity was not in thermal equilibrium with the mixing chamber in the first HAYSTAC data run, and this resulted in a contribution to the system noise temperature with a roughly Lorentzian profile centered on $\nu_{ci}$ in each spectrum. In the presence of this additional unknown noise $N_\text{cav}$, the $Y$-factor measurement associated with the $i$th spectrum measures not $(N_A)_{ij}+(N_\text{cav})_{ij}$ but rather $(N_A)_{ij} + Y_{ij}/(Y_{ij}-A_{ij})(N_\text{cav})_{ij}$, where $Y$ is the measured ratio of hot/cold noise power spectra and $A$ is the measured hot/cold gain ratio.\footnote{The additional factors multiplying the $N_\text{cav}$ term account for the fact that it contributes only to the cold load noise measurement, whereas $N_A$ contributes to the noise in both the hot load and the cold load; see also discussion in Ref.~\cite{NIM2017}.}

$N_A$ should be independent of the presence of the cavity mode in the spectrum, and empirically it also exhibits no systematic dependence on RF frequency. Thus we can break the degeneracy in $Y$-factor measurements around the TM$_{010}$ resonance by subtracting $(\bar{N}_A)_j$, the average receiver added noise obtained from off-resonance $Y$-factor measurements. By doing so we obtain an estimate of $N_\text{cav}$ in each $Y$-factor measurement throughout the data run, though this method implies that deviations from $\bar{N}_A$ across spectra are attributed instead to variation in $N_\text{cav}$.

We do expect $N_\text{cav}$ to vary across spectra due to variation in $Q$ and $\beta$. Moreover, the effective temperature of the cavity mode is determined by a competition between the walls, which are well coupled to the mixing chamber, and the rod, which was at a higher temperature throughout the first HAYSTAC data run; the relative strength of these contributions will depend on the shape of the cavity mode and thus on the mode frequency. 

Empirically, there were clearly correlations among the $N_\text{cav}$ profiles obtained from nearby $Y$-factor measurements, but no deterministic frequency dependence strong enough to justify any particular interpolation scheme. Thus, we simply set $T_{ij}$ for each spectrum at which we did not make a $Y$-factor measurement using the nearest measured value of $N_\text{cav}$. In Appendix~\ref{app:error} we estimate the uncertainty in our exclusion limit resulting from possible miscalibration of the noise temperature.

\subsection{Constructing the combined spectrum}\label{sub:combine}
We have shown that the rescaled spectrum IF bins corresponding to each RF bin are independent Gaussian random variables with the same mean (1 in the presence of a single-bin KSVZ axion and 0 in the absence of a signal) and different variances. To obtain the ML estimate of this mean value (see Appendix~\ref{app:mle}) we weight each bin by its inverse variance:
\begin{equation}
w_{ijk} = \frac{\Gamma_{ijk}(\sigma^\text{s}_{ij})^{-2}}{\sum_{i'}\sum_{j'}\Gamma_{i'j'k}(\sigma^\text{s}_{i'j'})^{-2}},
\label{eq:weights}
\end{equation}
where the denominator ensures that the weights are normalized.\footnote{Many of the expressions to follow contain sums over $i$ and $j$ in both the numerator and denominator. We will avoid cumbersome primes through slight abuse of notation by using the same indices $i$ and $j$ in both sums. $k$, which is not summed over, is understood to have the same value in the numerator and denominator. Sums whose upper and lower limits are elided are to be interpreted as running over all possible values of the index.} Then the ML estimate of the mean in each combined spectrum bin $k$ is given by the weighted sum of contributing bins: 
\begin{eqnarray}
\delta^\text{c}_k &&= \sum_i\sum_jw_{ijk}\delta^\text{s}_{ij} \nonumber\\
 &&= \frac{\sum_i\sum_j\Gamma_{ijk}\left(P_{ij}\delta^\text{p}_{ij}/k_BT_{ij}\Delta\nu_b(\sigma^\text{p}_{i})^2\right)}{\sum_i\sum_j\Gamma_{ijk}\left(P_{ij}/k_BT_{ij}\Delta\nu_b\sigma^\text{p}_{i}\right)^2}\label{eq:delta_c}.
\end{eqnarray}
The standard deviation of each bin in the combined spectrum is the quadrature weighted sum of contributing standard deviations:
\begin{eqnarray}
\sigma^\text{c}_k &&= \sqrt{\sum_i\sum_jw_{ijk}^2\left(\sigma^\text{s}_{ij}\right)^2} \nonumber\\
&&= \sqrt{\frac{\sum_i\sum_j\Gamma_{ijk}\left(\sigma^\text{s}_{ij}\right)^{-4}\left(\sigma^\text{s}_{ij}\right)^2}{\left[\sum_{i}\sum_{j}\Gamma_{ijk}(\sigma^\text{s}_{ij})^{-2}\right]^2}} \nonumber\\
\Rightarrow \sigma^\text{c}_k &&= \left[\sum_i\sum_j\Gamma_{ijk}\left(\frac{P_{ij}}{k_BT_{ij}\Delta\nu_b\sigma^\text{p}_{i}}\right)^2\right]^{-1/2}.\label{eq:sigma_c}
\end{eqnarray}

For each $k$, there are $m_k$ nonvanishing contributions to the sums in the expressions above. In the first HAYSTAC data run, typical values of $m_k$ ranged from 50 to 120 across the combined spectrum due to nonuniform tuning.\footnote{There are two $\sim$ MHz-width peaks in the distribution of $m_k$ with peak values of 150 and 200, due to scans in which the tuning rod was temporarily stuck at a single frequency. $m_k$ also drops precipitously around the frequency of the intruder mode where we cut spectra (Sec.~\ref{sub:badscans}) and at the edges of the scan range. On spectral scales small compared to the analysis band width, $m_k$ fluctuates by $\pm2$ due to the presence of missing bins in the processed spectra.}

Two numbers are required to characterize the combined spectrum at each frequency: $\delta^\text{c}_k$ and $\sigma^\text{c}_k$ describe respectively the actual power excess in each combined spectrum bin and the power excess we expect from statistical fluctuations. Absent any axion signals, each $\delta^\text{c}_k$ should be a Gaussian random variable drawn from a distribution with mean $\mu^\text{c}_k = 0$ and standard deviation $\sigma^\text{c}_k$. Thus the distribution of normalized bins
\begin{equation}
\frac{\delta^\text{c}_k}{\sigma^\text{c}_k} = \frac{\sum_i\sum_j\Gamma_{ijk}\left(P_{ij}\delta^\text{p}_{ij}/k_BT_{ij}\Delta\nu_b(\sigma^\text{p}_{i})^2\right)}{\sqrt{\sum_i\sum_j\Gamma_{ijk}\left(P_{ij}/k_BT_{ij}\Delta\nu_b\sigma^\text{p}_{i}\right)^2}}
\label{eq:ds_c}
\end{equation}
should be Gaussian with zero mean and unit variance; we can see in Fig.~\ref{fig:data_hist}(b) that this is indeed the case.\footnote{In practice $\delta^\text{c}_k/\sigma^\text{c}_k$ will still appear to have a standard normal distribution even in the presence of axion conversion, since $\mu^\text{c}_k\neq0$ in only a few bins.}

We can equivalently describe the combined spectrum by specifying the values of $\delta^\text{c}_k/\sigma^\text{c}_k$ and $R^\text{\,c}_k = \left(\sigma^\text{c}_k\right)^{-1}$ for each $k$. The normalization of the ML weights implies that, for a single-bin KSVZ axion at frequency $k'$, $\mu^\text{c}_{k'}=1$ and thus $E\big[\delta^\text{c}_{k'}/\sigma^\text{c}_{k'}\big]=R^\text{\,c}_{k'}$. Physically, $R^\text{\,c}_k$ is the SNR that a single-bin KSVZ axion \textit{would have} in the $k$th bin of the combined spectrum (whether or not such an axion exists). In terms of the SNR, Eq.~\eqref{eq:sigma_c} becomes
\begin{equation}
R^\text{\,c}_k = \sqrt{\sum_i\sum_j\Gamma_{ijk}\left(R^\text{\,s}_{ij}\right)^2}\label{eq:snr_c},
\end{equation}
which tells us that the SNR in each bin of the combined spectrum is simply the (unweighted) quadrature sum of the SNR across contributing bins. 

As discussed in Appendix~\ref{app:mle}, the ML estimate of the mean of a Gaussian distribution also has the smallest variance among unbiased estimates. The variance of the mean of a Gaussian distribution is simply proportional to the variance of the distribution, so equivalently ML weights yield the smallest $\sigma^\text{c}_k$ and thus the largest $R^\text{\,c}_k$ among all possible weights that do not systematically bias $\delta^\text{c}_k$. Thus, ML weighting is optimal for the haloscope analysis in a real physically intuitive sense.

\section{Combining bins horizontally}\label{sec:rebin}
The parameterization of the combined spectrum in terms of $\delta^\text{c}_k/\sigma^\text{c}_k$ and $R^\text{\,c}_k$ lends itself naturally to identifying axion candidates and setting exclusion limits, via the procedure outlined in Sec.~\ref{sec:candidates}. However, $R^\text{\,c}_k$ is the (unrealistically large) SNR for an axion signal confined to a single bin, whereas our goal here is to construct an analysis tailored to the detection of virialized axions with $\Delta\nu_a \gg \Delta\nu_b$. Thus, our next task is to determine an explicit expression for the grand spectrum $\delta^\text{g}_\ell/\sigma^\text{g}_\ell$ as a weighted sum of adjacent combined spectrum bins. As in Sec.~\ref{sec:rescale_combine}, we take the optimal weights to be those that yield the ML estimate of the mean grand spectrum power excess, after rescaling to make the expected excess due to axion conversion uniform across all contributing bins. The discussion above indicates that ML weights in the horizontal sum will maximize $R^\text{\,g}_\ell$, the SNR for a virialized axion signal concentrated in the $\ell$th grand spectrum bin. 

In the choice of ML weights for the vertical sums that define the combined spectrum, we have followed the published ADMX analysis procedure~\cite{ADMX2001}, albeit with a somewhat different approach for pedagogical purposes.\footnote{See also Refs.~\cite{ADMX2000,daw1998,yu2004,hotz2013} for different presentations of ML weighting in the ADMX analysis; note that there are a number of errors in the expressions corresponding to Eqs.\ \eqref{eq:delta_c} and \eqref{eq:sigma_c} in Refs.~\cite{ADMX2001}, \cite{ADMX2000}, and \cite{daw1998}.} In extending ML weighting to horizontal sums of adjacent bins in the combined spectrum, we are deviating from the procedure used by ADMX. We discuss the key differences between our present approach and the ADMX procedure further in Sec.~\ref{sub:grand_spectrum}.

Though the principles of ML estimation remain valid, horizontal sums differ from the vertical sums considered in Sec.~\ref{sec:rescale_combine} in two important respects. First, we can no longer assume that the bins in each sum are independent random variables; indeed, as noted in Sec.~\ref{sec:baseline}, we have reason to expect correlations on small spectral scales in the processed spectra, and thus also in the combined spectrum. ML estimation of the mean of a multivariate Gaussian distribution with arbitrary covariance matrix is in principle straightforward (see Appendix~\ref{app:mle}). In practice, it requires additional information about off-diagonal elements of the covariance matrix that are not as easily estimated as the variances. In the present analysis, we take ML weights that neglect correlations as approximations to the true ML weights, and define the horizontal sum using expressions appropriate for the uncorrelated case. We will quantify the effects of correlations in Sec.~\ref{sub:correlations}.

Second, independent of any subtleties involving correlations, we have some additional freedom in how we define the horizontal sum besides the choice of weights. The simplest approach is to define each grand spectrum bin as a ML-weighted sum of all bins within a segment of length $K\approx\Delta\nu_a/\Delta\nu_b$ in the combined spectrum, such that the segments corresponding to different grand spectrum bins do not overlap. The total number of grand spectrum bins is then $n^\text{g}\approx n^\text{c}/K$. The disadvantage of this approach is that the signal power will generally be split across multiple bins unless $\nu_a$ happens to line up with our binning. We need to introduce an attenuation factor $\eta_m$ (Sec.~\ref{sub:lineshape}) to account for the average effect of misalignment on the SNR.

We can minimize misalignment effects by allowing the segments of the combined spectrum corresponding to different grand spectrum bins to overlap: if each such segment is $K$ bins long, then the first grand spectrum bin will be a ML-weighted sum of the first through $K$th combined spectrum bins, the second grand spectrum bin will be a ML-weighted sum of the second through $(K+1)$th bins, and so on. But with $K\approx\Delta\nu_a/\Delta\nu_b$ this procedure implies a total of $n^\text{g}\approx n^\text{c}$ grand spectrum bins, and thus the number of statistical rescan candidates (Sec.~\ref{sec:candidates}) will be larger at any given sensitivity than in the non-overlapping case; equivalently the total integration time required to exclude axions of a given coupling will be longer.

The two approaches considered above may be regarded as limiting cases of a more general procedure in which we split the construction of the grand spectrum into two steps. First we take ML-weighted sums of adjacent bins in non-overlapping segments of the combined spectrum to yield a rebinned spectrum with resolution $\Delta\nu_r = K^\text{r}\Delta\nu_b$. Then we construct the grand spectrum via ML-weighted sums of adjacent bins in overlapping segments of length $K^\text{g}$ in the rebinned spectrum. $K^\text{r}$ and $K^\text{g}$ should be chosen so that the product $K^\text{r}K^\text{g}\approx\Delta\nu_a/\Delta\nu_b$; it should be emphasized that we have thus far cited only a very rough estimate for $\Delta\nu_a$, and we are free to choose $K^\text{r}$ and $K^\text{g}$ independently within a reasonable range. 

In the two-step procedure described above, the rebinned spectrum weights and grand spectrum weights are each obtained from the ML principle, but of course we must specify a supposed distribution of signal power before we can define ML weights. The $\ell$th grand spectrum bin should be a sum over bins in the rebinned spectrum frequency range $[\nu_\ell,\nu_{\ell+K^\text{g}-1}]$ weighted so that the SNR is maximized if $\nu_a\approx\nu_\ell$. We will articulate this condition more precisely in Sec.~\ref{sub:lineshape}, but we can already see that the grand spectrum weights will depend on the axion lineshape. 

The weights used to construct the rebinned spectrum cannot themselves depend on the lineshape: the above example demonstrates that any given $\nu_\ell$ will correspond to the axion mass in one grand spectrum bin and the tail of the axion power distribution in another. We thus define weights to yield the ML estimate of the mean power excess in each bin of the rebinned spectrum assuming the axion signal distribution is uniform across contributing combined spectrum bins. As we reduce $K^\text{r}$, the distribution of signal power on scales smaller than $\Delta\nu_r$ becomes more uniform, and we can also use a finer approximation to the axion lineshape in the grand spectrum weights. 

For the analysis of the first HAYSTAC data run we used $K^\text{r}=10$ and $K^\text{g}=5$, informed by the tradeoffs noted above. In the next section, we will briefly digress on the expected axion lineshape and its implications for the analysis. Then we will construct the rebinned spectrum in Sec.~\ref{sub:rebinned_spectrum} and the grand spectrum in Sec.~\ref{sub:grand_spectrum}.

\subsection{The expected axion signal lineshape}\label{sub:lineshape}
Experiments aiming to directly detect non-gravitational interactions of dark matter must make assumptions about the local dark matter mass and velocity distributions. Virialization of the dark matter in the galactic halo relates these two distributions. Searches typically assume a virialized halo which is moreover spherically symmetric and approximately isothermal, such that the dark matter velocity distribution is very nearly Maxwellian in the galactic rest frame. Such a pseudo-isothermal distribution~\cite{jimenez2003} is fully specified by the values of two parameters, which we can take to be the local density $\rho_a=0.45\text{ GeV/cm}^3$~\cite{[{Haloscope searches typically assume the local density is $0.45\text{ GeV/cm}^3$, while WIMP searches typically cite $0.3\text{ Gev/cm}^3$ instead. Both values fall within the range of recent measurements; see }][{}]read2014} and the local circular velocity $v_c=220\text{ km/s}$; the latter is the mode of the Maxwell-Boltzmann distribution. 

It is also possible that some fraction of the dark matter has not virialized due to cold, high-density streams of axions that fell into the galaxy relatively recently; such streams would manifest as sharp features in the spectrum of a putative haloscope signal. Fixing the values of the experimental parameters, a haloscope search specifically targeting non-virialized axions will generally be sensitive to smaller couplings $|g_\gamma|$ because the signal bandwidth is smaller, but the converse is not true: the sensitivity of a search that assumes virialization is not appreciably degraded if there is non-virialized structure in the true signal. In this sense virialization is a conservative assumption.\footnote{The orbital motion of the earth~\cite{turner1990} can also shift the frequency of a non-virialized axion signal by an amount comparable to its linewidth between repeated scans around the same frequency. Roughly speaking, searches for non-virialized signals of fractional width $\Delta\nu_a/\nu_a\lesssim10^{-7}$ must make further assumptions about the direction of the axion stream unless candidates were rescanned more frequently than once per week during the acquisition of the search data set, with correspondingly more stringent requirements on the frequency of rescans for narrower signals.}

For the present analysis we assume a fully virialized pseudo-isothermal halo, emphasizing that its chief virtues are simplicity and the absence of strong evidence for any particular alternative; see Ref.~\cite{ADMX2016} for a recent discussion of alternative halo models in the haloscope search. The form in which we save the HAYSTAC axion search data (Sec.~\ref{sub:data}) enables future searches for nonvirialized features with fractional width as small as $\Delta\nu_b/\nu_a\sim2\times10^{-8}$. 

The spectral shape of a haloscope signal is proportional to the axion kinetic energy distribution. For a pseudo-isothermal halo in the galactic rest frame, axion velocities obey a Maxwell-Boltzmann distribution, and the corresponding kinetic energies have a $\chi^2$ distribution of degree 3. As a function of the measured signal frequency $\nu\geq\nu_a$, this distribution is
\begin{equation}
f(\nu) = \frac{2}{\sqrt{\pi}}\sqrt{\nu-\nu_a}\left(\frac{3}{\nu_a\left<\beta^2\right>}\right)^{3/2}e^{-\frac{3(\nu-\nu_a)}{\nu_a\left<\beta^2\right>}},
\label{eq:f_dist}
\end{equation}
where $\left<\beta^2\right>=\left<v^2\right>/c^2$ and the second moment of the Maxwell-Boltzmann distribution is $\left<v^2\right> = 3v_c^2/2 = (270\text{ km/s})^2$. In a frame moving relative to the galactic rest frame, the dark matter velocity distribution is not in general Maxwellian. The motion of a terrestrial laboratory relative to the galactic halo is dominated by the orbital velocity of the sun about the center of the galaxy $v_s\approx v_c$. In the lab frame the spectrum of the axion signal thus becomes~\cite{turner1990}
\begin{eqnarray}
f'(\nu) = &&\frac{2}{\sqrt{\pi}}\left(\sqrt{\frac{3}{2}}\frac{1}{r}\frac{1}{\nu_a\left<\beta^2\right>}\right)\sinh\left(3r\sqrt{\frac{2(\nu-\nu_a)}{\nu_a\left<\beta^2\right>}}\right)\nonumber \\
&&\times\exp\left(-\frac{3(\nu-\nu_a)}{\nu_a\left<\beta^2\right>}-\frac{3r^2}{2}\right),
\label{eq:f_dist_2}
\end{eqnarray}
where $r=v_s/\sqrt{\left<v^2\right>}\approx\sqrt{2/3}$. Eq.~\eqref{eq:f_dist_2} is not a $\chi^2$ distribution, but is reasonably well approximated by Eq.~\eqref{eq:f_dist} with $\left<\beta^2\right>\rightarrow1.7\left<\beta^2\right>$; of course, it approaches Eq.~\eqref{eq:f_dist} in the limit $r\rightarrow0$. 

We used Eq.~\eqref{eq:f_dist} where we should have used Eq.~\eqref{eq:f_dist_2} in our original analysis.\footnote{We thank B. R. Ko for drawing our attention to this point.} Specific parameter values cited throughout Sec.~\ref{sec:rebin} and \ref{sec:candidates} assume Eq.~\eqref{eq:f_dist}, as this was used to derive the exclusion limit published in Ref.~\cite{PRL2017}, but we emphasize that the formal procedure outlined in the present paper is independent of any specific assumptions about the signal shape. If the spectrum of the axion signal is actually given by Eq.~\eqref{eq:f_dist_2}, our exclusion limit will be degraded by $\approx20\%$ (quantified more precisely in Appendix~\ref{app:axion_width}) due to the combination of an irreducible effect from the wider signal bandwidth and the fact that our analysis was not optimized for this wider signal, as future HAYSTAC analyses will be.

A haloscope analysis can ultimately depend on the spectral shape of the axion signal only through the grand spectrum weights, which in turn can only depend on slices of $f(\nu)$ integrated over the resolution of the rebinned spectrum $\Delta\nu_r=K^\text{r}\Delta\nu_b$. Thus we define the integrated signal lineshape to be
\begin{equation}
L_q(\delta\nu_r) = K^\text{g}\int_{\nu_a+\delta\nu_r+(q-1)\Delta\nu_r}^{\nu_a+\delta\nu_r+q\Delta\nu_r}f(\nu)\,\mathrm{d}\nu,
\label{eq:int_lineshape}
\end{equation}
where $q=1,\dots,K^\text{g}$, and the misalignment $\delta\nu_r$ is defined in the range $-z\Delta\nu_r < \delta\nu_r \leq (1-z)\Delta\nu_r$, with $0<z<1$. The value of $z$ should be chosen so that for any $\delta\nu_r$ in this range, $\eta_c(\delta\nu_r) = \sum_qL_q(\delta\nu_r)/K^\text{g}$ is larger than the value we would obtain by shifting the range over which the $q$ index is defined up or down by 1.\footnote{We might naively imagine a symmetric interval (corresponding to $z=0.5$) would be optimal in this sense. In practice, given the asymmetry of the axion lineshape, there will be more power in the $K^\text{g}$-bin sum if the lower bound of the integral in the $q=1$ bin is detuned below $\nu_a$ than at an equal detuning above $\nu_a$. This implies that we should consider $z>0.5$; the optimal value will depend on the choice of $K^\text{r}$ and $K^\text{g}$.} Physically, $\eta_c$ is the fraction of signal power contained within a grand spectrum bin; it approaches 1 independent of $\delta\nu_r$ for $K^\text{g}$ sufficiently large. At any fixed value of $K^\text{g}$, the sum also depends on $\delta\nu_r$ and thus on $K^\text{r}$.

We can gain some insight into the considerations that enter into the choice of $K^\text{r}$ and $K^\text{g}$ by imagining for the moment that we take the grand spectrum weights to be uniform, as in Ref.~\cite{ADMX2001}. Then, with $K^\text{r}=1$, $\eta_c\rightarrow1$ as $K^\text{g}$ increases, but the RMS noise power grows as $\sqrt{K^\text{g}}$, so the grand spectrum SNR ($\propto\eta_c/\sqrt{K^\text{g}}$) is maximized at a finite value of $K^\text{g}$. The SNR is relatively insensitive to $\delta\nu_r$ at $K^\text{r}=1$; as we increase $K^\text{r}$, keeping $K^\text{r}K^\text{g}$ fixed, $\eta_c$ remains unchanged in the best-case scenario $\delta\nu_r=0$, but larger misalignments are possible, so dependence of the SNR on $\delta\nu_r$ grows more pronounced.

In order to define ML weights for the grand spectrum (Sec.~\ref{sub:grand_spectrum}), we will need an expression for some ``typical'' lineshape $\bar{L}_q$ that is independent of misalignment. The best approach is to define $\bar{L}_q$ as the average of $L_q(\delta\nu_r)$ over the range in which $\delta\nu_r$ is defined.\footnote{$\bar{L}_q$ has no $\ell$ index because in practice we evaluated Eq.~\eqref{eq:int_lineshape} with $\nu_a = 5.75\text{ GHz}$ both in the limits of integration and within $f(\nu)$. It would be trivial to instead calculate the lineshape with $\nu_a=\nu_\ell$ in the $\ell$th grand spectrum bin, but the variation of the lineshape over the initial HAYSTAC scan range was negligible.} Then the misalignment attenuation can be defined as $\eta_m= \text{SNR}(\{\bar{L}_q\})/\text{SNR}(\{L_q(\delta\nu_r=0)\})$.\footnote{With this definition, $\eta_m$ is a useful figure of merit for comparing different values of $K^\text{r}$ and $K^\text{g}$, but we will not have to explicitly account for it in our analysis procedure, as the average effect of misalignment on the SNR is included in the definition of $\bar{L}_q$.} In the ML-weighted grand spectrum the SNR is no longer proportional to $\eta_c$ (indeed, it asymptotes to a constant value rather than degrading as we continue to increase $K^\text{g}$). However, the above prescription for defining $\eta_m$ still holds if we use the correct expression for the SNR [Eq.~\eqref{eq:phi_align} in Appendix~\ref{app:error}]. With $K^\text{r}=10$ and $K^\text{g}=5$, the optimal range for $\delta\nu_r$ is obtained for $z=0.7$, and the misalignment attenuation is $\eta_m=0.97$.

\subsection{Rebinning the combined spectrum}\label{sub:rebinned_spectrum}
After choosing the values of $K^\text{r}$ and $K^\text{g}$ to be used in the remainder of the analysis, we rescale the combined spectrum, taking $\delta^\text{c}_k \rightarrow (K^\text{r}K^\text{g})\delta^\text{c}_k$ and $\sigma^\text{c}_k \rightarrow (K^\text{r}K^\text{g})\sigma^\text{c}_k$. This rescaling leaves $\delta^\text{c}_k/\sigma^\text{c}_k$ formally unchanged and takes $R^\text{\,c}_k \rightarrow R^\text{\,c}_k/(K^\text{r}K^\text{g})$, just what we would have obtained had we normalized Eq.~\eqref{eq:power} to a more physically plausible fraction $1/(K^\text{r}K^\text{g})$ of the expected KSVZ signal power in the first place. After this rescaling we expect $\mu^\text{c}_{k'}=1$ if a KSVZ axion signal deposits a fraction $1/(K^\text{r}K^\text{g})$ of its power in the combined spectrum bin $k'$.

In Sec.~\ref{sub:combine} we wrote rather verbose expressions for Eqs.~\eqref{eq:delta_c} and \eqref{eq:sigma_c} to make the dependence on physically meaningful quantities such as $P_{ij}$ explicit. The ML-weighted sum can be written more succinctly in terms of
\begin{equation}
D^\text{c}_k = \frac{\delta^\text{c}_k}{(\sigma^\text{c}_k)^2} = \frac{1}{K^\text{r}K^\text{g}}\sum_i\sum_j\Gamma_{ijk}\frac{\delta^\text{s}_{ij}}{(\sigma^\text{s}_{ij})^2},
\label{eq:d_c}
\end{equation}
which is just the sum in the numerator of Eq.~\eqref{eq:delta_c} rescaled by $1/(K^\text{r}K^\text{g})$ as discussed above. Each $D^\text{c}_k$ is a Gaussian random variable with standard deviation $R^\text{\,c}_k$. We obtain the ML-weighted rebinned spectrum from
\begin{equation}
D^\text{r}_\ell = \sum_{k=k_i(\ell)}^{k_f(\ell)}D^\text{c}_k \label{eq:d_r}
\end{equation}
and
\begin{equation}
\left(R^\text{\,r}_\ell\right)^2 = \sum_{k=k_i(\ell)}^{k_f(\ell)}\left(R^\text{\,c}_k\right)^2,\label{eq:snr_r}
\end{equation}
where $k_i(\ell)=(\ell-1)K^\text{r}+1$, $k_f(\ell)=\ell K^\text{r}$, $\ell=1,\dots,n^\text{r}$, and $n^\text{r}\approx n^\text{c}/K^\text{r}\approx1.07\times10^5$ for the first HAYSTAC data run. 

In the absence of correlations between combined spectrum bins, $D^\text{r}_\ell$ is a Gaussian random variable with standard deviation $R^\text{\,r}_\ell$. Defining $\sigma^\text{r}_\ell = (R^\text{\,r}_\ell)^{-1}$ and $\delta^\text{r}_\ell = D^\text{r}_\ell(\sigma^\text{r}_\ell)^2$ as in the combined spectrum, it follows that each rebinned spectrum bin $\delta^\text{r}_\ell$ is a Gaussian random variable with standard deviation $\sigma^\text{r}_\ell$ (and mean $\mu^\text{r}_\ell=0$ in the absence of axion signals). Each $\delta^\text{r}_\ell$ is the ML-weighted estimate of the mean power excess in $K^\text{r}$ adjacent combined spectrum bins $\delta^\text{c}_k$ if the axion power distribution is uniform on scales smaller than $\Delta\nu_r$. More precisely, $\mu^\text{r}_{\ell'}=1$ if a KSVZ axion deposits a fraction $1/(K^\text{r}K^\text{g})$ of its power in each of the $K^\text{r}$ adjacent combined spectrum bins corresponding to the rebinned spectrum bin $\ell'$, and $R^\text{\,r}_{\ell'}$ is the SNR for such a signal. 

Neglecting small-scale variation in $R^\text{\,c}_k$, Eq.~\eqref{eq:snr_r} implies that the SNR in each bin of the rebinned spectrum has increased by $\sqrt{K^\text{r}}$. This is exactly what we should expect given that the signal power grows roughly linearly with bandwidth $\Delta\nu$ (for $\Delta\nu$ sufficiently small compared to $\Delta\nu_a$) and the RMS noise power grows as $\sqrt{\Delta\nu}$. Empirically, the RMS variation in $\sigma^\text{c}_k$ is typically $\lesssim1\%$ on 10-bin scales (and $\approx3\%$ on 50-bin scales), so the rebinned spectrum would not change much if we used uniform weights instead of ML weights. We will see in Sec.~\ref{sub:grand_spectrum} that ML weighting of the grand spectrum leads to a larger improvement relative to an unweighted analysis.

In the absence of correlations, each $\delta^\text{r}_\ell$ has standard deviation $\sigma^\text{r}_\ell$, so $\delta^\text{r}_\ell/\sigma^\text{r}_\ell$ should have a standard normal distribution, like the analogous quantity in the combined spectrum. Empirically, in the first HAYSTAC data run, $\delta^\text{r}_\ell/\sigma^\text{r}_\ell$ was Gaussian with standard deviation $\xi^\text{r}=0.98$. $\xi^\text{r}\neq1$ is a consequence of the fact that the expression for the variance of a sum of Gaussian random variables used in Eq.~\eqref{eq:snr_r} does not hold in the presence of correlations, as noted at the end of Sec.~\ref{sub:stats}.\footnote{A similar reduction in the standard deviation following a horizontal sum was observed in Ref.~\cite{yu2004}, pg. 122, and attributed to the baseline removal procedure, but not discussed further.} An analogous effect will arise in the construction of the grand spectrum, so we will defer further discussion of this point to Sec.~\ref{sub:correlations}.

\subsection{Constructing the grand spectrum}\label{sub:grand_spectrum}
To extend the ML-weighted horizontal sum further, we must account for the fact that, for any given value of $\nu_a$, the distribution of axion signal power in the $K^\text{g}$ rebinned spectrum bins containing most of the signal is nonuniform. Specifically, for a KSVZ axion with $\nu_a\approx\nu_{\ell'}$, we expect $\mu^\text{r}_{\ell'+q-1}=\bar{L}_q$ for $q=1,\dots,K^\text{g}$. As in Sec.~\ref{sec:rescale_combine}, we must rescale the contributing bins so that they all have the same mean power excess before defining ML weights. For the $\ell$th grand spectrum bin, the appropriate rescaling is obtained by dividing both $\delta^\text{r}_{\ell+q-1}$ and $\sigma^\text{r}_{\ell+q-1}$ by $\bar{L}_q$, or equivalently by multiplying both $D^\text{r}_{\ell+q-1}$ and $R^\text{\,r}_{\ell+q-1}$ by $\bar{L}_q$. The quantities of interest in the ML-weighted grand spectrum are then given by
\begin{equation}
R^\text{\,g}_\ell = \sqrt{\sum_q(R^\text{\,r}_{\ell+q-1}\bar{L}_q)^2}\label{eq:snr_g}
\end{equation}
and
\begin{equation}
\frac{\delta^\text{g}_\ell}{\sigma^\text{g}_\ell} = \frac{D^\text{g}_{\ell}}{R^\text{\,g}_\ell} = \frac{\sum_q D^\text{r}_{\ell+q-1}\bar{L}_q}{\sqrt{\sum_q(R^\text{\,r}_{\ell+q-1}\bar{L}_q)^2}}\label{eq:delta_sigma_g},
\end{equation}
for $\ell=1,\dots,n^\text{g}$, and $n^\text{g}\approx n^\text{r}$.

Neglecting effects of the SG filter stopband, each $\delta^\text{g}_\ell$ should be a Gaussian random variable with standard deviation $\sigma^\text{g}_\ell=(R^\text{\,g}_\ell)^{-1}$ and mean $\mu^\text{g}_\ell$. Our definition of $\bar{L}_q$ in Sec.~\ref{sub:lineshape} implies that $\mu^\text{g}_{\ell'}=1$ (equivalently, $E[\delta^\text{g}_{\ell'}/\sigma^\text{g}_{\ell'}]=R^\text{\,g}_{\ell'}$) for a KSVZ axion signal with average misalignment in bin $\ell'$.\footnote{Here and elsewhere in this paper, ``an axion signal in the grand spectrum bin $\ell'$'' should be taken as shorthand for the condition $-0.7\Delta\nu_r < \nu_{\ell'}-\nu_a < 0.3\Delta\nu_r$, where $\nu_\ell$ refers to the frequency at the lower edge of bin $\ell$. For detunings outside this range, the SNR will be larger in a different grand spectrum bin, and we will speak of the signal ``in'' that bin instead.} The small uncertainty in $\mu^\text{g}_{\ell'}$ associated with the range of possible misalignments will contribute to the uncertainty in our exclusion limit, discussed in Appendix~\ref{app:error}. Within $K^\text{g}$ bins of $\ell'$, $0<\mu^\text{g}_\ell<1$, because the overlapping horizontal sum correlates nearby grand spectrum bins.\footnote{It should be emphasized that these correlations are independent of, and would occur even in the absence of, the correlations between \textit{combined spectrum} bins responsible for $\xi^\text{r}\neq1$. The implications of these grand spectrum correlations for the analysis will be discussed further in Sec.~\ref{sub:target_confidence}.} Of course, $\mu^\text{g}_\ell=0$ for $|\ell-\ell'|\geq K^\text{g}$. 

Empirically, $\delta^\text{g}_\ell/\sigma^\text{g}_\ell$ [histogrammed in Fig.~\ref{fig:data_hist}(c)] has a Gaussian distribution with mean 0 and standard deviation $\xi=0.93$. We saw above that correlations within each bin of the rebinned spectrum already reduced the width of the histogram by a factor $\xi^\text{r}=0.98$, which implies that the reduction we can attribute specifically to correlations between \textit{different} rebinned spectrum bins is $\xi^\text{g}=\xi/\xi^\text{r}=0.95$.

Setting aside the issue of correlations, we can gain further insight into the properties of our ML-weighting horizontal sum by considering how it differs from the corresponding step in the ADMX haloscope analysis procedure. ADMX analyses tailored to the detection of virialized axions have consistently used $\Delta\nu_b/\Delta\nu_a$ approximately a factor of 10 larger than in the present analysis and $K^\text{r}=1$ (i.e., no rebinning after data combining). The original ADMX analysis~\cite{ADMX2001,ADMX2000,daw1998} took the grand spectrum to be an unweighted sum of $K^\text{g}=6$ combined spectrum bins. This is not quite the same as setting $\bar{L}_q=1$ in Eqs.~\eqref{eq:snr_g} and \eqref{eq:delta_sigma_g} because our sums are still ML-weighted by $(\sigma^\text{r}_\ell)^{-2}$ in this limit. However, as noted in Sec.~\ref{sub:rebinned_spectrum}, the variation in $\sigma^\text{c}_k$ on the relevant scales is small enough that in practice there is not much difference. 

Thus we will compare our ML analysis to the unweighted $K^\text{g}$-bin sum in the limit that $\sigma^\text{r}_\ell$ (equivalently $R^\text{\,r}_\ell$) is equal in all contributing bins. In this limit, the grand spectrum SNR may be written in the form
\begin{equation}
R^\text{\,g}_\ell=F\big(K^\text{g},\Delta\nu_r,\{\bar{L}_q\}\big)K^\text{g}\sqrt{\Delta\nu_r}R^\text{\,r}_\ell,
\label{eq:uw_limit}
\end{equation}
where we have introduced a figure of merit $F$ to encode the dependence of $R^\text{\,g}_\ell$ on $K^\text{r}$, $K^\text{g}$, and $\bar{L}_q$. It becomes apparent that $R^\text{\,g}_\ell$ only depends on these quantities through $F$ when we rewrite the rebinned spectrum SNR in the form 
\begin{equation}
R^\text{\,r}_\ell=[(P_\ell/K^\text{g})/(k_BT_\ell)]\sqrt{\tau/\Delta\nu_r},\label{eq:snr_r_approx}
\end{equation}
where $P_\ell$ ($T_\ell$) is an appropriately weighted average of the total axion conversion power (noise temperature) in all contributing processed spectrum bins.

For our ML-weighted analysis, we obtain an explicit expression for $F$ by comparing Eq.~\eqref{eq:uw_limit} to Eq.~\eqref{eq:snr_g}:
\begin{equation}
F_\text{ML} = \sqrt{\frac{1}{\Delta\nu_r}\sum_q(\bar{L}_q/K^\text{g})^2}\label{eq:fom_ml}.
\end{equation}
The figure of merit for an unweighted sum follows from Eq.~\eqref{eq:uw_limit} and $R^\text{\,g}_\ell=\sqrt{K^\text{g}}\eta_cR^\text{\,r}_\ell$ (see Sec.~\ref{sub:lineshape}):
\begin{equation}
F_\text{uw} = \frac{1}{\sqrt{K^\text{g}\Delta\nu_r}}\sum_q \bar{L}_q/K^\text{g}\label{eq:fom_rect}.
\end{equation}

For a meaningful comparison between analyses, we must assume the same underlying signal spectrum $f(\nu)$ in both cases. If we also assume that both analyses are characterized by the same values of $K^\text{g}$ and $\Delta\nu_r$, and thus the same $\bar{L}_q$, then $F_\text{uw}$ is just the mean of $\bar{L}_q$ multiplied by $(K^\text{g}\Delta\nu_r)^{-1/2}$, whereas $F_\text{ML}$ is the RMS of $\bar{L}_q$ times the same factor. Thus $F_\text{ML}\geq F_\text{uw}$ independent of any specific features of the lineshape; this is another way to understand the improvement in sensitivity from ML weighting.\footnote{Eq.~\eqref{eq:fom_ml} only quantifies the true improvement in the SNR from a ML analysis if our analysis has assumed the correct signal lineshape, but insofar as the true signal distribution is closer to the nominal lineshape than to a ``boxcar'' of width $K^\text{g}\Delta\nu_r$, the ML analysis will still be more sensitive than an unweighted sum.}

We can also use Eqs.~\eqref{eq:fom_ml} and \eqref{eq:fom_rect} to compare the sensitivity of analyses based on the same model $f(\nu)$ but characterized by different $\Delta\nu_r$ and/or $K^\text{g}$ and thus different $\bar{L}_q$; this is a convenient way to quantify the considerations discussed in Sec.~\ref{sub:lineshape}. For $f(\nu)$ given by Eq.~\eqref{eq:f_dist}, the improvement in the SNR from an optimal ML-weighted analysis relative to an optimal unweighted analysis is about 7.5\%.\footnote{Here ``optimal'' means the SNR is maximized with respect to $\Delta\nu_r$ and $K^\text{g}$ (or, for ML weighting, it is sufficiently close to its asymptotic value). The values of $\Delta\nu_r$ and $K^\text{g}$ adopted for the present analysis are not optimal in this sense, and indeed the SNR for our present analysis is only about 2\% better than the SNR in the optimal unweighted case. However, this optimization does not take into account the fact that the integration time required for rescans increases as we reduce $\Delta\nu_r$, as emphasized at the beginning of Sec.~\ref{sec:rebin}. A better comparison would consider the improvement in SNR for a ML analysis relative to an unweighted analysis that results in comparable total rescan time. Our present ML analysis has $11.5\%$ better SNR than the unweighted analysis with the same $\Delta\nu_r$ and $K^\text{g}$.}

In more recent ADMX analyses~\cite{ADMX2004,yu2004,hotz2013,lyapustin2015}, the grand spectrum is defined as a weighted sum of combined spectrum bins, with weights corresponding to the coefficients of a Wiener Filter (WF). In our notation, the WF weight for the bin $\delta^\text{r}_{\ell+q-1}$ is
\begin{equation}
u^\text{WF}_q = \frac{\bar{L}_q^2}{\bar{L}_q^2+(\sigma^\text{r}_{\ell+q-1})^2},
\label{eq:wf}
\end{equation}
up to a normalization factor. These weights are obtained as solutions to the least-squares minimization of the difference between the noisy observations $\delta^\text{r}_{\ell+q-1}$ and the mean power $\bar{L}_q$ independently in each bin. In the high-SNR limit $\sigma^\text{r}_{\ell+q-1}\ll \bar{L}_q$, $u^\text{WF}_q\rightarrow1$, whereas in the low-SNR limit, $u^\text{WF}_q\rightarrow (\bar{L}_q/\sigma^\text{r}_{\ell+q-1})^2$. In neither limit do they agree with the unnormalized ML weights,\footnote{Here we are comparing the coefficients of the bins $\delta^\text{r}_{\ell+q-1}$ in the ML and WF analyses. The ML weights are more properly defined as the coefficients of the rescaled bins $\delta^\text{r}_{\ell+q-1}/\bar{L}_q$. With this definition the numerator is $(\bar{L}_q/\sigma^\text{r}_{\ell+q-1})^2$, but there is no such rescaling step in the WF analysis.} $u^\text{ML}_q = \bar{L}_q/(\sigma^\text{r}_{\ell+q-1})^2$.

The origin of this discrepancy is the fact that, while the ML and WF schemes are both based on least-squares optimization, they are obtained by minimizing the mean squared error with respect to different quantities: the ML procedure yields the least-squares optimal estimate of the mean power excess in the (appropriately rescaled) contributing bins (and thus results in larger SNR than all other unbiased analyses, as noted in Sec.~\ref{sub:combine}), whereas the WF procedure yields the least-squares optimal estimates of the weights that most robustly undo the smearing of the axion lineshape due to the presence of noise. In our view, the ML scheme relates more directly to the fundamental quantities of interest in the haloscope search.

Finally, we briefly note one more practical difference between the WF and ML methods: the WF weights depend on the SNR, whereas the ML weights only depend on the shape of the axion signal independent of any overall normalization. In practice the WF should be evaluated at an estimate of the average threshold sensitivity $|g^\text{min}_\gamma|$ to be obtained from the analysis. In the high-SNR limit, the WF sum becomes unweighted, and the SNR improvement from ML weighting may be estimated from $F_\text{ML}/F_\text{uw}$ as noted above.

\subsection{Accounting for correlations}\label{sub:correlations}
In the discussion above we noted two distinct effects on the rebinned spectrum [Eqs.~\eqref{eq:d_r} and \eqref{eq:snr_r}] and grand spectrum [Eqs.~\eqref{eq:snr_g} and \eqref{eq:delta_sigma_g}] due to correlations between nearby combined spectrum bins. First, we have not used the correct expression for the variance of a weighted sum of correlated Gaussian random variables in Eqs.~\eqref{eq:snr_r} and \eqref{eq:snr_g}. Second, in the presence of correlations, the weights we have used are not actually the optimal ML weights. The former effect is responsible for $\xi^\text{r},\xi^\text{g}\neq1$; note that it is completely independent of whether or not the weights are optimal. We will consider the effect on the variance first; doing so will allow us to estimate the sum of off-diagonal elements in the relevant covariance matrices, and thus quantify the deviation from the optimal weights.

The most general expression for the variance of a weighted sum of $K$ Gaussian random variables $X_q$ is 
\begin{eqnarray}
\text{Var}\Big(\sum_qw_qX_q\Big) = \ &&\sum_q w_q^2\,\text{Var}(X_q) +\cdots \label{eq:var_general}\\
&&\cdots+2\sum_{q}\sum_{q'=1}^{q-1}w_qw_{q'}\text{Cov}(X_q,X_{q'}).\nonumber
\end{eqnarray}
We will apply this expression to obtain the correct variance $(\hat{\sigma}^{\text{g}}_\ell)^2$ of the $\ell$th grand spectrum bin. With $X_q=\delta^\text{r}_{\ell+q-1}/\bar{L}_q$ and the grand spectrum weights used in Sec.~\ref{sub:grand_spectrum}, we obtain
\begin{eqnarray}
\big(\hat{\sigma}^{\text{g}}_\ell\big)^2 = \big(\sigma^\text{g}_\ell\big)^2+2\big(\sigma^\text{g}_\ell\big)^4\sum\limits_{q=1}^{K^{g}}\sum\limits_{q'=1}^{q-1}\frac{\bar{L}_q\bar{L}_{q'}\Sigma^\text{r}_{\ell qq'}}{(\sigma^\text{r}_{\ell+q-1}\sigma^\text{r}_{\ell+q'-1})^2},\ \ \ \ \ \ \label{eq:sigma_g_hat}
\end{eqnarray}
where $\Sigma^\text{r}_{\ell qq'} = \text{Cov}(\delta^\text{r}_{\ell+q-1},\delta^\text{r}_{\ell+q'-1})$, and the factor of $(\sigma^\text{g}_\ell)^4$ multiplying the second term comes from the normalization of the ML weights. The analogous expression for the correct variance of the $\ell$th rebinned spectrum bin is
\begin{equation}
\big(\hat{\sigma}^{\text{r}}_\ell\big)^2 = \big(\sigma^\text{r}_\ell\big)^2+2\big(\sigma^\text{r}_\ell\big)^4\sum\limits_{k=k_i(\ell)}^{k_f(\ell)}\sum\limits_{k'=k'_i(\ell)}^{k-1}\frac{\Sigma^\text{c}_{kk'}}{(\sigma^\text{c}_{k}\sigma^\text{c}_{k'})^2}, \label{eq:sigma_r_hat}
\end{equation}
with $\Sigma^\text{c}_{kk'} = \text{Cov}\big(\delta^\text{c}_{k},\delta^\text{c}_{k'}\big)$. 

Having established the requisite formalism, we can now ask whether taking correlations into account explains the observed reduction of the grand spectrum and rebinned spectrum standard deviations. We see immediately that $\hat{\sigma}^{\text{g}}_\ell$ can be smaller than $\sigma^{\text{g}}_\ell$ if the sum over off-diagonal elements of the covariance matrix is on average slightly negative. Formally the ratio $\hat{\sigma}^{\text{g}}_\ell/\sigma^{\text{g}}_\ell$ is frequency-dependent, but if nonzero $\Sigma^\text{r}_{\ell qq'}$ is a consequence of the stopband properties of the SG filter, we should expect the correlation matrix $\rho^\text{r}_{\ell qq'}=\Sigma^\text{r}_{\ell qq'}/(\sigma^\text{r}_{\ell+q-1}\sigma^\text{r}_{\ell+q'-1})$ to depend only on the bin spacing $\Delta q= q-q'$. Analogous arguments also apply to the ratio $\hat{\sigma}^{\text{r}}_\ell/\sigma^{\text{r}}_\ell$. Thus we expect
\begin{equation}
\xi^\text{g}=\hat{\sigma}^{\text{g}}_\ell/\sigma^{\text{g}}_\ell \label{eq:xi_g}
\end{equation}
and
\begin{equation}
\xi^\text{r}=\hat{\sigma}^{\text{r}}_\ell/\sigma^{\text{r}}_\ell\label{eq:xi_r}
\end{equation}
in the case of filter-induced correlations. 

We used a simulation to show that the observed values of $\xi^\text{r}$ and $\xi^\text{g}$ are indeed fully explained by processed spectrum correlations imprinted by the SG filter. Each iteration in the $\xi^\text{g}$ simulation generates a set of $m$ 14020-bin Gaussian white noise spectra with mean 1 and standard deviation $\sigma^\text{p}=1/\sqrt{\Delta\nu_b\tau}$, multiplies each spectrum by a random sample baseline derived from data, then uses the baseline removal procedure described in Sec.~\ref{sec:baseline} to obtain a set of simulated processed spectra.\footnote{The sample baselines used here and in the simulation described in Sec.~\ref{sub:axion_atten} were each obtained by applying a high-order SG filter (as in Sec.~\ref{sub:badbins}) to the average of about 50 consecutive raw spectra after removing contaminated bins.} The $m$ processed spectra are averaged without weighting or offsets to obtain a single simulated combined spectrum, in which we average non-overlapping 10-bin segments. We calculate the product of each pair of bins with $0 \leq \Delta q \leq 5$ in the simulated rebinned spectrum. Averaging each such product over $\approx500$ iterations of the simulation, we obtain reasonably precise estimates of $(\sigma^{\text{r}}_\ell)^2$ and $\Sigma^\text{r}_{\ell qq'}$ for each bin $\ell$ in the rebinned spectrum. Then we calculate $\sigma^{\text{g}}_\ell$ and $\hat{\sigma}^{\text{g}}_\ell$ from Eqs.~\eqref{eq:snr_g} and \eqref{eq:sigma_g_hat}, and $\xi^\text{g}$ from Eq.~\eqref{eq:xi_g}. 

We find that $\xi^\text{g}=0.95$ is constant throughout the analysis band, independent of $m$ for values ranging from $m=1$ out to at least $m=400>\text{max}(m_k)$ and independent of $\tau$ out to at least $\tau=900\text{ s}$.\footnote{Our simulation and Eq.~\eqref{eq:xi_g} measure $\xi^\text{g}$ rather than $\xi=\xi^\text{g}\xi^\text{r}$ because we use $\sigma^{\text{r}}_\ell$ rather than $\hat{\sigma}^{\text{r}}_\ell$ in Eqs.~\eqref{eq:sigma_g_hat} and \eqref{eq:snr_g}. Note also that $m_k$ is itself an upper bound on the averaging in each bin, because contributing spectra are not uniformly weighted.} From an analogous simulation to quantify the effects of correlations on the rebinned spectrum we obtain a constant $\xi^\text{r}=0.98$. To verify that the implementation of the simulation was correct, we calculate the same quantities from the simulated Gaussian white noise spectra directly (bypassing the steps where we imprint and then remove the baseline); we obtain $\xi^\text{g}=\xi^\text{r}=1$ as expected for this null test.

These results demonstrate conclusively that the observed values of $\xi^\text{r}$ and $\xi^\text{g}$ depend only on the stopband properties of the SG filter. Fig.~\ref{fig:filter} indicates that the filter-induced negative correlations increase at larger bin separations, consistent with the empirical result $1-(\xi^\text{g})^2>1-(\xi^\text{r})^2$. The explicit demonstration that $\xi^\text{g}$ and $\xi^\text{r}$ are independent of $m$ is critical because in the real data $m_k$ varies throughout the combined spectrum: $m$-independence implies that nonuniform weighting and frequency offsets between processed spectra will not affect our results. We conclude that $\xi^\text{r}$ and $\xi^\text{g}$ are frequency-independent, as indeed the numerical agreement between the simulated and observed values already indicates.\footnote{The values of $\xi^\text{g}$ and $\xi^\text{r}$ obtained from the real data were unchanged when we divided the axion search dataset in half in various ways (winter/summer, high/low RF frequency, upper/lower half of analysis band) and constructed the grand spectrum separately from each subset of the data.}

It follows that each grand spectrum bin $\delta^\text{g}_\ell$ is a Gaussian random variable with standard deviation 
\begin{equation}
\tilde{\sigma}^\text{g}_\ell \, = \, \xi\sigma^\text{g}_\ell \, = \, \xi^\text{g}\xi^\text{r}\sigma^\text{g}_\ell.
\label{eq:sigma_g_tilde}
\end{equation}
and mean $\mu^\text{g}_\ell=0$ in the absence of axion signals. Now let us suppose there exists a KSVZ axion in bin $\ell'$ of the grand spectrum. If the only effect of the imperfect SG filter stopband were to correlate the statistical fluctuations of the noise in nearby bins, we would still have $\mu^\text{g}_{\ell'}=1$, since the mean of a weighted sum of Gaussian random variables is independent of whether or not they are correlated. 

However, the imperfect SG filter stopband will also lead to slight attenuation of any locally correlated power excess (e.g., an axion signal) in the raw spectra, so we should expect $\mu^\text{g}_{\ell'}=\eta'<1$. It follows that $\delta^\text{g}_{\ell'}/\tilde{\sigma}^\text{g}_{\ell'}$ is a Gaussian random variable with standard deviation 1 and mean
\begin{equation}
\tilde{R}^\text{\,g}_{\ell'} = \eta'/\tilde{\sigma}^\text{g}_{\ell'} = \eta R^\text{\,g}_{\ell'},
\label{eq:snr_g_tilde}
\end{equation}
where $\eta=\eta'/\xi$. Thus we see that filter-induced \textit{signal} attenuation $\eta'$ actually only reduces the SNR by the smaller factor $\eta$, because the RMS fluctuations of the noise power within the axion bandwidth are also reduced. The procedure we use to quantify $\eta$ is described in detail in Sec.~\ref{sub:axion_atten}; though formally Eq.~\eqref{eq:snr_g_tilde} allows $\eta>1$, we will find that $\eta<1$, indicating that the net effect is indeed reduction of the SNR.

Finally, we can return to the second effect of correlations neglected in the construction of the grand spectrum: in the presence of correlations, neither the rebinned spectrum weights nor the grand spectrum weights are actually the true ML weights. We are now equipped to show that in practice deviations from the optimal weights are negligibly small in both cases. 

We noted in Appendix~\ref{app:mle} that the true ML weights in the presence of correlations are sums over rows of the inverse covariance matrix. Applying the approximation in Eq.~\eqref{eq:weights_cor},\footnote{It can be shown using Eq.~\eqref{eq:sigma_g_hat} that the average of the off-diagonal elements of the correlation matrix is $1.5\big[\big(\xi^\text{g}\big)^2-1\big]/(K^\text{g}-1)\approx-0.035$, where the numerical factor is due to lineshape weighting. Thus a first-order approximation is appropriate.} we find that the (properly normalized) true ML weights for the grand spectrum are
\begin{eqnarray}
\tilde{w}_{\ell q} &&= \frac{\big(\sigma^\text{g}_\ell\big)^2}{2-\big(\xi^\text{g}\big)^2}\Bigg[\frac{\bar{L}_q^2}{\big(\sigma^\text{r}_{\ell+q-1}\big)^2} - \sum_{q'\neq q}\frac{\bar{L}_q\bar{L}_{q'}\Sigma^\text{r}_{\ell qq'}}{\big(\sigma^\text{r}_{\ell+q-1}\sigma^\text{r}_{\ell+q'-1}\big)^2}\Bigg] \nonumber \\
&&= w_{\ell q}^0 + \delta w_{\ell q}. \label{eq:w_tilde}
\end{eqnarray}
Up to an overall change in the normalization, $w_{\ell q}^0=w_{\ell q}$, the ML weights in the absence of correlations. The mean value of $\delta w_{\ell q}$ just compensates for this rescaling such that $\tilde{w}_{\ell q}$ remain normalized. The typical change in the relative weighting is given by the standard deviation of $\delta w_{\ell q}$, which is easy to calculate given the covariances obtained in our simulation: we find that the RMS fractional change in the weights is about 5\%. 

The resulting fractional change in $\delta^\text{g}_\ell$ will be much smaller because it is the average of $K^\text{g}$ 5\% deviations that are mutually negatively correlated (because the weights remain normalized). Thus, the systematic effect from neglecting correlations in the grand spectrum ML weights is small compared to the sources of error we consider in Appendix~\ref{app:error}; the analogous effect in the rebinned spectrum is smaller still due to the smaller value of $1-(\xi^\text{r})^2$.

\section{Candidates and Exclusion}\label{sec:candidates}
Via the procedure described in the previous sections, we have condensed our axion search data into the $2n^\text{g}$ numbers $\delta^\text{g}_\ell/\tilde{\sigma}^\text{g}_\ell$ and $\tilde{R}^\text{\,g}_\ell$. The statistical fluctuations of the total noise power result in a standard normal distribution for the corrected grand spectrum $\delta^\text{g}_\ell/\tilde{\sigma}^\text{g}_\ell$ in the absence of axion signals, and a KSVZ axion signal in a particular bin $\ell'$ would displace the mean of $\delta^\text{g}_{\ell'}/\tilde{\sigma}^\text{g}_{\ell'}$ by $\tilde{R}^\text{\,g}_{\ell'}$. Now we will explain how we use these quantities to interrogate the presence of axion conversion power in our scan range and derive an exclusion limit if there are no persistent signals.

It should be emphasized that we have no \textit{a priori} knowledge of which bin $\ell'$ (if any) corresponds to the axion mass, and the only qualitative difference between an axion signal and a positive excess power fluctuation in any given bin is that a true signal should be persistent across different scans at the same frequency. Thus the best we can do is set a threshold $\Theta$ and define any bin with $\delta^\text{g}_\ell/\tilde{\sigma}^\text{g}_\ell\geq\Theta$ as a \textit{rescan candidate}. In the absence of grand spectrum correlations, we would expect
\begin{equation}
\hat{S}=n^\text{g}\big[1-\Phi(\Theta)\big]
\label{eq:candidates}
\end{equation}
such rescan candidates from statistics alone, where $\Phi(x)$ is the cumulative distribution function of the standard normal distribution. We can then collect sufficient data at each rescan frequency to reproduce the sensitivity in the initial scan (Sec.~\ref{sub:rescan_daq}), and thereby distinguish any real axion signal from statistical fluctuations (Sec.~\ref{sub:rescan_analysis}). 

In light of the above discussion, our proximate task is to determine an appropriate value for $\Theta$. To simplify matters, let us first assume $\tilde{R}^\text{\,g}_\ell=R_T$ is constant throughout the scan range. Perhaps the simplest choice of threshold is $\Theta=R_T$. Taking  $n^\text{g}\approx1.07\times10^5$ for the first HAYSTAC data run and assuming for now that $R_T=5$, we obtain $\hat{S}=0.03$; thus any bin exceeding the threshold is extremely unlikely to be a statistical fluctuation. The problem with this choice of threshold becomes clear when we suppose there is an axion signal with SNR $R_T$ in some bin $\ell'$: then $\delta^\text{g}_{\ell'}/\tilde{\sigma}^\text{g}_{\ell'}$  is a Gaussian random variable with mean $R_T$ and standard deviation 1. $\Theta=R_T$ is a poor choice of threshold because the probability that $\delta^\text{g}_{\ell'}/\tilde{\sigma}^\text{g}_{\ell'}\geq\Theta$ is only 50\%. 

\begin{figure}[t]
\includegraphics[width=0.5\textwidth]{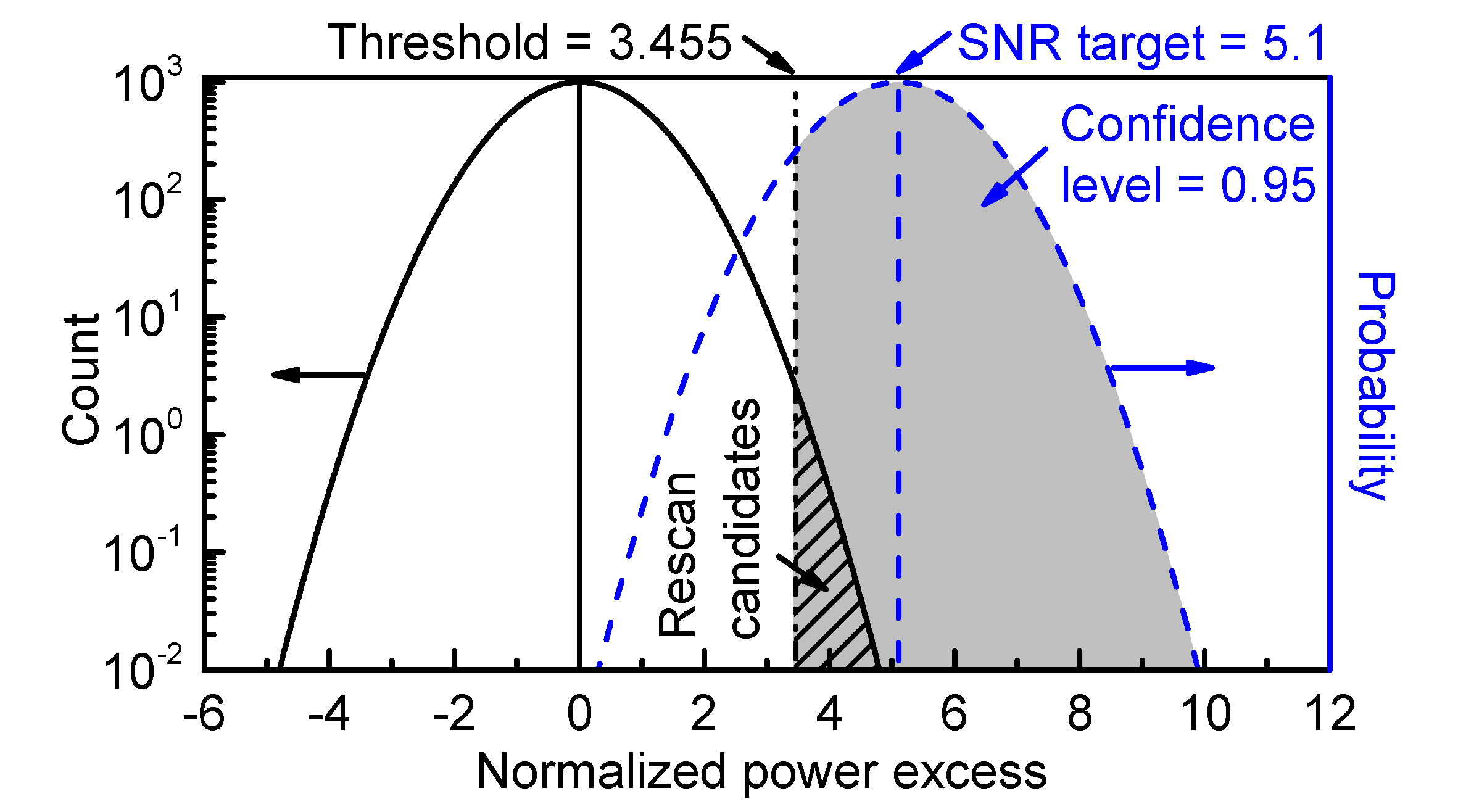}
\caption{\label{fig:confidence} Schematic illustration of the relationship between the SNR target $R_T$, rescan threshold $\Theta$, rescan yield $\hat{S}$, and initial scan confidence level $c_1$. The vertical axis on the left applies to the solid black curve representing the expected standard normal distribution of grand spectrum bins $\delta^\text{g}_\ell/\tilde{\sigma}^\text{g}_\ell$; the integral is the total number of grand spectrum bins $n^\text{g}$. The vertical axis on the right applies to the dashed blue curve, which is a normalized Gaussian distribution with unit standard deviation and mean $R_T$, and represents the expected distribution of excess power $\delta^\text{g}_{\ell'}/\tilde{\sigma}^\text{g}_{\ell'}$ in a single grand spectrum bin $\ell'$ containing an axion signal. The threshold $\Theta$ (dot-dashed vertical line) intersects both distributions: $\hat{S}$ (hatched region) is the integral of the grand spectrum distribution above $\Theta$, and $c_1$ (gray shaded region) is the integral of the signal distribution above $\Theta$.}
\end{figure}

For arbitrary $\Theta$ (again assuming a signal with SNR $R_T$ in bin $\ell'$), the probability that $\delta^\text{g}_{\ell'}/\tilde{\sigma}^\text{g}_{\ell'}\geq\Theta$ in the presence of noise is called the axion search confidence level. If we require a confidence level $\geq c_1$ for the initial scan, the appropriate threshold is
\begin{equation}
\Theta=R_T - \Phi^{-1}(c_1),
\label{eq:theta}
\end{equation}
and the expected rescan yield $\hat{S}$ follows from Eq.~\eqref{eq:candidates}. The relationship between all of these quantities is illustrated in Fig.~\ref{fig:confidence}. In Sec.~\ref{sub:target_confidence} we will see that grand spectrum correlations modify the expected rescan yield slightly, so we should actually expect $\bar{S}<\hat{S}$ candidates.

In the above discussion we assumed constant SNR throughout the scan range, when in fact $\tilde{R}^\text{\,g}_\ell$ varied significantly on scales $\gtrsim1$~MHz in the first HAYSTAC data run, with typical values between 0.7 and 1.2, due to nonuniform tuning and frequency-dependence of the cavity $Q$, form factor, etc. Recall that $\tilde{R}^\text{\,g}_\ell$ is the SNR for an axion signal with photon coupling $|g_\gamma|=|g^\text{KSVZ}_\gamma|$, and as we emphasized in Sec.~\ref{sub:rescale}, our decision to normalize Eq.~\eqref{eq:U_0} to the KSVZ coupling specifically was completely arbitrary. To obtain a frequency-independent threshold, we can simply define
\begin{equation}
G_\ell=\left(R_T/\tilde{R}^\text{\,g}_\ell\right)^{1/2},
\label{eq:g_ell}
\end{equation}
from which it follows that $R_T$ is the SNR for an axion with frequency-dependent coupling
\begin{equation}
|g^\text{min}_\gamma|_\ell=G_\ell|g^\text{KSVZ}_\gamma|.
\label{eq:g_min}
\end{equation}

Eqs.~\eqref{eq:theta} -- \eqref{eq:g_min} completely determine the confidence level at which we can exclude axions as a function of the two-photon coupling $|g_\gamma|$ in each bin of the grand spectrum.\footnote{In contrast, in most ADMX analyses~\cite{ADMX2001,ADMX2000,daw1998,ADMX2004,yu2004} the confidence level is obtained from Monte Carlo. The Monte Carlo procedure involves constructing a mock grand spectrum containing a large number of simulated axion signals with known SNR $R_T$, setting a threshold $\Theta$, and defining $c_1$ as the fraction of simulated axions flagged as rescan candidates; the simulation may be repeated many times to determine the behavior of $c_1$ as a function of $R_T$ and/or $\Theta$. This more involved approach was originally adopted to circumvent the effects of correlations on the horizontal sum, which we have shown we can quantify.} By varying $\Theta$ we can adjust the tradeoff between $\bar{S}$ (which determines the total time we need to spend acquiring rescan data) and $|g^\text{min}_\gamma|_\ell$, the minimum coupling to which our search is sensitive.\footnote{A coupling $|g_\gamma|_\ell>|g^\text{min}_\gamma|_\ell$ corresponds to a signal with $\text{SNR} > R_T$. At any given threshold $\Theta$, a result $\delta^\text{g}_{\ell'}/\tilde{\sigma}^\text{g}_{\ell'}<\Theta$ implies that axions with mass $\nu_{\ell'}$ and coupling $|g^\text{min}_\gamma|_{\ell'}$ are excluded with confidence $c_1$, and axions with the same mass but larger coupling are excluded at higher confidence.} The validity of these expressions hinges crucially on our ability to regard each grand spectrum bin as a sample drawn from a Gaussian distribution with known mean and standard deviation. We demonstrated in Sec.~\ref{sub:correlations} that we are justified in treating any bin that does not contain an axion signal in this way. In Sec.~\ref{sub:axion_atten}, we will show that we can also quantify the mean and standard deviation for any bin containing an axion signal, thus validating the above procedure. Then we will return to the choice of threshold in Sec.~\ref{sub:target_confidence}.

\subsection{SG filter-induced attenuation}\label{sub:axion_atten}
In Sec.~\ref{sub:correlations} we claimed that with a KSVZ axion signal in the grand spectrum bin $\ell'$, $\delta^\text{g}_{\ell'}/\tilde{\sigma}^\text{g}_{\ell'}$ is a Gaussian random variable with mean given by Eq.~\eqref{eq:snr_g_tilde} and standard deviation 1. Let us consider each of the claims here more carefully. In writing Eq.~\eqref{eq:snr_g_tilde} we have implicitly assumed that $\eta$ is frequency-independent. While we could of course write a similar expression with $\eta\rightarrow\eta_\ell$, the utility of Eq.~\eqref{eq:snr_g_tilde} lies in the fact that we only need to specify a single correction factor to know the SNR in each bin. It is reasonable to expect $\eta'$ (and thus $\eta$) to be frequency-independent, as $\eta'\neq1$ is ultimately a consequence of the same imperfect SG filter stopband attenuation that led to frequency-independent $\xi\neq1$. We will see more directly that $\eta$ is constant below.

In claiming that the distribution of excess power about the mean value $\tilde{R}^\text{\,g}_{\ell'}$ is Gaussian with standard deviation 1, we are only assuming that the statistical fluctuations of the total noise power in any given bin are independent of whether or not that bin also includes excess power due to axion conversion. This is certainly a valid assumption for the raw data. We quantify $\eta$ using a simulation which will also demonstrate explicitly that this assumption still holds in the grand spectrum.

The simulation we use to quantify $\eta$ begins by defining a set of $m$ uniformly spaced simulated mode frequencies $\nu_{ci}$ and a frequency axis for a 14020-bin spectrum with resolution $\Delta\nu_b=100$~Hz centered on each mode frequency. With a tuning step size of $1.402\text{ MHz}/m$, the low-frequency end of the last spectrum lines up with the high-frequency end of the first, and $m_k$ (the number of spectra contributing to the $k$th combined spectrum bin) will vary from 1 to $m$ over the tuning range. Each spectrum is initialized to the expected signal power for an axion with coupling $|g_\gamma|$ and mass $\nu_a$ near the middle of the simulated frequency range. The signal power in the $j$th bin of the $i$th spectrum is proportional to the integral of Eq.~\eqref{eq:f_dist} over an interval $\Delta\nu_b$ around the RF frequency $\nu_k$ for which $\Gamma_{ijk}=1$, multiplied by the \textit{inverse} of the rescaling factor defined in Sec.~\ref{sub:rescale}. For simplicity we take $Q_{Li}$, $C_i$, $\beta_i$, and $T_{ij}$ to be the same for each spectrum $i$, so that variation in the rescaling factor only comes from the $j$-dependence of $T_{ij}$ and the Lorentzian mode profile.

After the initialization described above, each iteration of the simulation adds simulated Gaussian white noise with mean 1 and standard deviation $\sigma^\text{p}$ to each spectrum, and sends the full set of spectra through two analysis pipelines in parallel. The ``standard'' analysis multiplies each spectrum by a random sample baseline (see Sec.~\ref{sub:correlations}), then applies the baseline removal procedure of Sec.~\ref{sec:baseline} to obtain simulated processed spectra, and finally combines the simulated spectra both vertically and horizontally, following the procedure of Sec.~\ref{sec:rescale_combine}-\ref{sec:rebin}, to obtain a simulated grand spectrum. The ``ideal'' analysis is identical except that it bypasses the steps that imprint and then remove the baseline; thus we expect no effects associated with the SG filter in the ideal grand spectrum.

From each iteration, we record the values of the normalized power excess $\delta^\text{g}_{\ell}/\sigma^\text{g}_{\ell}$ (not $\delta^\text{g}_{\ell}/\tilde{\sigma}^\text{g}_{\ell}$) and the uncorrected SNR $R^\text{\,g}_{\ell}$ in $\approx 2K^\text{g}$ bins around $\nu_{\ell'}\approx\nu_{a}$ in both the standard and ideal grand spectra. We also record the value of $\delta^\text{g}_{\ell}/\sigma^\text{g}_{\ell}$ in a few other bins far from $\nu_{\ell'}$ in different parts of the standard grand spectrum. The coupling $|g_\gamma|$ is chosen to yield $R^\text{\,g}_{\ell'}\approx5$ for $m\approx200$. We let the simulation run for $\sim10^4$ iterations, after which we can histogram the distribution of any of the recorded bins across iterations. 

\begin{figure}[t]
\includegraphics[width=0.5\textwidth]{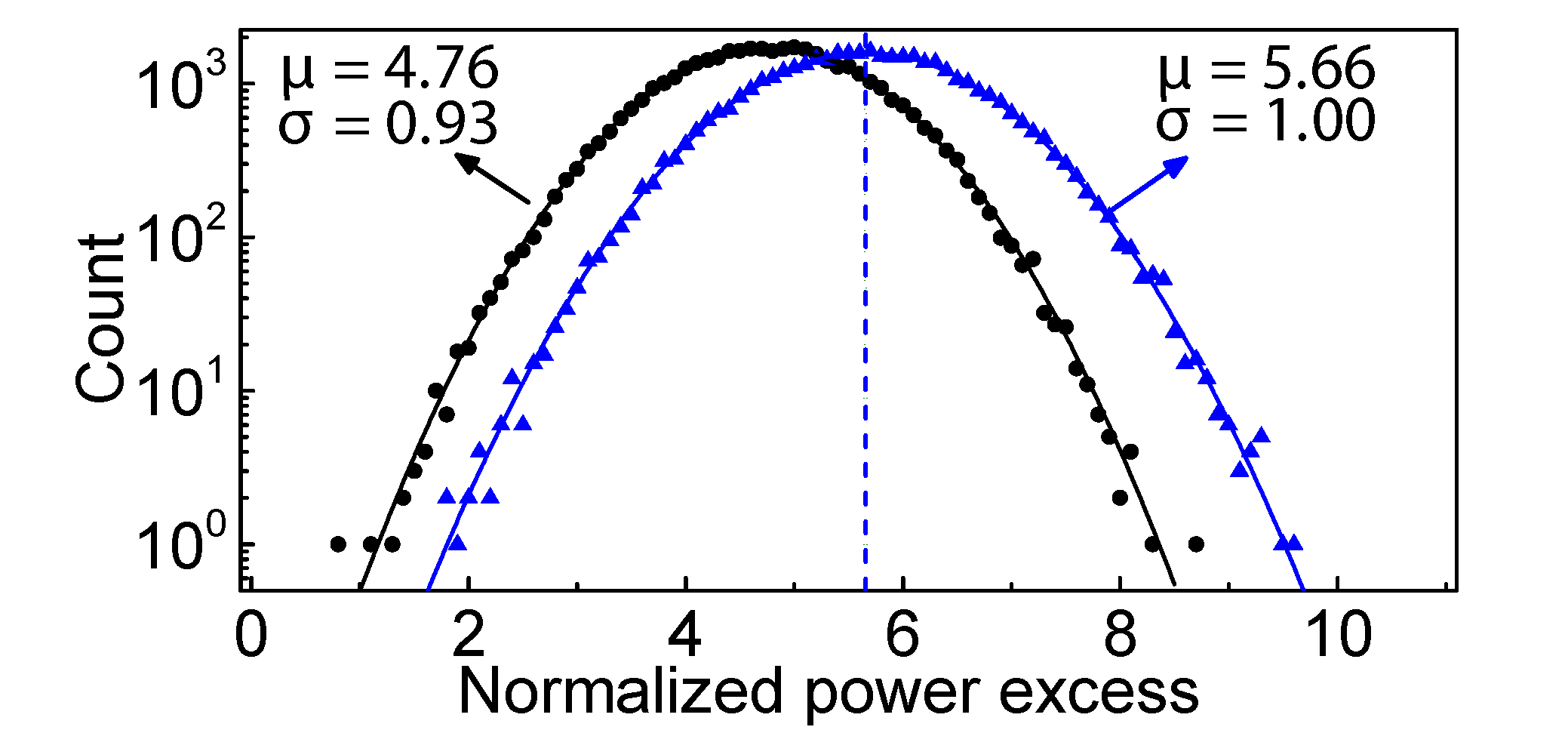}
\caption{\label{fig:simu_hist} The results of a simulation to quantify the filter-induced attenuation $\eta$ of the axion search SNR for a simulated axion signal with $\nu_a=\nu_{\ell'}$ in the presence of Gaussian white noise. The grand spectrum power excess $\delta^\text{g}_{\ell'}/\sigma^\text{g}_{\ell'}$ is obtained in each iteration of the simulation using two different analysis pipelines. For each analysis, we have histogrammed the distribution of excess power across iterations of the simulation (data points) and fit the histogram with a Gaussian (solid curves); the best-fit mean and standard deviation are shown on the plot. When the contributing spectra are combined and rebinned directly (blue triangles), the distribution is Gaussian with standard deviation 1 and mean equal to the calculated SNR $R^\text{\,g}_{\ell'}$ (indicated by the dashed line). When each contributing spectrum is scaled by an empirical baseline and the standard analysis procedure is then applied (black circles), the distribution is still Gaussian but with a smaller standard deviation $\xi=0.93$, equal to the value obtained in real data. The ratio of the two mean values is $\eta'$, from which we obtain $\eta=\eta'/\xi=0.90$.}
\end{figure}

We are primarily interested in comparing the excess power distribution in bin $\ell'$ of the standard grand spectrum to the excess power distribution in the same bin of the ideal grand spectrum. This comparison is shown in Fig.~\ref{fig:simu_hist}. We see that in the ideal grand spectrum, the fluctuations of the noise power in the bin $\ell'$ containing an axion signal are Gaussian with standard deviation $\sigma^\text{g}_{\ell'}$, as they would be in any other bin; we can also see that our standard analysis procedure correctly calculates the SNR $R^\text{\,g}_{\ell}$ in the absence of SG filter effects.\footnote{That is, $E[\delta^\text{g}_{\ell'}/\sigma^\text{g}_{\ell'}]_\text{i} = (R^\text{\,g}_{\ell'})_\text{i} = (R^\text{\,g}_{\ell'})_\text{s}$, where the subscripts ``s'' and ``i'' refer to the standard and ideal analyses, respectively. The calculated values of $(R^\text{\,g}_{\ell})_\text{i}$ and $(R^\text{\,g}_{\ell})_\text{s}$ are nearly equal in each bin $\ell$ because they only depend on the measured data through the distribution of processed spectrum standard deviations (Sec.~\ref{sub:stats}), which is changed only very marginally by the presence of the SG filter.}

In the standard grand spectrum, we find that the fluctuations of the noise power in bin $\ell'$ are still Gaussian, with a reduced standard deviation $\tilde{\sigma}^\text{g}_{\ell'}=\xi\sigma^\text{g}_{\ell'}$ and $\xi=0.93$ as in real data. We also obtain Gaussian fluctuations with standard deviation $\tilde{\sigma}^\text{g}_{\ell}$ in other bins $\ell$ far from the axion mass. This provides strong evidence for the assertion that \textit{each} $\delta^\text{g}_{\ell}$ is a Gaussian random variable with standard deviation $\tilde{\sigma}^\text{g}_{\ell}$, whether or not bin $\ell$ contains an axion signal. Since we histogrammed $\delta^\text{g}_{\ell'}/\sigma^\text{g}_{\ell'}$ rather than $\delta^\text{g}_{\ell'}/\tilde{\sigma}^\text{g}_{\ell'}$ to more directly see the effects of the SG filter on $\sigma^\text{g}_{\ell'}$, the ratio of the two bin $\ell'$ excess power distributions measures $\eta'$ rather than $\eta$; formally, $E[\delta^\text{g}_{\ell'}/\sigma^\text{g}_{\ell'}]_\text{s}/E[\delta^\text{g}_{\ell'}/\sigma^\text{g}_{\ell'}]_\text{i}=(\mu^\text{g}_{\ell'})_\text{s}(R^\text{\,g}_{\ell'})_\text{s}/(R^\text{\,g}_{\ell'})_\text{i}=\eta'$. Dividing the value of $\eta'$ obtained this way by $\xi$ we find $\eta=0.90$.

This result for $\eta$ is independent of $m$ out to at least $m=400$ (c.f. the analogous result for $\xi$ from the simulation described in Sec.~\ref{sub:correlations}). It also does not change if we vary $|g_\gamma|^2$ by $\pm50\%$; this linearity implies that we do not have worry about the simulation reproducing the precise value of $R_T$ to be used in the analysis. Finally, $\eta$ is independent of the misalignment of $\nu_a$ relative to the grand spectrum binning: with arbitrary misalignment $E[\delta^\text{g}_{\ell'}/\sigma^\text{g}_{\ell'}]_\text{i}\neq R^\text{\,g}_{\ell'}$, but $E[\delta^\text{g}_{\ell'}/\sigma^\text{g}_{\ell'}]_\text{s}$ always changes by the same factor.\footnote{For the simulation plotted in Fig.~\ref{fig:simu_hist}, we set $\nu_a$ to coincide with a bin boundary in the rebinned spectrum, and used $L_q(\delta\nu_r=0)$ rather than $\bar{L}_q$ in the grand spectrum weights. This choice made it simpler to confirm $E[\delta^\text{g}_{\ell'}/\sigma^\text{g}_{\ell'}]_\text{i} = R^\text{\,g}_{\ell'}$ and thereby verify the correct implementation of the analysis procedure (recall that with the lineshape $\bar{L}_q$, $E[\delta^\text{g}_{\ell'}/\sigma^\text{g}_{\ell'}]_\text{i} = R^\text{\,g}_{\ell'}$ if we average over the range of possible misalignments, but is not necessarily true for any given misalignment). We confirmed that we obtain the same value of $\eta$ using $\bar{L}_q$ in the grand spectrum weights.}

Taken together, the results of the simulation are entirely consistent with the interpretation of $\eta\neq1$ as a result of the imperfect stopband attenuation of the SG filter. Thus we conclude that Eq.~\eqref{eq:snr_g_tilde} correctly describes the SNR in each bin of the grand spectrum. Although we have seen that filter-induced attenuation is a small effect, we may still ask whether we can avoid this slight SNR degradation by using different SG filter parameters. This question is explored further in Appendix~\ref{app:sg_params}. 

\subsection{Setting the threshold}\label{sub:target_confidence}
We now return to the question of how we set appropriate values for $c_1$ and $\Theta$. In many subfields of particle physics it is conventional to cite parameter exclusion limits at $90\%$ or $95\%$ confidence. For the analysis of the first HAYSTAC data run we set $c_1=0.95$, for which Eq.~\eqref{eq:theta} becomes $R_T - \Theta = 1.645$.

\begin{figure}[t]
\includegraphics[width=0.5\textwidth]{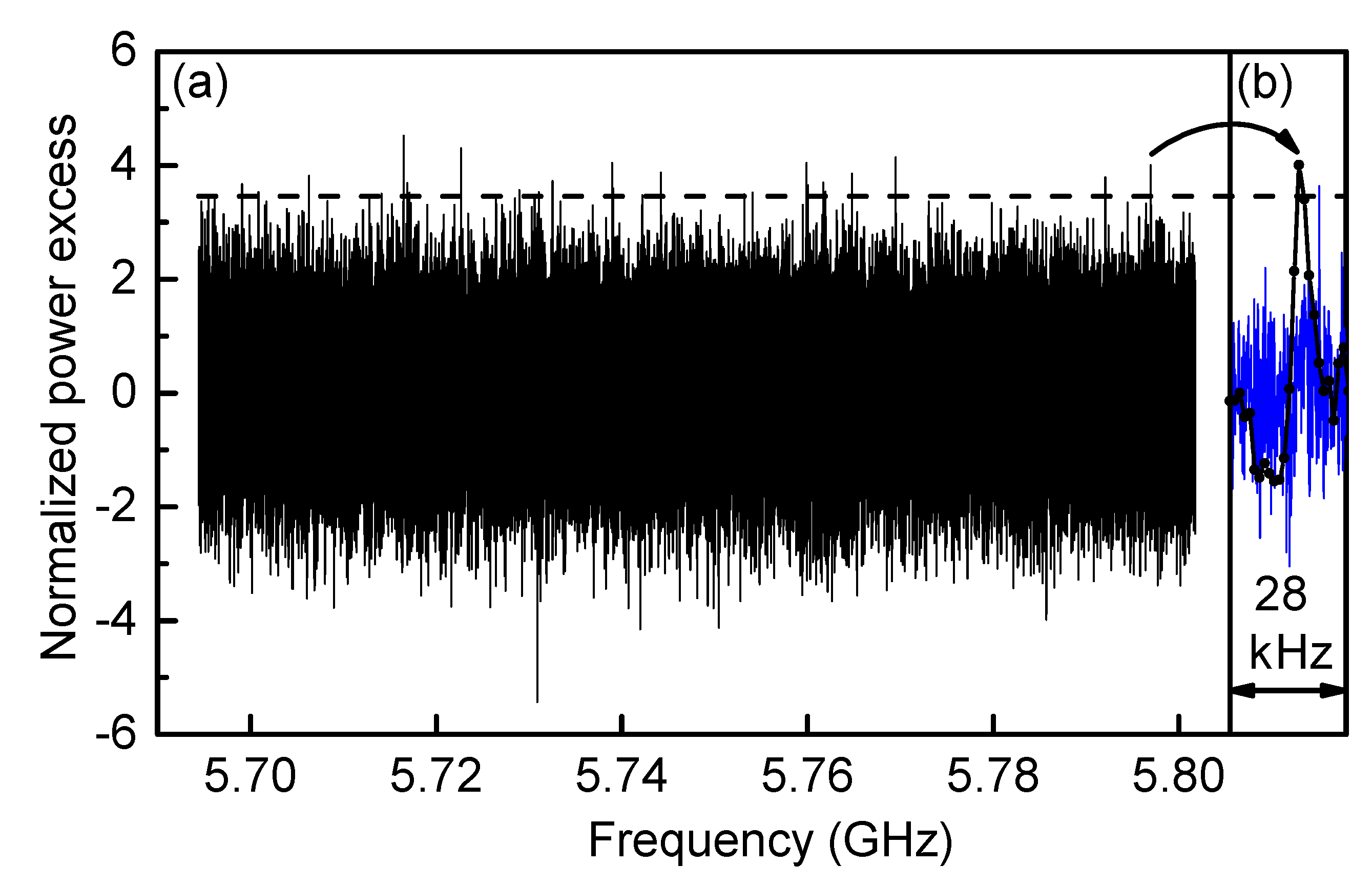}
\caption{\label{fig:threshold} (a) The corrected grand spectrum $\delta^\text{g}_\ell/\tilde{\sigma}^\text{g}_\ell$ plotted as a function of frequency $\nu_\ell$ along with the (frequency-independent) threshold $\Theta=3.455$. The 28 rescan candidates are those bins for which $\delta^\text{g}_\ell/\tilde{\sigma}^\text{g}_\ell\geq\Theta$; some are hard to see because of the finite line thickness. (b) In black, the corrected grand spectrum in a small region around the highest-frequency rescan candidate. The vertical scale is the same as in (a) and the horizontal scale has been expanded by a factor of $\sim500$. In blue, the combined spectrum $\delta^\text{c}_k/\sigma^\text{c}_k$ in the same frequency range. As expected, the large power excess at the candidate frequency in the grand spectrum is due to $\sim K^\text{r}K^\text{g}$ consecutive combined spectrum bins in which the power excess is on average slightly positive, rather than a few combined spectrum bins with extremely high power excess.}
\end{figure}

Given a value for $c_1$, the considerations that enter into the choice of $\Theta$ are best illustrated with an explicit example. For the first HAYSTAC data run we chose $\Theta=3.455$, corresponding to a threshold SNR of $R_T=5.1$. $S=28$ grand spectrum bins exceeded this threshold and were flagged as rescan candidates. The corrected grand spectrum $\delta^\text{g}_\ell/\tilde{\sigma}^\text{g}_\ell$ and threshold $\Theta$ are shown in Fig.~\ref{fig:threshold}. Visual inspection suffices to demonstrate qualitatively the important point that many of the candidates are quite marginal; more precisely 11 of the 28 candidates exceed the threshold by less than $\Delta\Theta=0.1$, implying that we could have eliminated all of these candidates at the cost of a $\Delta G_\ell/G_\ell = [(R_T+\Delta\Theta)/R_T]^{1/2}-1\approx 1\%$ degradation of our exclusion limit. Conversely, reducing the threshold by 0.1 would have improved the exclusion limit by 1\% at the cost of 10 additional rescan candidates.

Of course, this strong dependence of the rescan yield on the threshold is just what we expect from Gaussian noise statistics.\footnote{One consequence of the sensitivity of $S$ to small changes in $\Theta$ is that the rescan lists for even relatively similar analyses (characterized by e.g., slightly different choices of $K^\text{r}$ and/or $K^\text{g}$, or WF instead of ML weights) typically only overlap by $\sim60-80\%$.} It is common for haloscope searches to set $R_T=5$ in estimates of the sensitivity that can be achieved with a given set of design parameters, but there is nothing special about this choice. In principle, $\Theta$ (and thus $R_T$) should be chosen to optimize the coupling sensitivity at fixed \textit{total} integration time (initial scan plus rescans). For any haloscope detector using a coherent receiver, rescans are intrinsically less efficient than the initial scan, so the time spent on rescans should be a small fraction of the time spent acquiring the initial scan data.\footnote{This is because each measurement improves the SNR in $\sim\Delta\nu_c/\Delta\nu_a$ non-overlapping grand spectrum bins simultaneously. In the initial scan each of these bins is relevant whereas in rescans we only care about the SNR within $K^\text{g}$ bins of each candidate. In practice the discrepancy is further exacerbated by the fact that rescans are more difficult to fully automate than the continuous initial scan, and thus have worse live-time efficiency.} By this criterion, the optimal threshold is higher still than the value $\Theta=3.455$ we adopted for the first HAYSTAC data run.

Thus far in this section we have discussed the real data rescan yield $S(\Theta)$ without reference to any theoretical model. To confirm that we have obtained a rescan yield consistent with statistics, we must take into account the fact that any two grand spectrum bins $\ell$ and $\ell'$ will be positively correlated if $|\ell-\ell'|\leq K^\text{g}-1$ because the segments of the rebinned spectrum contributing to the bins $\ell$ and $\ell'$ will overlap. These correlations imply that both real axion signals and statistical fluctuations in the excess power are likely to result in several adjacent bins exceeding the threshold. We should not define all such bins as rescan candidates because they are largely redundant. Thus we add bins to the list of rescan candidates in order of decreasing excess power, and remove $K^\text{g}-1$ bins on either side of each candidate from consideration before moving on to the next candidate. The values of $S(\Theta)$ cited above were obtained using this procedure, which was originally proposed by Ref.~\cite{daw1998}. 

Recall that $\hat{S}(\Theta)$ defined by Eq.~\eqref{eq:candidates} describes the expected rescan yield for a grand spectrum whose bins are samples drawn from a standard normal distribution: it does not depend on whether or not nearby bins are correlated provided $n^\text{g}$ is much larger than the correlation length. Thus, Eq.~\eqref{eq:candidates} would correctly describe the expected rescan yield if we did not exclude the correlated bins around each candidate; given that we do exclude these bins, we should actually expect a rescan yield $\bar{S}(\Theta)<\hat{S}(\Theta)$. Note also that though the presence of grand spectrum correlations affects the rescan yield, it does not affect the initial scan confidence level $c_1$.\footnote{For any value of $\nu_a$ within our scan range, there will be some grand spectrum bin $\ell'$ in which the SNR is maximized, and the best limits we can set will come from this bin. $R_T$ is the SNR in bin $\ell'$ if the axion has mass $\nu_a$ and coupling $|g_\gamma| = |g^\text{min}_\gamma|_{\ell'}$ (up to an uncertainty quantified in Appendix~\ref{app:error}); it follows from Eq.~\eqref{eq:theta} that if bin $\ell'$ does not exceed the threshold $\Theta$, we can exclude such axions with confidence $c_1$. The non-observation of excess power above the threshold in adjacent correlated bins just gives us an additional, strictly less restrictive constraint on the coupling of the axion with mass $\nu_a$.} Our procedure for cutting correlated bins from the rescan yield \textit{will} affect the rescan analysis procedure, discussed in Sec.~\ref{sub:rescan_analysis}.

We obtain the $\Theta$-dependence of the expected rescan yield $\bar{S}(\Theta)$ from a simple simulation. We generate a simulated rebinned spectrum containing Gaussian white noise, apply the ML-weighted sum of Sec.~\ref{sub:grand_spectrum} to obtain a simulated grand spectrum, and then flag rescan candidates with the same procedure used for real data, cutting $K^\text{g}-1$ bins on either side of each candidate. We repeat this simulation with different values of $\Theta$ between 2.3 and 4.3, and then repeat it $\approx50$ times at our chosen value of $\Theta=3.455$ to obtain a range of probable values for $\bar{S}$.

From this simulation we obtain $\bar{S}(\Theta)$ consistently smaller than $\hat{S}(\Theta)$ as expected: at $\Theta=3.455$, $\hat{S}=29.5$, $\bar{S} = 24\pm5$ and $S=28$.\footnote{The fact that $S$ is closer to $\hat{S}$ than $\bar{S}$ at $\Theta=3.455$ is just a fluke made possible by the small candidate statistics at such a large threshold. At $\Theta=2.5$, for example, we would have $S=396$, $\bar{S}=372$, and $\hat{S}=588$. Note that for any value of $\Theta$, $\hat{S} > \bar{S} > \hat{S}/K^\text{g}$, where the latter is the rescan yield we would obtain from $n^\text{g}/K^\text{g}\approx n^\text{c}$ uncorrelated bins (this was also noted by Ref.~\citep{daw1998}). The second inequality gets at the reason (anticipated in Sec.~\ref{sec:rebin}) that we did not take $K^\text{r}=1$ and $K^\text{g}=50$ in constructing the grand spectrum: the number of rescan candidates would be much larger at comparable sensitivity even after we ensure that no two candidates fall within $K^\text{g}$ bins of each other.} We conclude that the observed rescan yield $S$ is consistent with statistics -- by itself this result does not disfavor the hypothesis that any of our candidates could be a real axion signal, since the expected variation in $\bar{S}$ is larger than one, and we expect at most one axion in the data set. To settle the question one way or another, we now turn to the acquisition and analysis of rescan data around each candidate.

\section{Rescan data and analysis}\label{sec:rescan}
Three numbers are required to fully characterize each of the $S=28$ candidates obtained from the initial scan data set: the signal frequency $\nu_{\ell(s)}$, the threshold coupling $G_{\ell(s)}$ (relative to the KSVZ coupling), and the properly normalized power excess $\delta^\text{g}_{\ell(s)}/\tilde{\sigma}^\text{g}_{\ell(s)}$, where $\ell(s)$ is the index of the grand spectrum bin that exceeded the threshold and $s=1,\dots,S$. Only the first two quantities explicitly appear in our subsequent analysis (though of course the power excess determines whether any given bin is flagged as a rescan candidate in the first place).

To establish whether any of our rescan candidates is persistent, we must first determine for each candidate the rescan time $\tau^*_s$ required to obtain SNR $R^*_T$ for an axion signal at frequency $\nu_{\ell(s)}$ with coupling $G_{\ell(s)}$. Then we can acquire rescan data at each candidate frequency. The considerations that enter into these steps are described in Sec.~\ref{sub:rescan_daq}.

We can imagine two alternative approaches to processing the rescan data. One possibility is to process the rescan and initial scan data sets together to produce a single combined spectrum, from which we obtain a modified grand spectrum by following the procedure in Sec.~\ref{sec:rebin}. The extra integration time at each candidate frequency implies that each $\tilde{R}^\text{\,g}_{\ell(s)}$ will increase by roughly a factor of $\sqrt{2}$. Since we are interested in probing the same value of $G_{\ell(s)}$, we can impose a higher threshold $\Theta^*_{\ell(s)}$ around each candidate. We can thus ensure that a real axion signal exceeds the new threshold with some desired confidence $c_2$, while simultaneously greatly reducing the probability that a statistical fluctuation does so.

Alternatively, we can process the rescan data separately, following the procedure of Sec.~\ref{sec:baseline} -- \ref{sec:rebin} to produce a rescan grand spectrum, and leaving the initial scan grand spectrum unchanged. The rescan data set should allow us to set a \textit{coincidence threshold} $\Theta^*_{\ell(s)}$ around each candidate frequency which a real axion signal should exceed with confidence $c_2$. If $c_2\approx c_1$, we do not expect $\Theta^*_{\ell(s)}$ to be substantially greater than $\Theta$ in this case, so the probability that a statistical fluctuation exceeds the threshold in any given bin will not change, but it is much less likely that this should happen in any of the same bins as in the initial scan.

If no changes to the analysis procedure are required for rescan data, these two approaches are completely equivalent. Here we take ``separate processing'' approach, which is conceptually cleaner in that we process spectra together whenever we want to \textit{improve} the coupling sensitivity $|g^\text{min}_\gamma|$ and separately when we want to \textit{reproduce} the coupling sensitivity of a previous scan. As we will see in Sec.~\ref{sub:rescan_analysis}, the rescan analysis differs from the initial scan analysis in a few crucial respects, such that we must use separate processing to obtain correct expressions for the coincidence thresholds $\Theta^*_{\ell(s)}$.

\subsection{Rescan data acquisition}\label{sub:rescan_daq}
The most efficient way to acquire rescan data at the candidate frequency $\nu_{\ell(s)}$ is to take one long measurement with the axion-sensitive cavity mode fixed at frequency $\nu_{cs}\approx\nu_{\ell(s)}$. We can calculate the integration time $\tau^*_s$ required to obtain SNR $R^*_T$ by starting with an expression analogous to \eqref{eq:snr_r_approx} and using Eqs.~\eqref{eq:power}, \eqref{eq:noise}, \eqref{eq:uw_limit}, \eqref{eq:snr_g_tilde}, and \eqref{eq:g_ell}. The result is
\begin{equation}
\tau^*_s= \frac{1}{1-\varepsilon}\Bigg[\frac{R^*_Tk_BT_sH\big(\delta\nu_{as}\big)}{\eta^*F_\text{ML}G^2_{\ell(s)}U_0\nu_{cs}C_sQ_{Ls}\frac{\beta_s}{1+\beta_s}}\Bigg]^2,\label{eq:tau_*}
\end{equation}
where $\eta^*=0.76$ is the filter-induced attenuation for the rescan analysis (see Sec.~\ref{sub:rescan_analysis}), the noise temperature $T_s$ is evaluated in the middle of the analysis band, and we have lumped all dependence on the detuning $\delta\nu_{as}=\nu_{cs}-\nu_{\ell(s)}$ into the factor $H(\delta\nu_{as})$ normalized so that $H(0)=1$; we have also assumed that only a fraction $1-\varepsilon$ of the integration time at each cavity setting $\nu_{cs}$ will contribute to improving the SNR at the candidate frequency.

Eq.~\eqref{eq:tau_*} indicates that in order to know how long to integrate at each candidate frequency, we must estimate the values of the parameters $Q_{Ls}$, $\beta_{s}$, $C_s$, and $T_s$ (see Sec.~\ref{sub:rescale}) and the detuning $\delta\nu_{as}$ between the mode and candidate frequencies. If the true value any of these parameters during the rescan measurement deviates from the value we assume in the calculation of $\tau^*_s$, the true SNR $\hat{R}^*_{\ell(s)}$ calculated from the rescan data (see Sec.~\ref{sub:rescan_analysis}) will deviate from the target value $R^*_T$. 

This observation motivates the question of what nominal value to assign to $R^*_T$ in Eq.~\eqref{eq:tau_*} -- there is no \textit{a priori} reason we must set $R^*_T= R_T$. Note that $\hat{R}^*_{\ell(s)}\neq R^*_T$ for any given candidate is not a problem provided that the probability $p_s$ of a statistical fluctuation exceeding the corresponding coincidence threshold $\Theta^*_{\ell(s)}$ remains $\ll 1$. This probability may be roughly estimated as
\begin{equation}
p_s=n_K\big[1-\Phi\big(\Theta^*_{\ell(s)}\big)\big],\label{eq:p_s}
\end{equation}
where
\begin{equation}
\Theta^*_{\ell(s)} =\hat{R}^*_{\ell(s)} - \Phi^{-1}(c_2)\label{eq:theta_*}
\end{equation}
and we have defined an effective number of independent bins $1<n_K<2K^\text{g}-1$ to account for the fact that we will reject the hypothesis of an axion in bin $\ell(s)$ only if $\delta^{\text{g}*}_\ell/\tilde{\sigma}^{\text{g}*}_\ell$ exceeds the appropriate coincidence threshold in neither the original bin $\ell(s)$ nor any of the $(K^\text{g}-1)$ correlated bins on either side (see discussion in Sec.~\ref{sub:rescan_analysis}). $n_K = 1$ $(n_K = 2K^\text{g}-1)$ would correspond to treating the $2K^\text{g}-1$ bins associated with each candidate as perfectly correlated (uncorrelated); the appropriate value is clearly somewhere in between these two extremes.

We would like to demand that $\sum_sp_s \ll 1$ in order to avoid a second round of rescans in the absence of axion signals. If we assume for now that $\Theta^*_{\ell(s)}$ will not vary too much around the nominal value obtained by taking $\hat{R}^*_{\ell(s)}\rightarrow R^*_T$ in Eq.~\eqref{eq:theta_*}, we should set 
\begin{equation}
R^*_T = \Phi^{-1}\Big(1-\Big[\sum_sp_s/(S\times n_K)\Big]\Big)+\Phi^{-1}(c_2).\label{eq:snr_t_*}
\end{equation}
For the first HAYSTAC analysis, we estimated $n_K\approx K^\text{g}$ and demanded that simultaneously $\sum_sp_s\leq0.05$ and $c_2=0.95$; with these choices, Eq.~\eqref{eq:snr_t_*} yields $R^*_T=5.03$ (equivalently, $\Theta_{\ell(s)}^*\approx3.28$).

Next we need to specify how we evaluate the other parameters that enter into the calculation of $\tau^*_s$. For each candidate we set $T_s$ by averaging $T_{ij}$ over all initial scan $Y$-factor measurements $i$, and evaluating the average in the IF bin $j$ corresponding to the cavity resonance. The form factor $C_s$ and the unloaded cavity quality factor $Q_{0s}$ depend deterministically on the cavity frequency and thus are easy to accurately estimate; $Q_{Ls}=Q_{0s}/(1+\beta_s)$ then follows from our ability to control the cavity-receiver coupling $\beta$ by adjusting the antenna insertion. We set $\beta_s=2$ in Eq.~\eqref{eq:tau_*} for each candidate, to match the average value of $\beta_i$ throughout the initial scan.\footnote{$\beta\approx2$ is optimal for a continuous data run because it maximizes the scan rate for a given sensitivity $|g_\gamma|$~\cite{NIM2017}. For a rescan measurement in which we only care about the SNR in a few bins around $\nu_{\ell(s)}$, critical coupling ($\beta=1$) is better if $\delta\nu_{as}\approx0$ and $N_\text{cav}=0$. However, with $N_\text{cav}\neq0$ the system noise temperature also depends on $\beta$: with $\beta_s=1$, $T_s$ would systematically underestimate the true noise temperature in the rescan measurement.}

The detuning $\delta\nu_{as}$ is trivial to measure but hard to control precisely, due to the mode frequency drifts discussed in Sec.~\ref{sub:badscans} and the backlash inevitably present in any mechanical tuning system. In practice we acquired the rescan data starting with the highest-frequency candidate and tuning down: for each candidate, we tuned the TM$_{010}$ mode $\approx 100 - 200$~kHz above $\nu_{\ell(s)}$ and waited 20 minutes for the mode frequency to settle before starting the measurement. We proceeded with the measurement only if $|\delta\nu_{as}|<150$~kHz after this interval. We set $\delta\nu_{as}=0$ for each candidate in Eq.~\eqref{eq:tau_*} for simplicity; since the cavity will be overcoupled and the noise temperature also decreases for $\delta\nu_{as}\neq0$, $\tau^*_s$ is not too sensitive to small detunings. 

Another potentially more serious consequence of mode frequency drift is that for any given $s$, some or all of the processed spectrum bins contributing to the grand spectrum bin $\ell(s)$ may happen to coincide with a region of the analysis band contaminated by IF interference. We saw in Sec.~\ref{sub:badbins} that 11\% of analysis band bins were contaminated in this way -- thus there is a non-negligible chance that $\hat{R}^*_{\ell(s)}$ will be substantially smaller than the target value $R^*_T$ due to missing bins. 

We mitigate this effect by splitting the total integration time $\tau^*_s$ required for each candidate across 10 cavity noise measurements of duration $\tau^*_s/10$, and step the LO and JPA pump frequencies together by 1~kHz (without tuning the cavity mode) between measurements. On average, we expect the candidate to fall in a contaminated part of the analysis band in about 1 of 10 such measurements: thus we set $\varepsilon=0.1$ in Eq.~\eqref{eq:tau_*}.

Finally, unlike the experimental parameters discussed above, $\eta^*$ and $F_\text{ML}$ depend on fixed parameters of the rescan analysis procedure and cannot change from one rescan measurement to the next. We will see in Sec.~\ref{sub:rescan_analysis} that while  $F_\text{ML}$ will not change in the rescan analysis, $\eta^*$ will not in general be equal to $\eta$ and thus should be estimated in advance to avoid systematically biasing $\tau^*_s$. 

Applying Eq.~\eqref{eq:tau_*} to the 28 rescan candidates from the first HAYSTAC data run, we obtained rescan times $\tau^*_s$ ranging from 5.8 hours (corresponding to $G_{\ell(s)}=2.72$) to 17.9 hours ($G_{\ell(s)}=2.03$). We had $|\nu_{\ell(s)}-\nu_{\ell(s+1)}|<200$~kHz for 3 of the 27 pairs of adjacent candidates: thus there is a 100~kHz range for $\nu_{cs}$ in which the condition $|\delta\nu_{as}|<150$~kHz can be satisfied simultaneously for both candidates. In each of these cases, we acquired rescan data for both candidates together, taking the larger of the two calculated integration times (which were generally very similar). Thus we made 25 rescan measurements, for a total of 282 hours of rescan time (c.f. $M\tau=1692$~hours of initial scan time).\footnote{We will use $s$ to index quantities $a_s$ associated with each rescan measurement as well as quantities $b_s$ associated with each candidate, with the implicit understanding that in three cases, we will have $a_s=a_{s+1}$ but $b_s\neq b_{s+1}$}

At each iteration $s$, after tuning the cavity to the appropriate frequency $\nu_{cs}$ and setting $\beta_s\approx2$, we used a LabVIEW program to make 10 cavity noise measurements and acquire auxiliary data. Each cavity noise measurement was saved as an averaged power spectrum with frequency resolution $\Delta\nu_b$ as in the initial scan. The auxiliary data at each iteration comprised VNA measurements of the cavity mode in transmission and the JPA gain profile both before and after the set of cavity noise measurements, a VNA measurement of the cavity mode in reflection, and a $Y$-factor measurement. 

We use this auxiliary data to quantify $\hat{R}^*_{\ell(s)}$ as described in Sec.~\ref{sub:rescan_analysis}, and also to flag and cut anomalous iterations as in Sec.~\ref{sub:badscans}. Unlike in the initial scan analysis we must repeat any iterations we cut at this stage, to ensure that we have meaningful data around each rescan candidate. In the first HAYSTAC data run we had to repeat 6 of our 25 rescan measurements, in each case because of excessive mode frequency drift $|\nu_{c1}-\nu_{c2}|>130$~kHz.\footnote{Mode frequency drifts were generally larger than in the initial scan due to a combination of much larger tuning steps between iterations and much longer integration times. Four rescan measurements had drifts below 130 kHz but above the more conservative 60 kHz threshold used in the initial scan. The range over which the mode drifted was roughly centered on the candidate frequency $\nu_{\ell(s)}$ in these cases, so the systematic deviation from the correct ML weight for any processed spectrum bin contributing to the combined spectrum around $\nu_{\ell(s)}$ will be quite small. To bound this error we can consider the more extreme case where the mode frequency initially coincides with the candidate frequency and then drifts away slowly over the 10 subsequent measurements: with the maximum allowed drift and the minimum cavity bandwidth $\Delta\nu_{cs}$, the RMS fractional deviation from the true combined spectrum ML weights is 13\%. As noted in Sec.~\ref{sub:correlations}, the systematic effect on the combined spectrum bin values $\delta^{\text{c}*}_k$ and the SNR $R^{\text{\,c}*}_k$ will be much smaller.}

\subsection{The rescan analysis}\label{sub:rescan_analysis}
Once we have acquired a complete rescan dataset, the next step is to process and combine all rescan power spectra to produce a single rescan grand spectrum. We begin by truncating each of our 250 rescan spectra as in Sec.~\ref{sec:baseline}, normalizing each spectrum to the average baseline from the initial scan analysis, and using the list defined in Sec.~\ref{sub:badbins} to cut bins contaminated by IF interference from each spectrum.

Next we must use an SG filter to remove the residual baseline from each spectrum. At this stage it becomes important that $\tau^*_s/10>\tau$ even for the smallest value of $\tau^*_s$ obtained from Eq.~\eqref{eq:tau_*}; moreover the residual baselines for the 10 spectra from each iteration will be very similar, since we do not tune the cavity or rebias the JPA between power spectrum measurements. Thus, although the total averaging at each candidate frequency in the rescan data is comparable to the total averaging at that frequency in the initial scan, we should expect the amplitude (relative to $\sigma^\text{p}$) of any small-scale systematic structure in the rescan processed spectra to be enhanced by a factor $\sim\sqrt{\tau^*_s/\tau}$ if we use the same SG filter parameters as in the initial scan (see also discussion in Appendix~\ref{app:sg_params}).

We have seen in the preceding sections that the statistics of the initial scan spectra are Gaussian at each stage of the processing, and in particular that the narrowing of the histogram of normalized grand spectrum bins $\delta^\text{g}_\ell/\sigma^\text{g}_\ell$ is completely explained by the stopband properties of the SG filter with parameters $d=4$ and $W=500$. This good agreement between the observed and expected statistics indicates that the amplitude of any small-scale systematic structure in the initial scan processed spectra must be $\ll\sigma^\text{p}$.

The observation that baseline systematics will grow coherently over at least the single-spectrum integration time (and likely over the full rescan integration time) indicates that we cannot necessarily assume systematic structure will remain negligibly small in the rescan processed spectra. Studies of the effects of SG filters on simulated Gaussian white noise indicate that the parameters $d$ and $W$ used in the initial scan would produce unacceptable deviations from Gaussianity if applied to the rescan analysis. Thus we used an SG filter with $d^*=6$ and $W^*=300$ for the rescan analysis instead; Fig.~\ref{fig:filter} suggests that with these parameters we should expect $\xi^*<\xi$ and $\eta^*<\eta$, and we will see below that this is indeed the case. 

After applying the SG filter with parameters $d^*$ and $W^*$ to each rescan spectrum, we verify that the bins in each of the 10 processed spectra at iteration $s$ have the expected Gaussian distribution with mean 0 and standard deviation $\sigma^{\text{p}*}_s = 1/\sqrt{\Delta\nu_b\tau^*_s/10}$. We then rescale the spectra to obtain a mean power excess of 1 in any rescaled spectrum bin in which a KSVZ axion deposits a fraction $1/(K^\text{r}K^\text{g})$ of its total conversion power. Formally, the required rescaling is given by Eqs.~\eqref{eq:delta_s} and \eqref{eq:sigma_s}, with the additional factor of $1/(K^\text{r}K^\text{g})$ discussed at the beginning of Sec.~\ref{sub:rebinned_spectrum} absorbed into the definition of the signal power. Values for the factors in Eq.~\eqref{eq:power} and Eq.~\eqref{eq:noise} are obtained from the auxiliary data at each rescan measurement via the procedure described in Sec.~\ref{sub:rescale}; unlike in the initial scan analysis, no interpolation is required for $T^*_{sj}$ because we made a $Y$-factor measurement at each rescan iteration.\footnote{The $j$-dependent quantities in the rescaling factor should more properly be written with an additional index $a=1,\dots,10$ to account for the fact that the LO frequency varies across the 10 spectra at each iteration $s$. Apart from this small frequency offset, the rescaling factor is the same for all the spectra at a given iteration $s$.}

We then follow the procedure of Sec.~\ref{sub:combine} to construct a single ML-weighted combined spectrum from the set of 250 rescaled spectra. The frequency axis for the rescan combined spectrum extends from the smallest candidate frequency minus 651 kHz (i.e., half the analysis band) to the largest candidate frequency plus 651 kHz: there are thus formally a total of $1.02\times10^6$ combined spectrum bins, though about 70\% of these bins are empty because we only took data around candidate frequencies. The typical spacing between candidate frequencies is such that most (non-empty) combined spectrum bins $k$ are obtained by averaging only the $m_k=10$ spectra from a single rescan measurement. But the formal procedure of Sec.~\ref{sub:combine} also correctly treats the cases where adjacent candidates are sufficiently close that spectra from different iterations overlap, and thus $m_k>10$. As expected, the distribution of combined spectrum bins $\delta^{\text{c}*}_k/\sigma^{\text{c}*}_k$ is Gaussian with mean 0 and standard deviation 1.

Finally, we follow the procedure of Sec.~\ref{sub:rebinned_spectrum} and Sec.~\ref{sub:grand_spectrum} to obtain the rescan grand spectrum. Since we want to reproduce the initial scan sensitivity without changing any assumptions about the axion signal, we should use the same values of $K^\text{r}$, $K^\text{g}$, and $\bar{L}_q$ in Eqs.~\eqref{eq:d_r}, \eqref{eq:snr_r}, \eqref{eq:snr_g}, and \eqref{eq:delta_sigma_g}. However, we should expect $\xi^{\text{r}*}\neq\xi^\text{r}$ and $\xi^{\text{g}*}\neq\xi^\text{g}$ because we have used a different SG filter. Empirically, the distribution of rebinned spectrum bins $\delta^{\text{r}*}_\ell/\sigma^{\text{r}*}_\ell$ is Gaussian with mean 0 and standard deviation $\xi^{\text{r}*}=0.96$, and the distribution of grand spectrum bins $\delta^{\text{g}*}_\ell/\sigma^{\text{g}*}_\ell$ is Gaussian with mean 0 and standard deviation $\xi^*=0.83$, implying $\xi^{\text{g}*}=\xi^*/\xi^{\text{r}*}=0.86$.

As in the initial scan analysis, we are ultimately interested in the quantities $\delta^{\text{g}*}_\ell/\tilde{\sigma}^{\text{g}*}_\ell=\delta^{\text{g}*}_\ell/\big(\xi^*\sigma^{\text{g}*}_\ell\big)$ and $\tilde{R}^{\text{\,g}*}_\ell=\eta^*R^{\text{\,g}*}_\ell$ that have been corrected for filter effects. As before, we obtain the value of $\xi^*$ directly from the data; the value of $\eta^*$ can only be obtained from simulation, but the common origin of $\xi^*$ and $\eta^*$ and good agreement between the observed and simulated values of $\xi^*$ gives us confidence that we have applied the appropriate correction factor.

\begin{figure*}[t]
\includegraphics[width=0.8\textwidth]{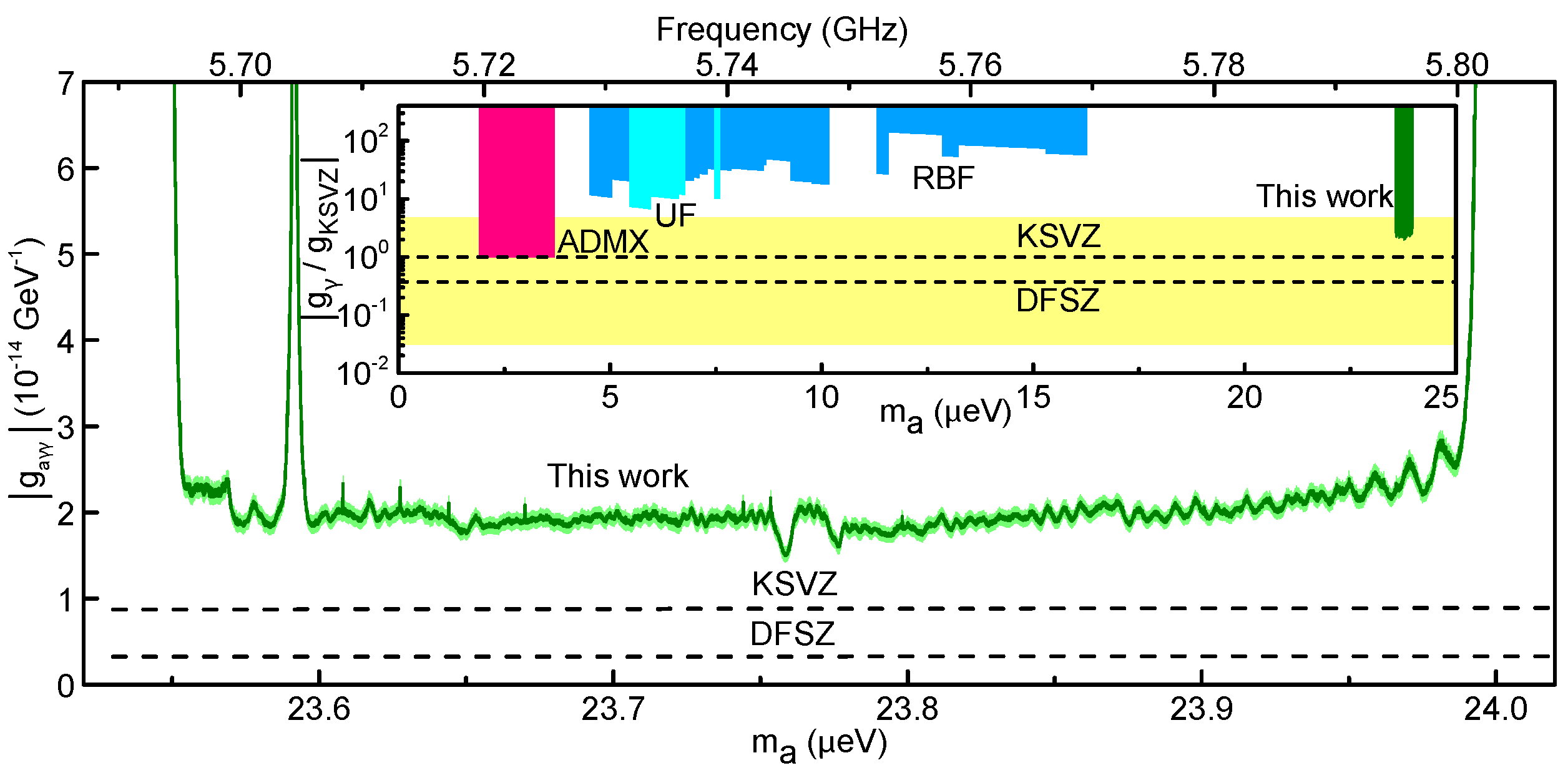}
\caption{\label{fig:exclusion} The exclusion limit from the first HAYSTAC data run at 95\% confidence (see discussion in text), reprinted from Ref.~\cite{PRL2017}. The light green shaded region is a rough estimate of our uncertainty, discussed in Appendix~\ref{app:error}. The large notch near 5.704 GHz is the result of cutting spectra around the intruder mode discussed in Sec.~\ref{sub:badscans}. The narrow notches correspond to frequencies at which synthetic axion signals were injected during the winter scan (see Appendix~\ref{app:fake_axions}). The inset shows our limit (green) together with previous haloscope limits from ADMX (magenta, Refs.~\cite{ADMX1998,ADMX2002,ADMX2004,ADMX2010,ADMX2016}, $\text{C.L.} \geq 90\%$) and early experiments at Brookhaven (RBF, blue, Refs.~\cite{RBF1987,RBF1989}, $\text{C.L.} = 95\%$) and the University of Florida (UF, cyan, Ref.~\cite{UF1990}, $\text{C.L.} = 95\%$). For uniformity of presentation, both RBF and UF limits have been rescaled to $\rho_a = 0.45$~GeV/cm$^3$ from their original published values, where $\rho_a = 0.3$~GeV/cm$^3$ was used. The axion model band~\cite{cheng1995} is shown in yellow.}
\end{figure*}

We validate the observed value of $\xi^*$ and measure $\eta^*$ using simulations very similar to the ones described in Sec.~\ref{sub:correlations} and Sec.~\ref{sub:axion_atten}, respectively. Formally, the rescan simulations only differ in two respects: we multiply each simulated white noise spectrum by the same sample baseline instead of a random sample baseline, and we assign the same mode frequency to each spectrum in the simulation to quantify $\eta^*$.\footnote{We did not assign frequency offsets to the spectra in the simulation used to measure $\xi$ in the initial scan analysis (see Sec.~\ref{sub:correlations}). The fact that we nonetheless obtained the same value of $\xi$ as in real data indicates that small changes in the baseline shape (associated with tuning the cavity or rebiasing the JPA) can suppress the growth of small-scale systematics substantially, even without frequency offsets between spectra.} We reproduced the observed values of $\xi^{\text{r}*}$ and $\xi^*$ and obtained $\eta^*=0.76$ from these simulations; we verified that the results in each case were independent of the number of averages $m$ and integration time $\tau$, at least for $m\tau\leq20$~hours, and thus independent of frequency (see discussion in Sec.~\ref{sub:correlations}).

At this point, we have obtained an explicit expression for the SNR $\tilde{R}^{\text{\,g}*}_{\ell}$ for a KSVZ axion signal in each bin $\ell$ of the rescan grand spectrum, whereas we care about the SNR for an axion signal with the threshold coupling $|g^\text{min}_\gamma|_\ell$. Naively we only care about evaluating the SNR in the $S$ bins $\ell(s)$ that passed the threshold in the initial scan, as the hypothesis that an axion signal with coupling $|g^\text{min}_\gamma|_\ell$ is present in any other bin has already been excluded with confidence $c_1$. 

The presence of grand spectrum correlations complicates this picture slightly. If several adjacent bins pass the threshold together, we associate the candidate with the bin whose power excess was largest, but in the presence of fluctuations the bin with larger power excess does not necessarily have the largest SNR. Thus it is possible in principle that the rescan candidate we have associated with bin $\ell(s)$ actually corresponds to an axion signal in any of the $2K^\text{g}-1$ grand spectrum bins $\ell'(s)$ correlated with $\ell(s)$. To be conservative we require \textit{each} such hypothesis be rejected with confidence $c_2$ before we can reject the candidate. The above discussion implies that we should define
\begin{equation}
\hat{R}^*_{\ell'(s)} = G^2_{\ell'(s)}\tilde{R}^{\text{\,g}*}_{\ell'(s)}\label{eq:hat_snr_*}
\end{equation}
with $\ell'(s)$ defined in the range $[\ell(s)-(K^\text{g}-1),\ell(s)+(K^\text{g}-1)]$. Values of $\hat{R}^*_{\ell'(s)}$ in the first HAYSTAC data run ranged from 4.26 to 7.19, with an average of 5.19.\footnote{The SNR was consistently above 6.4 for all the bins associated with two adjacent candidates that were separated in frequency by only 270~kHz: this was above our threshold for acquiring data for both candidates together, but still close enough that the integration at each candidate contributed significantly to the SNR for the other. The average SNR among all other candidates was 5.09, close to our target value $R^*_T=5.03$. The RMS variation in $\hat{R}^*_{\ell'(s)}$ among the bins $\ell'(s)$ associated with each candidate $s$ was typically less than 1\%, but was $\sim5\%$ in a few cases where the candidate frequency was close to a region of the grand spectrum with reduced exposure due to missing bins.} The effects of uncertainty in the factors used to calculate the rescan SNR are discussed in Appendix~\ref{app:error}.

The appropriate coincidence threshold $\Theta^*_{\ell'(s)}$ for each bin correlated with each candidate is then obtained by using Eq.~\eqref{eq:hat_snr_*} in Eq.~\eqref{eq:theta_*} with the substitution $\ell(s)\rightarrow\ell'(s)$. In the first HAYSTAC data run, $\delta^{\text{g}*}_{\ell'(s)}/\tilde{\sigma}^{\text{g}*}_{\ell'(s)}$ did not exceed $\Theta^*_{\ell'(s)}$ for any of the bins $\ell'(s)$ associated with any of our $S=28$ rescan candidates.\footnote{Had we observed a small number of persistent candidates, we could easily have subjected them to an unambiguous test by repeating the rescan measurement with different applied magnetic fields. It is difficult to imagine any instrumental systematic capable of mimicking the $B_0^2$ scaling of the axion signal power.} The final result of the first HAYSTAC data run is thus a limit on the axion-photon coupling $|g_\gamma|$.

\section{Conclusion}\label{sec:conclusion}
The absence of any persistent candidates in the first HAYSTAC data run implies that $|g^\text{min}_\gamma|_\ell$ given by Eq.~\eqref{eq:g_min} may be interpreted as an exclusion limit on the dimensionless coupling $|g_\gamma|$ in each bin $\ell$ in our initial scan range. The corresponding limit on the physical coupling $|g_{a\gamma\gamma}|=|g_\gamma|\alpha/(\pi\Lambda^2)m_a$ that appears in the Lagrangian is plotted in Fig.~\ref{fig:exclusion}. Assuming an axion signal lineshape described by Eq.~\eqref{eq:f_dist}, we excluded $|g_{\gamma}|\geq2.3\times|g^\text{KSVZ}_\gamma|$ on average over the mass range $23.55 < m_a < 24.0~\mu$eV. 

What confidence should we ascribe to the exclusion of axions with the threshold coupling $|g^\text{min}_\gamma|_\ell$? Following Ref.~\cite{daw1998}, we initially chose $c_1=c_2=0.95$ to ensure the product $c_1c_2\geq0.9$, and interpreted this product as the net confidence level. But this interpretation is overly conservative, because we only acquired and analyzed rescan data at frequencies that exceeded the initial scan threshold. The hypothesis of an axion signal with the threshold coupling in any given bin is excluded with confidence $c_1$ if that bin did not exceed the initial scan threshold. In the bins correlated with each candidate, the appropriate confidence level is the conditional probability that a true axion signal would fail to exceed the coincidence threshold, having already exceeded the initial scan threshold; since the two scans are independent, this probability is just $c_2$. Thus, our result $|g^\text{min}_\gamma|_\ell$ is properly interpreted as an exclusion limit at 95\% confidence.\footnote{We can equivalently interpret this result as a marginally more sensitive exclusion limit at lower confidence. Our threshold coupling at 90\% confidence would be smaller by a factor of $\big[\big(R_T-\Phi^{-1}(0.95)+\Phi^{-1}(0.9)\big)/R_T\big]^{1/2}\approx0.964$. HAYSTAC collaborators are also working on developing a Bayesian approach to the haloscope search analysis, which should offer an alternative prescription for defining rescan candidates and establishing confidence levels.}

In this paper, we have described in detail the analysis procedure used to derive the first limits on cosmic axions from the HAYSTAC experiment. We have cited specific examples from the analysis of our first data run, but our formal procedure may easily be adapted to the analysis of data from other haloscope detectors. 

Throughout the preceding sections we have specifically emphasized our use of Savitzky-Golay filters to remove individual spectral baselines, our quantitative understanding of how filtering affects the statistics of the spectra, and our consistent application of maximum-likelihood weights to both the ``vertical'' sum of overlapping spectra and the ``horizontal'' sum of adjacent bins in the combined spectrum. All of these were innovations of our approach to the haloscope search analysis; taken together, they enable us to calculate our search sensitivity with minimal input from simulation, and obtain the relationship between sensitivity and confidence level directly from statistics. 

With the results of the first HAYSTAC data run we demonstrated that a sufficiently low-noise experiment can reach the QCD axion model band for $m_a > 20~\mu$eV, despite the unfavorable scaling of the haloscope signal power with increasing frequency~\cite{NIM2017}. A second run to extend this coverage is presently underway, with an improved thermal link to the tuning rod (and thus significantly reduced $N_\text{cav}$) and a new piezoelectric actuator with more reliable mechanical performance; these upgrades will be described in a forthcoming publication along with new results from the experiment. Ongoing cavity and amplifier R\&D by members of the HAYSTAC collaboration also indicates several promising avenues for further improving the scan rate and extending the haloscope technique to still higher frequencies~\cite{NIM2017,lamoreaux2013}. 

\begin{acknowledgments}
This work was supported by the National Science Foundation, under grants PHY-1362305 and PHY-1607417, and by the Heising-Simons Foundation under grants 2014-181, 2014-182, and 2014-183. We thank Ana Malagon and Dan Palken for fruitful discussions, and Marguerite Epstein-Martin and Miguel Goncalves for contributions to the project.
\end{acknowledgments}

\appendix
\section{Notation}\label{app:notation}
Tab.~\ref{tab:notation} summarizes the notation used in the formal description of the HAYSTAC analysis procedure above, and indicates where each commonly used symbol is first introduced in the text. We have omitted quantities which are not referenced outside the subsection in which they are defined, and haloscope physics parameters for which we have followed the standard notation in the field.
\begin{table}[h]
\centering
\begin{tabular}{|c|c|c|}
\hline
\rule{0pt}{2.8ex} \textbf{Symbol} & \textbf{Meaning} & \textbf{Introduced} \\ \hline
\rule{0pt}{2.8ex}$\Delta\nu_b$ & Raw spectrum resolution & Sec.~\ref{sub:data} \\ \hline
\rule{0pt}{2.8ex}$\delta\nu_a$ & Axion/cavity detuning & Sec.~\ref{sub:data}  \\ \hline
\rule{0pt}{2.8ex}$d$ & SG filter polynomial order & Sec.~\ref{sub:sg_filter}  \\ \hline
\rule{0pt}{2.8ex}$W$ & SG filter window size &  Sec.~\ref{sub:sg_filter} \\ \hline
\rule{0pt}{2.8ex}$M$ & Total no.~spectra &  Sec.~\ref{sub:stats} \\ \hline
\rule{0pt}{2.8ex}$n^\text{y}$ & No.~bins in the $\text{y}$ spectrum &  Sec.~\ref{sub:stats} \\ \hline
\rule{0pt}{2.8ex}$\delta^\text{y}_{x}$, $\delta^\text{y}_{xz}$ & Single-bin power excess & Sec.~\ref{sub:stats} \\ \hline
\rule{0pt}{2.8ex}$\sigma^\text{y}_x$, $\sigma^\text{y}_{xz}$ & St.~dev.~of power excess &  Sec.~\ref{sub:stats} \\ \hline
\rule{0pt}{2.8ex}$m_k$ & No.~spectra contributing to & Sec.~\ref{sec:rescale_combine} \\ 
\rule{0pt}{2.8ex} & combined spectrum bin $k$ &  \\ \hline
\rule{0pt}{2.8ex}$R^\text{y}_x$, $R^\text{y}_{xz}$ & Haloscope SNR &  Sec.~\ref{sub:rescale} \\ \hline
\rule{0pt}{2.8ex}$\mu^\text{y}_x$ & Mean power excess &  Sec.~\ref{sub:combine} \\ \hline
\rule{0pt}{2.8ex}$K^\text{r}$, $K^\text{g}$ & No.~bins in rebinned/grand &  Sec.~\ref{sec:rebin} \\ 
\rule{0pt}{2.8ex} & spectrum horizontal sum &  \\ \hline
\rule{0pt}{2.8ex}$\Delta\nu_r$ & Rebinned spectrum resolution & Sec.~\ref{sec:rebin} \\ \hline
\rule{0pt}{2.8ex}$\delta\nu_r$ & Axion/bin boundary detuning & Sec.~\ref{sub:lineshape} \\ \hline
\rule{0pt}{2.8ex}$L_q(\delta\nu_r)$ & Axion signal lineshape &  Sec.~\ref{sub:lineshape} \\ \hline
\rule{0pt}{2.8ex}$\bar{L}_q$ & Misalignment-averaged lineshape &  Sec.~\ref{sub:lineshape} \\ \hline
\rule{0pt}{2.8ex}$\eta_c$ & Signal fraction contained & Sec.~\ref{sub:lineshape} \\
\rule{0pt}{2.8ex} & in a grand spectrum bin &  \\ \hline
\rule{0pt}{2.8ex}$\eta_m$ & Signal attenuation & Sec.~\ref{sub:lineshape} \\ 
\rule{0pt}{2.8ex} & from average misalignment &  \\ \hline
\rule{0pt}{2.8ex}$D^\text{y}_x$ & ML-weighted power excess & Sec.~\ref{sub:rebinned_spectrum} \\ \hline
\rule{0pt}{2.8ex}$\xi^\text{r}$, $\xi^\text{g}$, $\xi$ & Correlation-induced $\sigma$~reduction & Sec.~\ref{sub:rebinned_spectrum} \\ \hline
\rule{0pt}{2.8ex}$F_\text{ML}$, $F_\text{uw}$ & Horizontal sum figure of merit &  Sec.~\ref{sub:grand_spectrum} \\ \hline
\rule{0pt}{2.8ex}$\tilde{\sigma}^\text{y}_x$ & Corrected power excess st.~dev. &  Sec.~\ref{sub:correlations} \\ \hline
\rule{0pt}{2.8ex}$\tilde{R}^\text{y}_x$ & Corrected haloscope SNR &  Sec.~\ref{sub:correlations} \\ \hline
\rule{0pt}{2.8ex}$\eta'$ & Correlation-induced signal loss & Sec.~\ref{sub:correlations}  \\ \hline
\rule{0pt}{2.8ex}$\eta$ & SNR reduction from correlations & Sec.~\ref{sub:correlations}  \\ \hline
\rule{0pt}{2.8ex}$\Theta$ & Rescan threshold &  Sec.~\ref{sec:candidates} \\ \hline
\rule{0pt}{2.8ex}$\hat{S}$ & Expected rescan yield without & \\
\rule{0pt}{2.8ex} & grand spectrum correlations &  Sec.~\ref{sec:candidates} \\ \hline
\rule{0pt}{2.8ex}$R_T$ & Threshold SNR target &  Sec.~\ref{sec:candidates} \\ \hline
\rule{0pt}{2.8ex}$c_1$ & Initial scan confidence level &  Sec.~\ref{sec:candidates} \\ \hline
\rule{0pt}{2.8ex}$G_\ell$ & Normalized axion coupling &  Sec.~\ref{sec:candidates} \\ \hline
\rule{0pt}{2.8ex}$S$ & Observed rescan yield &  Sec.~\ref{sub:target_confidence} \\ \hline
\rule{0pt}{2.8ex}$\bar{S}$ & Expected rescan yield &  Sec.~\ref{sub:target_confidence} \\ \hline
\rule{0pt}{2.8ex}$c_2$ & Rescan confidence level &  Sec.~\ref{sec:rescan} \\ \hline
\end{tabular}
\caption{\label{tab:notation} Commonly used notation.}
\end{table}

In Tab.~\ref{tab:notation}, the roman superscript $\text{y}$ represents one of the spectrum labels defined in Tab.~\ref{tab:labels} and the italic subscripts $x$ and $z$ represent indices defined in Tab.~\ref{tab:indices}. We adhere to these conventions throughout the text: for example, $R^\text{\,s}_{ij}$ denotes the SNR for an axion confined to the $j$th bin of the $i$th rescaled spectrum, and the index $k$ always runs from $1$ to $n^\text{c}$, the number of bins in the combined spectrum. Primed indices (e.g., $k'$, $\ell'$) are used throughout the text to single out a particular bin containing a putative axion signal; in Sec.~\ref{sub:correlations} a primed index is also used to specify the second bin in an expression for the covariance of nearby bins in a given spectrum.
\begin{table}[h]
\centering
\begin{tabular}{|c|c|c|}
\hline
\rule{0pt}{2.8ex} \textbf{Superscript} & \textbf{Meaning} & \textbf{Introduced} \\ \hline
\rule{0pt}{2.8ex}p & processed spectrum & Sec.~\ref{sub:stats} \\ \hline
\rule{0pt}{2.8ex}s & rescaled spectrum & Sec.~\ref{sub:rescale}\\ \hline
\rule{0pt}{2.8ex}c & combined spectrum & Sec.~\ref{sub:combine}\\ \hline
\rule{0pt}{2.8ex}r & rebinned spectrum & Sec.~\ref{sub:rebinned_spectrum}\\ \hline
\rule{0pt}{2.8ex}g & grand spectrum & Sec.~\ref{sub:grand_spectrum}\\ \hline
\end{tabular}
\caption{\label{tab:labels} Spectrum labels.}
\end{table}

\begin{table}[h]
\centering
\begin{tabular}{|c|c|c|}
\hline
\rule{0pt}{2.8ex} \textbf{Index} & \textbf{Meaning} & \textbf{Introduced} \\ \hline
\rule{0pt}{2.8ex}$i$ & indexes spectra & Sec.~\ref{sub:stats} \\ \hline
\rule{0pt}{2.8ex}$j$ & indexes IF bins & Sec.~\ref{sub:stats}\\ \hline
\rule{0pt}{2.8ex}$k$ & indexes 100~Hz RF bins & Sec.~\ref{sub:combine}\\ \hline
\rule{0pt}{2.8ex}$\ell$ & indexes 1~kHz RF bins & Sec.~\ref{sub:rebinned_spectrum}\\ \hline
\rule{0pt}{2.8ex}$q$ & indexes lineshape elements & Sec.~\ref{sub:lineshape}\\ \hline
\rule{0pt}{2.8ex}$s$ & indexes rescan candidates & Sec.~\ref{sec:rescan}\\ \hline
\end{tabular}
\caption{\label{tab:indices} Indices.}
\end{table}
Finally, in Sec.~\ref{sec:rescan}, we use the superscript $^*$ to denote previously defined quantities whose values differ in the rescan analysis.

\section{Maximum Likelihood Estimation}\label{app:mle}
Taking the discussion at the beginning of Sec.~\ref{sec:rescale_combine} as motivation, we will assume we have $m$ independent Gaussian random variables $y_k$ drawn from distributions with the same mean $\mu$ but different variances $\sigma_k^2$. We are interested in finding an estimate of $\mu$ that maximizes the likelihood function, which is just the joint probability distribution of the observations $y_k$ considered as a function of $\mu$:
\begin{equation}
\mathcal{L}(\mu)=\exp\left(-\frac{1}{2}\sum_k\left(\frac{y_k-\mu}{\sigma_k}\right)^2\right).\label{eq:likelihood_ind}
\end{equation}
We can equivalently maximize $\log\mathcal{L}$, since the logarithm is monotonically increasing. So we take
\begin{equation*}
\frac{\mathrm{d}}{\mathrm{d}\mu}\log\mathcal{L}=\sum_k\left(\frac{y_k-\mu}{\sigma_k^2}\right)=0.
\end{equation*}
Solving for $\mu$ yields
\begin{equation}
\mu=\frac{\sum_k y_k/\sigma_k^2}{\sum_k 1/\sigma_k^2},
\label{eq:ml_ind}
\end{equation}
which may be compared to Eq.~\eqref{eq:delta_c}.

If our observations are not independent but rather correlated, Eq.~\eqref{eq:likelihood_ind} should be replaced with
\begin{equation}
\mathcal{L}(\mu)=\exp\left(-\frac{1}{2}\left(\mathbf{y}-\mu\mathbf{i}\right)^\intercal\mathbf{\Sigma}^{-1}\left(\mathbf{y}-\mu\mathbf{i}\right)\right)\label{eq:likelihood_cor},
\end{equation}
where $\mathbf{i}$ is the $m$-vector $(1,1,\dots,1)$, and $\mathbf{\Sigma}$ is the covariance matrix whose diagonal elements are $\sigma_k^2$. Maximizing with respect to $\mu$ we obtain
\begin{equation}
\mu=\frac{\mathbf{y}^\intercal\mathbf{\Sigma}^{-1}\mathbf{i}}{\mathbf{i}^\intercal\mathbf{\Sigma}^{-1}\mathbf{i}}.
\label{eq:ml_cor}
\end{equation}
We see that the (unnormalized) ML weight for each $y_k$ is a sum over the $k$th row of $\mathbf{\Sigma}^{-1}$. A useful approximation to this sum for sufficiently small correlations is
\begin{equation}
\sum_{k}\big(\mathbf{\Sigma}^{-1}\big)_{kk'} \approx \frac{1}{\sigma^2_{k'}}\left[1 - \sum_{k\neq k'}\frac{\Sigma_{kk'}}{\sigma^2_{k}}\right],
\label{eq:weights_cor}
\end{equation}
where we have neglected all terms that are higher than first order in the ratio of any off-diagonal element to any diagonal element of $\mathbf{\Sigma}$; to first order the normalization is then just the sum of Eq.~\eqref{eq:weights_cor} over $k'$. In Sec.~\ref{sec:rebin} we consider ML weighting in the presence of small correlations. We continue to use Eq.~\eqref{eq:ml_ind} rather than Eq.~\eqref{eq:ml_cor}, and argue in Sec.~\ref{sub:correlations} that deviations from the true optimal weights are acceptably small.

The ML estimate of the mean of a multivariate Gaussian distribution with arbitrary covariance matrix $\mathbf{\Sigma}$ can also be obtained from a least-squares perspective. To see this, consider a linear regression model $\mathbf{y} = \mu\mathbf{x} + \boldsymbol{\epsilon}$, where we would like to estimate the slope $\mu$ in the presence of noise $\boldsymbol{\epsilon}$, assumed to be drawn from a Gaussian distribution with zero mean and covariance matrix $\mathbf{\Sigma}$. The generalized least squares (GLS) estimate of $\mu$ is the value that minimizes the mean squared error
\begin{equation}
\chi^2(\mu) = \frac{1}{m}\left(\mathbf{y}-\mu\mathbf{x}\right)^\intercal\mathbf{\Sigma}^{-1}\left(\mathbf{y}-\mu\mathbf{x}\right).
\end{equation}
For $\mathbf{x}=\mathbf{i}$, $\chi^2(\mu)\propto\log\mathcal{L}$, so the estimate that extremizes either criterion will also extremize the other. This equivalence between the ML and GLS methods requires only that the statistics of the underlying noise distribution be Gaussian, and this condition will always be satisfied in our haloscope analysis. It can be proved that the variance of the GLS estimator is smaller than the variance of any other unbiased linear estimator~\cite{GLS1935}.

Finally, we note as an aside that if we allow the elements of $\mathbf{x}$ to vary, and take $\mathbf{\Sigma}$ to be diagonal for simplicity, the least squares estimate of $\mu$ becomes
\begin{equation}
\mu=\frac{\sum_k x_ky_k/\sigma_k^2}{\sum_k (x_k/\sigma_k)^2},
\label{eq:ls_x}
\end{equation}
The elements of $x_k$ here play the role of the rescaling factor discussed in Sec.~\ref{sub:rescale}; thus from a least-squares perspective the rescaling of the spectra need not be regarded as a distinct step of the analysis procedure. We stick to the ML perspective in the text to emphasize the value of using units in which the expected axion conversion power is 1, and thus the $R=\sigma^{-1}$ correspondence has an intuitive interpretation.

\section{Optimizing SG filter parameters}\label{app:sg_params}
We discussed the optimization of the SG filter parameters $d$ and $W$ briefly at the end of Sec.~\ref{sub:sg_filter}, but it is instructive to revisit this question after having observed the filter-induced narrowing $\xi$ of the distribution of grand spectrum bins (Sec.~\ref{sub:correlations}) and the filter-induced attenuation of the SNR (Sec.~\ref{sub:axion_atten}). Fig.~\ref{fig:filter} indicates that reducing $d/W$ moves the 3~dB point of the SG filter down towards larger spectral scales and increases the stopband attenuation on the small spectral scales of interest ($\leq K^\text{r}K^\text{g}$ bins). Thus we should expect $\xi,\eta\rightarrow1$ as we reduce $d/W$. 

However, as noted in Sec.~\ref{sub:sg_filter}, reducing the 3~dB point of the SG filter invariably moves progressively larger-amplitude components of the baseline from the filter's passband into its stopband. This claim implicitly assumes that the power spectrum \textit{of the residual baseline} falls off monotonically towards smaller spectral scales, and we can confirm this empirically: on small spectral scales the residual baseline power spectrum follows a power law distribution with spectral index $\alpha\approx-2$.

The largest-amplitude baseline component that is not removed by the SG filter (and thus remains in the processed spectra) will coincide with the first zero of the filter's transfer function; let us call the corresponding bin separation $\kappa$. As we reduce $d/W$ at fixed integration time $\tau$ (or increase $\tau$ for a given filter), the baseline amplitude $a(\kappa)$ will grow relative to the statistical fluctuations $\sigma^\text{p}=1/\sqrt{\Delta\nu_b\tau}$. For $a(\kappa)/\sigma^\text{p}$ sufficiently large, the distribution of processed spectrum bins $\delta^\text{p}_{ij}$ will appear non-Gaussian. Of course, each bin in each processed spectrum is still a Gaussian random variable with standard deviation $\sigma^\text{p}$; the apparent breakdown of Gaussianity just indicates that $\mu^\text{p}_{ij}=0$ for each bin $j$ has become a poor approximation given our failure to completely remove the spectral baseline.

Even if the distribution of $\delta^\text{p}_{ij}$ exhibits no signs of non-Gaussianity, $a(\kappa)\neq0$ implies \textit{positive} correlations in the processed spectra on scales $\leq\kappa/2$; since $\kappa>K^\text{r}K^\text{g}$, this effect tends to counteract the negative correlations due to the SG filter stopband alone (i.e., independent of the spectrum of the baseline). In other words, systematic effects due to the shape of the baseline grow coherently in the horizontal sum over adjacent bins. They can also grow coherently in the vertical sum if the $m_k$ contributing spectra have small detunings and similar baselines, as in the rescan data set (Sec~\ref{sub:rescan_analysis}).

Thus, we find that unless $a(\kappa) \ll 1/\sqrt{\Delta\nu_b\tau}$, $\xi$ and $\eta$ will depend on the integration time $\tau$ (and possibly also on $m_k$). The simulations discussed in Secs.~\ref{sub:correlations}, \ref{sub:axion_atten}, and \ref{sub:rescan_analysis} demonstrate that we are safely in the $a(\kappa) \ll 1/\sqrt{\Delta\nu_b\tau}$ regime with filter parameters $d,W$ ($d^*,W^*$) for the initial scan (rescan) analysis.

\section{Parameter uncertainties}\label{app:error}
The corrected grand spectrum SNR $\tilde{R}^\text{\,g}_\ell$ depends on many measured parameters whose uncertainties we have thus far ignored. Here we will quantify the effects of these uncertainties on the analysis, but first we note that there is potential for terminological confusion because ``confidence level'' is a generic statistical term often used to quantify uncertainty. The axion search confidence level $c_1$ we have defined in this paper is the probability that an axion with SNR $R_T$ in any given grand spectrum bin will exceed the threshold -- since the value of $R_T$ is not fixed by measurement, $c_1$ is completely independent of parameter uncertainties. Rather, uncertainty in $\tilde{R}^\text{\,g}_\ell$ translates [via Eqs.~\eqref{eq:g_ell} and \eqref{eq:g_min}] into uncertainty in the threshold coupling $|g^\text{min}_\gamma|_\ell$ for which we obtain SNR $R_T$ in each bin $\ell$.

We can estimate the size of the fractional uncertainty $\delta|g^\text{min}_\gamma|/|g^\text{min}_\gamma|$ in a typical grand spectrum bin by first noting that
\begin{equation}
|g^\text{min}_\gamma| \propto \left(\frac{T_\text{eff}}{\eta\phi(\delta\nu)\eta_LC_{010}}\right)^{1/2},\label{eq:g_error}
\end{equation}
where we have elided factors without uncertainty and quantities like $Q_L$ and $B_0$ that are easily measured with fractional uncertainty $\leq1\%$, and introduced an effective noise temperature $T_\text{eff}$ and a function $\phi(\delta\nu)$ discussed below. It is easy to estimate the error in the factor $\eta_L$ introduced in Sec.~\ref{sub:rescale} to quantify loss between the cavity and JPA. We estimated this loss to be $-0.60 \pm 0.15$~dB, which implies $\delta\eta_L/\eta_L\approx3.5\%$.

The filter-induced attenuation $\eta$ and cavity mode form factor $C_{010}$ are both obtained from simulation, and thus estimating the uncertainty in these parameters is not necessarily straightforward. Nonetheless, our result for $\eta$ is very robust against changes in the parameters of the simulation (see discussion in Sec.~\ref{sub:axion_atten}), and this implies a fractional uncertainty of $\delta\eta/\eta\lesssim1\%$ which we can safely neglect. We did not include uncertainty in $C_{010}$ in our error budget because we did not have a reliable way to quantify it. Preliminary field profiling measurements suggest that the simulated form factors are reliable to better than $10\%$, so a careful treatment of the form factor uncertainty would likely change our final result $\delta|g^\text{min}_\gamma|/|g^\text{min}_\gamma|\approx4\%$ by at most a factor of 2 and possibly much less.

In the denominator of Eq.~\eqref{eq:g_error} we have defined
\begin{equation}
\phi(\delta\nu)=\sqrt{\sum_q\bar{L}_qL_q(\delta\nu)/\big(K^\text{g})^2}\label{eq:phi_align}
\end{equation} 
to encode the dependence of the SNR on the misalignment $\delta\nu$ of the axion mass relative to the lower edge of the grand spectrum bin in which the SNR is maximized (see discussion in Sec.~\ref{sub:lineshape}). The misalignment attenuation $\eta_m\approx\bar{\phi}/\phi(0)$, where $\bar{\phi}$ is the mean value of $\phi(\delta\nu)$ over the range of possible misalignments; note also the formal similarity of Eq.~\eqref{eq:phi_align} to Eq.~\eqref{eq:fom_ml}.\footnote{Formally, $\eta_m$ as defined in Sec.~\ref{sub:lineshape} is obtained by replacing each $L_q(\delta\nu)$ by its average value $\bar{L}_q$ and then normalizing to $\phi(0)$, which is not quite the same because $\phi$ is not linear in $L_q$. In practice, the difference is negligible.} With the misalignment error $\delta\phi$ defined as the standard deviation of $\phi(\delta\nu)$ over this same range, we obtain $\delta\phi/\bar{\phi}\approx2\%$.

Finally, in any given grand spectrum bin, the effective noise temperature $T_\text{eff}$ is formally given by a ML-weighted average of $T_{ij}$ across all contributing processed spectrum bins. Since we are only interested in estimating the typical fractional uncertainty in the noise temperature, we make the same approximation we used to set $T_s$ in the calculation of the rescan time in Sec.~\ref{sub:rescan_daq}: we average $T_{ij}$ over all spectra and evaluate it in the IF bin $j$ corresponding to the middle of the analysis band, where the ML weight is largest.

Taking a typical cavity frequency $\nu_c=5.75$~GHz in the middle of the first HAYSTAC scan range, we can then write $T_\text{eff}=h\nu_c[N_T + N_\text{cav} + N_A]$; the reader is referred to Sec.~\ref{sub:rescale} for the definition of these additive contributions and to Ref.~\cite{NIM2017} for detailed discussion of the noise calibration procedure. Briefly, we obtain $N_A=1.35\pm0.05$ quanta from off-resonance $Y$-factor measurements and $N_\text{cav}=1.00\pm0.17$ quanta from the average of all $Y$-factor measurements during the data run. Even allowing for a $\pm20$~mK uncertainty in the calibration of the mixing chamber thermometer, the uncertainty in $N_T=0.63$ remains negligibly small, in part because the nominal HAYSTAC operating temperature $T_C=127$~mK is sufficiently far into the Wien limit that $N_T$ depends only weakly on the physical temperature, and in part because errors in different contributions to the total noise $T_\text{eff}$ are somewhat anti-correlated. Negative correlations arise because increasing any of the additive terms in $T_\text{eff}$ while holding the others constant would reduce the measured value of the hot/cold noise power ratio $Y$.

Adding the uncertainties cited in the above paragraph in quadrature and using $k_BT_\text{eff}\approx3h\nu_c$ we obtain $\delta T_\text{eff}/T_\text{eff}\approx 6\%$. This estimate (dominated by the variation in measurements of $N_\text{cav}$) is conservative in that we have neglected the fact that $\delta N_A$ and $\delta N_\text{cav}$ are negatively correlated, and because we have included the RMS systematic variation of $N_\text{cav}$ across the tuning range in the ``uncertainty'' $\delta N_\text{cav}$. Miscalibration of the still thermometer would need to be larger than $\pm20$ mK to affect our estimate of $\delta T_\text{eff}$.

Combining the results of the preceding paragraphs, we obtain
\begin{equation*}
\frac{\delta |g^\text{min}_\gamma|}{|g^\text{min}_\gamma|} \approx \sqrt{\Bigg(\frac{1}{2}\frac{\delta T_\text{eff}}{T_\text{eff}}\Bigg)^2 + \Bigg(\frac{1}{2}\frac{\delta\phi}{\bar\phi}\Bigg)^2 + \Bigg(\frac{1}{2}\frac{\delta\eta_L}{\eta_L}\Bigg)^2} \approx4\%.
\end{equation*}
This result (represented by the light green shaded region in Fig.~\ref{fig:exclusion}) should be interpreted as a rough estimate of the uncertainty in our exclusion limit, not a formal $1\sigma$ error bar on the threshold coupling $|g^\text{min}_\gamma|_\ell$ in each bin.

We should also consider the effects of miscalibrating the SNR in the rescan analysis. We can distinguish between ``global'' effects (e.g., overall miscalibration of the system noise temperature or uncertainty in $\eta_L$) and effects confined to the rescan analysis (e.g., miscalibration of $\eta^*$ or mode frequency drifts in particular rescan measurements). The former affect $\tilde{R}^{\text{\,g}*}_\ell$ and $\tilde{R}^\text{\,g}_\ell$ in the same way: thus they do not change the candidate SNR $\hat{R}^*_{\ell'(s)}$ obtained from Eq.~\eqref{eq:hat_snr_*}, and cannot change the results of the rescan analysis.

Conversely, miscalibration of $\tilde{R}^{\text{\,g}*}_{\ell}$ relative to $\tilde{R}^\text{\,g}_{\ell}$ around any given candidate $s$ implies that we have either underestimated or overestimated $\hat{R}^*_{\ell'(s)}$, which in turn implies that the coincidence thresholds $\Theta^*_{\ell'(s)}$ we imposed on the bins correlated with $\ell(s)$ were either unnecessarily low or too high. Clearly, the latter possibility is the one that should concern us: it implies that relative miscalibration of the rescan SNR can cause the probability that we miss a real persistent signal to exceed $1-c_2$.

Empirically, in the first HAYSTAC data run, we could reduce each $\hat{R}^*_{\ell'(s)}$ by 17\% before any of the $(2K^\text{g}-1)S$ bins we examined exceeded the corresponding threshold.\footnote{The first bin to do so had $\delta^{\text{g}*}_{\ell}/\tilde{\sigma}^{\text{g}*}_{\ell}=2.7$. Among $S\times n_K$ independent bins, we expect 0.5 bins with power excess this large, so the observation of one should not be surprising.} All of the parameter uncertainties whose contributions to $\delta |g^\text{min}_\gamma|/|g^\text{min}_\gamma|$ we have considered in this section are global effects to which the coincidence thresholds are insensitive. We conclude that miscalibration of $\tilde{R}^{\text{\,g}*}_{\ell}$ relative to $\tilde{R}^\text{\,g}_{\ell}$ by more than 17\% is extremely unlikely. A more formal way to account for the possibility of relative miscalibration is to require a rescan confidence level $c_2>c_1$; we will adopt this approach in future HAYSTAC analyses.

\section{Effects of a wider lineshape}\label{app:axion_width}
As noted in Sec.~\ref{sub:lineshape}, the analysis presented in this paper has assumed the spectral distribution of axion conversion power is given by Eq.~\eqref{eq:f_dist} instead of Eq.~\eqref{eq:f_dist_2}, but we should actually expect the latter distribution in a terrestrial experiment if the halo axions are fully virialized with a pseudo-isothermal density profile and RMS velocity $\sqrt{\left<v^2\right>}=270$~km/s.

\begin{figure}[t]
\includegraphics[width=0.5\textwidth]{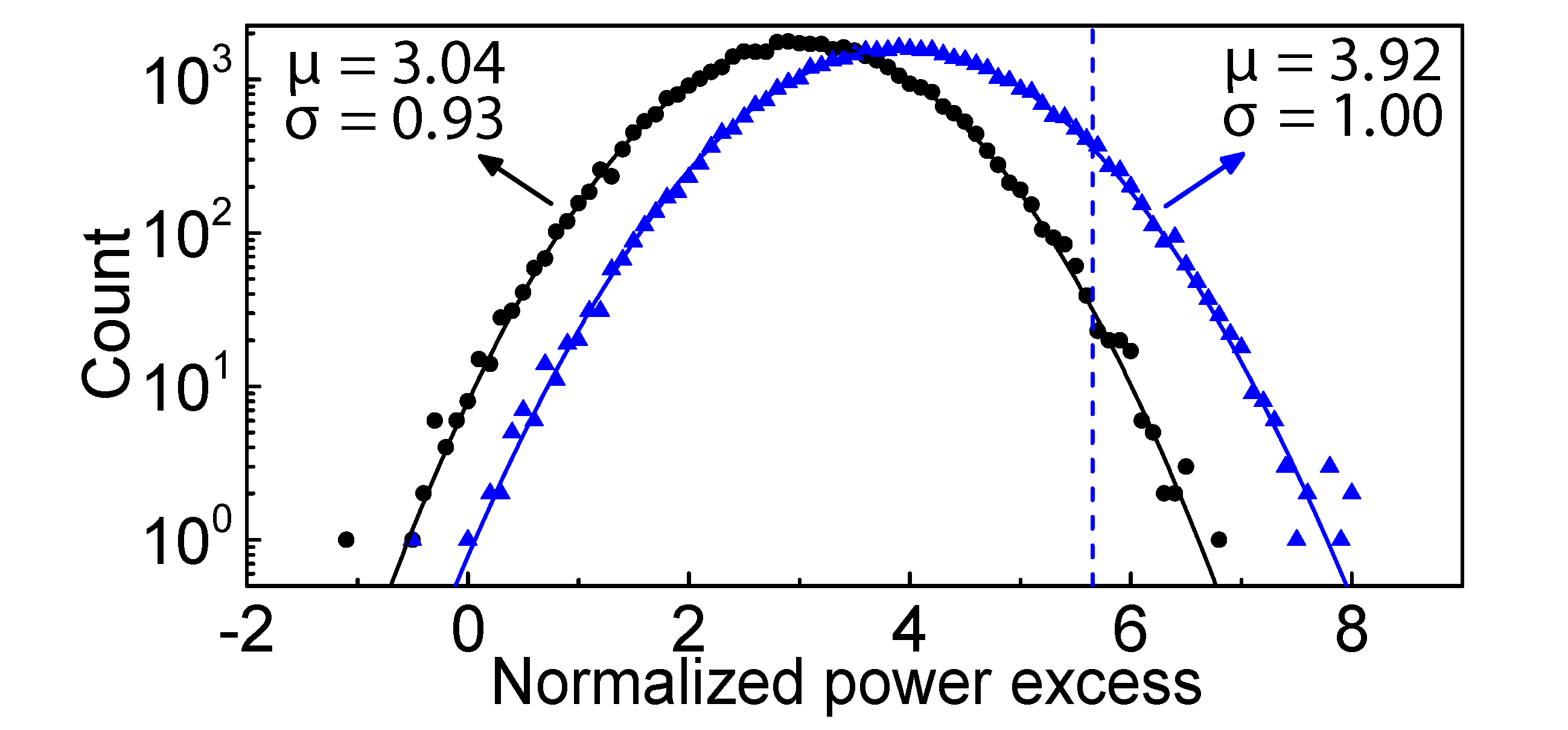}
\caption{\label{fig:wide_axion} The results of a simulation to quantify the reduction in SNR for an axion signal with the wider lineshape of Eq.~\eqref{eq:f_dist_2}. As in Fig.~\ref{fig:simu_hist}, the distribution of excess power $\delta^\text{g}_{\ell'}/\sigma^\text{g}_{\ell'}$ in a grand spectrum bin $\ell'$ containing an axion signal is histogrammed over iterations of the simulation; the two histograms correspond to two different analysis pipelines. Parameters obtained from Gaussian fits to the both histograms are displayed on the plot. The SNR $R^\text{\,g}_{\ell'}=5.66$ calculated assuming the narrower lineshape of Eq.~\eqref{eq:f_dist} is indicated by the dashed vertical line. With an analysis that neglects SG filter effects (blue triangles), the distribution is Gaussian with standard deviation 1 and mean smaller than $R^\text{\,g}_{\ell'}$ by a factor $\zeta_0=0.69$. From the analysis that takes into account effects of the SG filter (black circles), we obtain an additional SNR attenuation factor $\eta_\text{lab}=0.84$. The net reduction of the corrected grand spectrum SNR $\tilde{R}^\text{\,g}_{\ell'}$ is thus $\zeta=(\eta_\text{lab}/\eta)\zeta_0=0.64$.}
\end{figure}

To quantify the degradation of our exclusion limit $|g^\text{min}_\gamma|_\ell$ for an axion signal with the lab frame spectral distribution $f'(\nu)$, we repeated the simulation of Sec.~\ref{sub:axion_atten}, using Eq.~\eqref{eq:f_dist_2} instead of Eq.~\eqref{eq:f_dist} for the simulated axion signal but leaving the lineshape $\bar{L}_q$ used in both the ``standard'' and ``ideal'' analysis pipelines unchanged. As in Sec.~\ref{sub:axion_atten}, the main results of the simulation are two histograms (corresponding to the two analysis pipelines) representing the excess power distribution in the grand spectrum bin $\ell'$ best aligned with the simulated axion signal. These histograms are plotted in Fig.~\ref{fig:wide_axion}.

We see that the mean value of the ideal analysis histogram $E[\delta^\text{g}_{\ell'}/\sigma^\text{g}_{\ell'}]_\text{i}$ is no longer equal to the calculated SNR $R^\text{\,g}_{\ell'}$ represented by the dashed vertical line. This is unsurprising, as $R^\text{\,g}_{\ell'}$ is still calculated using the lineshape $\bar{L}_q$ obtained by integrating Eq.~\eqref{eq:f_dist}. Thus, neglecting SG filter effects, the ratio 
\begin{equation}
\zeta_0=E[\delta^\text{g}_{\ell'}/\sigma^\text{g}_{\ell'}]_\text{i}/R^\text{\,g}_{\ell'} = 0.69
\end{equation}
quantifies the reduction in SNR we should expect when we use an analysis optimized for signals with spectral distribution $f(\nu)$ to search for signals governed by the wider lab frame distribution $f'(\nu)$.

Next we can consider how $\zeta_0$ is modified by the imperfect SG filter stopband. From the width of the histogram obtained from the standard analysis, we obtain $\xi=0.93$, as we should expect given that we have not changed the parameters of the horizontal sum. Comparing the two histograms in Fig.~\ref{fig:wide_axion}, we obtain $\eta_\text{lab}=E[\delta^\text{g}_{\ell'}/\sigma^\text{g}_{\ell'}]_\text{s}/\big(\xi E[\delta^\text{g}_{\ell'}/\sigma^\text{g}_{\ell'}]_\text{i}\big) = 0.83$ [c.f. $\eta=0.90$ obtained in Sec.~\ref{sub:axion_atten} assuming the narrower distribution $f(\nu)$]. The result $\eta_\text{lab}<\eta$ is also expected, as the SG filter stopband attenuation gets worse towards larger spectral scales (See Fig.~\ref{fig:filter}). The net reduction of the corrected KSVZ SNR $\tilde{R}^\text{\,g}_{\ell'}$ is thus
\begin{equation}
\zeta=(\eta_\text{lab}/\eta)\zeta_0=0.64.
\end{equation}
Equivalently, at fixed $R_T$, $|g_\gamma|$ is increased by a factor $1/\sqrt{\zeta}=1.25$. Since we cannot change the threshold in a reanalysis of a completed run without acquiring more rescan data, we conclude that our published exclusion limit $|g^\text{min}_\gamma|_\ell$ is degraded by 25\% for axion signals with spectrum given by Eq.~\eqref{eq:f_dist_2}. The modified limits still cut into the allowed parameter space for viable KSVZ and DFSZ models~\cite{cheng1995,kim1998}; thus the qualitative conclusions of Ref.~\cite{PRL2017} remain unchanged.

It should be emphasized that the value of $\zeta_0$ derived from simulation above arises from the combination of two conceptually distinct effects. First, $f'(\nu)$ is wider than $f(\nu)$, and thus any analysis assuming the former will be less sensitive for a given noise temperature. Second, our analysis used values of $K^\text{g}$ and $\bar{L}_q$ appropriate for the distribution $f(\nu)$, so the horizontal sum is not optimally weighted if the true signal spectrum is $f'(\nu)$. With $K^\text{r}=10$, $K^\text{g}=7$, and $f(\nu)\rightarrow f'(\nu)$ in Eq.~\eqref{eq:int_lineshape}, we can obtain $\zeta_0=0.78$ analytically using Eq.~\eqref{eq:fom_ml}; simulation confirms this value and indicates that $\eta_\text{lab}$ is unchanged. Thus we should expect $\zeta=0.72$ for an analysis optimized for the wider signal distribution, or equivalently $|g_\gamma|$ larger than our present limit by 18\%, up to changes in other factors affecting the SNR. 

\section{Synthetic axion injections}\label{app:fake_axions}
In Fig.~\ref{fig:exclusion} we can see seven small notches in which $|g^\text{min}_\gamma|_\ell$ increases sharply over a very narrow range. These notches arise because we injected synthetic axion signals into the cavity at ten random frequencies during the initial data acquisition period in winter 2016, and cut data around each such signal before combining data from the winter and summer runs. 

In the two lowest-frequency notches, $|g^\text{min}_\gamma|_\ell$ increases by about a factor of $2^{1/4}$ because roughly half the data contributing to the SNR at these frequencies was acquired during the winter run. At higher frequencies, a larger fraction of the data came from the summer run, and thus the depth of the notches gets progressively smaller. In particular, the effects of cutting data from the winter run around two injected signals above 5.76~GHz are not visible at the resolution of Fig.~\ref{fig:exclusion}. The last injected signal happened to fall in the range where we cut spectra around an intruder mode (see Sec.~\ref{sub:badscans}), so it is also not visible in Fig.~\ref{fig:exclusion}.

The procedure we used to generate axion-like signals in HAYSTAC is summarized in Refs.~\cite{PRL2017} and \cite{NIM2017}. Our goal in injecting these signals into the experiment was not to demonstrate an alternative approach to calibrating the search sensitivity, as obtaining sufficiently good statistics would entail polluting our spectrum with a large number of synthetic axions. Instead, we used synthetic signal injections as a simple fail-safe check on our data acquisition and analysis procedures, to verify that faint narrowband signals injected into the cavity did indeed result in large excess power in the expected grand spectrum bins. 

We decided on a nominal signal power of $10^{-22}$~W, roughly equal to the expected conversion power for an axion with $|g_\gamma|=4|g^\text{KSVZ}_\gamma|$ and sufficiently far above our target sensitivity to allow us to immediately establish the presence or absence of excess power with only a single pass over the tuning range. Due to a miscalculation, we set the power lower than this by a factor of 2.5 for the three highest-frequency signals, and moreover the exposure was lowest at these frequencies in the winter run: thus the expected SNR for these three signals was $\approx1.5$. We observed excess power consistent with this estimate (though of course also consistent with the absence of a signal) at these three frequencies. After correcting the signal power, we observed $\delta^\text{g}_\ell/\tilde{\sigma}^\text{g}_\ell > 5$ in all bins corresponding to the remaining injected signals.

Having demonstrated to our satisfaction that our analysis procedure can detect real axion-like signals, we opted not to inject signals during the summer run. Before constructing the combined spectrum used in the final analysis, we cut RF bins around each injected signal in which we expect more than 1\% of the peak power given the measured signal lineshape.

\begin{figure}[t]
\includegraphics[width=0.5\textwidth]{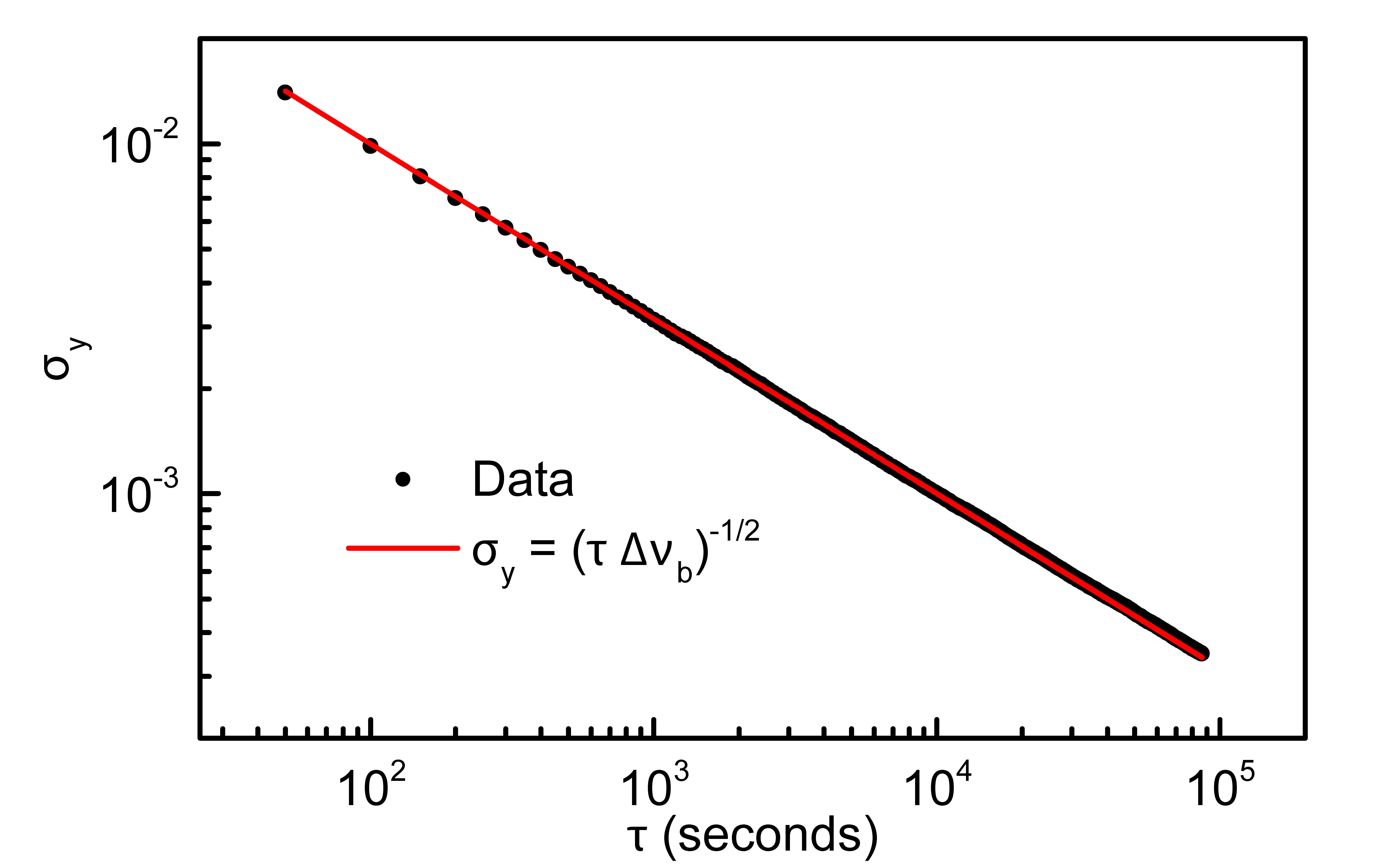}
\caption{\label{fig:avar} Results of a direct measurement of the scaling of RMS noise $\sigma_y$ with integration time $\tau$ in HAYSTAC, demonstrating that $\sigma_y\propto\tau^{-1/2}$ as expected out to at least 24 hours.}
\end{figure}

\section{Scaling with integration time}\label{app:avar}
On paper the expected SNR in a haloscope search is $\propto \sqrt{\tau}$, due to the $\tau^{-1/2}$ scaling of the RMS noise power in each bin expected from Gaussian statistics. The observed standard normal distribution of the combined spectrum power excess $\delta^\text{c}_k/\sigma^\text{c}_k$ in both the initial scan and rescan analyses implicitly indicates that the RMS noise continues to decrease in this way with increasing averaging. We also demonstrated more directly that this $\tau^{-1/2}$ scaling holds for real data out to $\tau>\text{max}(\tau^*_s)$ with a dedicated measurement described below.

For this measurement, we acquired 24 hours of noise data at a single frequency with the JPA gain maintained by feedback as in the data run. The data was saved to disk as a set of 17280 raw spectra obtained from $\tau_0=5$ s of averaging each. In offline analysis we removed bins contaminated by known  IF interference, divided by the average baseline as in Sec.~\ref{sec:baseline}, and averaged every 10 adjacent spectra. We used this set of $m=1728$ averaged spectra to probe the behavior of the RMS noise $\sigma_y$ as a function of the integration time $\tau_k=10k\tau_0$, for $k=1,\dots,m$.

To measure $\sigma_y(\tau)$, we apply a Savitzky-Golay filter with parameters $d^*$ and $W^*$ to each of the $m$ averages. Then for each $k=1,\dots,m$ we average $k$ filtered spectra and take $\sigma_y(\tau_k)$ to be the sample standard deviation of all bins in this $k$-spectrum average. We expect $\sigma_y(\tau)=1/\sqrt{\Delta\nu_b\tau}$ -- formally  $\sigma_y=\sigma^\text{p}$ considered as a function of the integration time $\tau$; we call this quantity $\sigma_y$ in analogy to the Allan deviation, a time-domain measure of the dependence of the RMS noise on $\tau$. The measured values of $\sigma_y(\tau_k)$ (plotted in Fig.~\ref{fig:avar}) exhibit this expected behavior out to at least $\tau=24$~hours. 

%

\end{document}